\documentclass[aps,prd,floatfix,10pt,nofootinbib,notitlepage]{revtex4-1}
\usepackage{hyperref}
\hypersetup{
pdfpagemode=UseOutlines,      
pdfstartview=Fit,             
pdffitwindow=true,            
pdfpagelayout=TwoColumnsRight,
pdftoolbar=true,              
pdfmenubar=true,              
bookmarksopen=false,          
bookmarksnumbered=true,       
colorlinks=true,              
pdfauthor={Florent Michel, Renaud Parentani},     
pdftitle={},    
pdfcreator=PDFLaTeX,          %
pdfproducer=PDFLaTeX,         %
linkcolor=blue,               
urlcolor=blue,                
anchorcolor=black,            
citecolor=black,              
frenchlinks=false,             
pdfborder={0 0 0}             
}

\usepackage{graphicx}		
\usepackage[caption=false]{subfig} 
\usepackage{amsmath,amssymb,amsfonts,amsthm}
\usepackage[utf8]{inputenc}   
\usepackage[english]{babel}
\usepackage{color}

   \usepackage{multido}

\newcommand{\nn}{\nonumber\\}
\newcommand{\be}{\begin{eqnarray}} 
\newcommand{\ee}{\end{eqnarray}}
\newcommand{\om}{\ensuremath{\omega}}

\newcommand{\pd}{\ensuremath{\partial}}

\newcommand{\lp}{\ensuremath{\left(}}
\newcommand{\rp}{\ensuremath{\right)}}

\newcommand{\Fmax}{\ensuremath{F_{\rm max}}}
\newcommand{\ommin}{\ensuremath{\om_{\rm min}}}

\newcommand{\eq}[1]{Eq.~(\ref{#1})}

\newcommand{\Fig}[1]{Fig.~\ref{#1}}
\newcommand{\fig}[1]{Fig.~\ref{#1}}

\newcommand{\beq}{\begin{equation}}
\newcommand{\eeq}{\end{equation}}
\newcommand{\bsub}{\begin{subequations}}
\newcommand{\esub}{\end{subequations}}
\newcommand{\bi} {\begin{itemize}}
\newcommand{\ei} {\end{itemize}}
\newcommand{\ben} {\begin{enumerate}}
\newcommand{\een} {\end{enumerate}}
\newcommand{\bmat} {\begin{pmatrix}}
\newcommand{\emat} {\end{pmatrix}} 
\newcommand{\bal} {\begin{aligned}}
\newcommand{\eal} {\end{aligned}}
\newcommand{\btab}{\begin{tabular}}
\newcommand{\etab}{\end{tabular}}

\begin{abstract}
We study and numerically compute the scattering coefficients of shallow water waves blocked by a stationary counterflow. 
When the flow is transcritical, the coefficients closely follow Hawking's prediction according to which black holes should emit a thermal spectrum. We study how the spectrum deviates from thermality when reducing the maximal flow velocity, with a particular attention to subcritical flows since these have been recently used to test Hawking's prediction. 
For such flows, we show that the emission spectrum is strongly suppressed, and that its Planckian character is completely lost. For low frequencies, we also show that the scattering coefficients are dominated by elastic hydrodynamical channels. Our numerical results reproduce rather well the observations made by S. Weinfurtner {\it et al.} in the Vancouver experiment. Nevertheless, we propose a new interpretation of what has been observed, as well as new experimental tests. 

\end{abstract}

\begin{document}
\selectlanguage{english} 

\title{Probing the thermal character of analogue Hawking radiation for shallow water waves?}

\author{Florent Michel}
\email{michel@clipper.ens.fr}
\affiliation{Laboratoire de Physique Th\'eorique, CNRS UMR 8627, B\^atiment 210, Universit\'e Paris-Sud 11, 91405 Orsay Cedex, France}
\author{Renaud Parentani}
\email{renaud.parentani@th.u-psud.fr}
\affiliation{Laboratoire de Physique Th\'eorique, CNRS UMR 8627, B\^atiment 210, Universit\'e Paris-Sud 11, 91405 Orsay Cedex, France}

\date{\today}

\maketitle

\section{Introduction}

According to Unruh's Letter~\cite{Unruhprl81} {\it Experimental Black Hole Evaporation?} it should be possible to use fluids to test the Hawking prediction~\cite{Hawking75} that black holes spontaneously emit a steady thermal flux. This remark rests on the fact that the wave equation governing the propagation of long wave length density perturbations in an inhomogeneous flow has the form of the d'Alembert equation in a curved space-time metric. As a result, in a transonic stationary flow, i.e., when the velocity $v$ of the flow crosses the speed of low-frequency waves $c$, the wave equation is identical to that of a scalar field in a black hole metric. Therefore the coefficients governing the scattering of density perturbations should show the mode amplification which is at the root of the Hawking effect. However, this strict correspondence breaks down due to the fact that the scattering involves short wave length modes~\cite{Jacobson-prd91}, the propagation of which is dispersive and thus no longer governed by the d'Alembert equation. To identify what could be the consequences of such dispersive effects, Urunh~\cite{Unruhprd95} numerically solved a dispersive wave equation which governs the propagation in an analogue black hole flow. When there is a neat separation between the short dispersive length scale and the surface gravity scale which fixes the Hawking temperature, he found no significant deviation of the spectral properties of the scattering coefficients. This second work therefore indicates that one may experimentally test the Hawking prediction in dispersive media, when some conditions are met.

This analogy is not restricted to density perturbations in fluids. In fact, surface waves propagating on top of a water flow in a flume also constitute a nice example~\cite{Schutzhold-U_2002}. Following this work, several experiments have been recently conducted to observe the conversion of shallow water waves (i.e., long wave lengths) into deep water waves (i.e., short wave lengths) which occurs near a blocking point~\cite{Nice,Unruh2010}. This process is the time reversed of the Hawking one, and the effective space-time metric near the blocking point is that of a white hole. To have a close analogy with black hole physics, the background flow that engenders this metric should be transcritical, {\it i.e.} the flow velocity $v$ should cross $c$. In the hydrodynamic language, the Froude number $F = v/c$ should become larger than 1. However in the experiments~\cite{Nice,Unruh2010}, the flows apparently possessed “no phase velocity horizon,” as they were globally subcritical. Yet, some mode conversion was clearly observed. In addition, when measuring the relative amplitudes of the scattered waves for different frequencies, Weinfurtner \textit{et al.} observed a “thermal law” in agreement with Hawking's prediction. To understand these observations, dispersion must play an important role. As a result, one significantly distances oneself from the relativistic settings Hawking used. 

Following~\cite{Unruhprd95} the consequences of short distance dispersion have received a lot of attention~\cite{Broutetal95,CorleyJac96,Corley97,Rivista05, MacherB/W,Scott-review, ACSFRP}, and by now there is a fair understanding of the spectral deviations due to dispersion when the flow is {\it significantly} transcritical. 
Comparatively, much less attention has been devoted to the cases where $F$ barely crosses 1, or does not cross it at all. 
In~\cite{finazziRP-proceedings}, it was shown that the Planckianity of the spectrum is progressively lost when $F$ barely crosses 1. When $F$ no longer crosses 1, it was also found that there is a new critical frequency $\omega_{\rm min}$ below which a new scattering channel opens up, and above which the spectrum closely resembles that found when $F$ barely crossed 1. These results have been derived with a superluminal dispersion, as that found in atomic Bose gases, but because of the symmetry between sub- and superluminal dispersion, see Sec.~III~E in \cite{ACRPFS}, they also apply to subluminal dispersion, as can be verified~\cite{Scott-thesis,Scott-review}.

The main objective of the present work is to complete these analyses by focusing on the class of flows used in the recent experiments~\cite{Nice,Unruh2010} so as to obtain a better understanding of what has been observed. To this end, we first consider monotonic flows in which $F$ either barely crosses $1$, or remains subcritical. We then study the scattering in nonmonotonic flows which either possess a pair of black and white horizons, or where the maximal value of $F < 1$ is reached at the top of an obstacle. The last case is the closest to those realized in~\cite{Unruh2010}, and our numerical results concerning the scattering coefficients closely reproduce what has been observed. However, our analysis also confirms the aforementioned results of~\cite{finazziRP-proceedings,Scott-review} that the Planckianity is lost for these subcritical flows, whereas the authors of~\cite{Unruh2010} observed a ``thermal law.'' This apparent contradiction triggered our interest and the forthcoming analysis. As we shall see, its resolution involves hydrodynamic modes which dominate below the critical frequency $\omega_{\rm min}$.

The effects of dispersion shall be computed in two different manners, along the lines of~\cite{2regimesFinazzi,Scott-review}. First, we numerically obtain the spectral properties in flows where the spatial gradient of the water height $h(x)$ is small when compared to the dispersive length scale, i.e., $\partial_x h \ll 1$. Second, by algebraic techniques, i.e., mode matching, we compute the Bogoliubov coefficients in the steep regime where the water depth is piecewise constant. Even though 
this regime is \textit{a priori} very far from the experimental setups, we shall establish that it governs some of the spectral properties in the zero-frequency limit of smooth profiles. The background flows shall also be described at two different levels. In most of this work, for simplicity and clarity, we work with water height profiles $h(x)$ chosen from the outset. Amongst these, we shall briefly consider profiles that are modulated by an undulation~\cite{Coutant_on_Undulations}, i.e., a zero-frequency mode with a large amplitude, since these were systematically observed in~\cite{Nice,Unruh2010}. We shall see that the main properties of the scattering coefficients are not significantly affected by this additional feature of the background flow. In Appendix~\ref{sub:NL}, we study profiles which result from integrating the nonlinear hydrodynamical equations. We shall see that the resulting spectra closely resemble those obtained by the first approach, thereby justifying it \textit{a posteriori}. 

A word of caution is perhaps necessary to conclude this Introduction. Our treatment is based on two main approximations: that of an ideal and irrotational fluid, and that based on a low-order expansion of the dispersion relation. To estimate the errors induced by these approximations is not an easy task, as it would require a precise description of the background flow, including the effects of viscosity and vorticity, and using as well the full dispersion relation, perhaps including surface tension, see~\cite{Germain5, Germain7}. Yet, we believe our description captures the essential aspects of the scattering in the flows of~\cite{Nice,Unruh2010}. We thus expect that its main predictions will be qualitatively correct, in particular, the strong suppression of the low-frequency spectrum.  

The paper is organized as follows. In Sec.~\ref{sec:GPAS}, we present the wave equation and background flows we use. We also compute the critical frequencies which separate the various regimes. In Sec.~\ref{sec:SP}, we solve the wave equation numerically for transcritical and subcritical flows, and determine which observables are, or are not, sensitive to the fact that $F$ crosses one.  We conclude in Sec.~\ref{sec:concl}. In Appendix~\ref{sub:NL} we solve the nonlinear hydrodynamic equations to relate the shape of the free surface to that of the obstacle, before solving the wave equation in the resulting flow. Appendix~\ref{App:gradino} is devoted to the steep horizon limit.

\section{General properties and settings} 
\label{sec:GPAS}

In this section, we review the key concepts that enter into the calculation of the scattering coefficients of shallow water waves when they are blocked by a counterflow. Since these concepts are now well established, we shall be rather brief. The reader is invited to consult Refs.~\cite{Schutzhold-U_2002,Unruh2012,Coutant_on_Undulations} for the derivation of the wave equation and the description of its properties. The general properties of the scattering coefficients of dispersive waves in transcritical flows are explained in detail in Refs.~\cite{MacherB/W,Scott-review,ACRPFS}.

\subsection{Wave equation and dispersion relation} 

We consider irrotational laminar flows of an inviscid, ideal, incompressible fluid in an elongated flume. All dependences in the horizontal direction perpendicular to the flow are neglected. The propagation of linear surface waves is governed by~\cite{Schutzhold-U_2002,Unruh2012,Coutant_on_Undulations} 
\be \label{eq:waveeq}
\left[ \left(\partial _t+\partial _xv\right)\left(\partial _t+v\partial _x\right) -i g \partial _x \tanh \left(-i h \partial _x\right)\right] \phi = 0,
\ee 
where $v(x,t)$ is the horizontal component of the flow velocity, $h(x,t)$ the background fluid depth, and $g$ the gravitational acceleration. The field $\phi$ is the perturbation of the velocity potential. 
It is related to the linear variation  of the water depth $\delta h$ through
\be \label{eq:dh}
\delta h(t,x) = -\frac{1}{g}\left(\partial _t+ v \partial _x\right)\phi.
\ee 
For the sake of simplicity, in \eq{eq:waveeq} we neglected the contributions of the vertical velocity of the free surface. This would have the effect of adding to the gravitational acceleration $g$ a term associated with the centrifugal acceleration of a fluid particle at the surface~\cite{Unruh2012,Coutant_on_Undulations}. For the flow of \cite{Unruh2010}, we found that its maximum value is $\sim 0.03 \, g$. For the flows considered in Appendix~\ref{sub:NL}, it is smaller than $\sim 0.01 \, g$. We therefore expect that neglecting this term will not significantly affect the scattering coefficients. 

We also assume that the flow is stationary, so that we can work with (complex) stationary waves $e^{- i\omega t}\phi_\omega(x)$ with fixed laboratory frequency $\om$. We then expand the dispersive term of~\eq{eq:waveeq} to lowest nontrivial order in $h \partial_x$, assuming higher-order terms play no significant role in the determination of the scattering coefficients. When this is the case, these can be correctly obtained by solving
\be \label{eq:om}
\left[ 
\left(-i \omega +\partial _xv\right)\left(-i \omega +v\partial_x\right) 
- g \partial_x 
h \partial_x 
-\frac{g}{3}\partial_x 
\left(h \partial_x \right)^3 
\right]
\phi_\omega 
= 0 .
\ee
Notice that the ordering of $h(x)$ and $\partial_x$ has been preserved. This is important when considering the steep regime limit where $\partial_x h \gg 1$. This interesting limit, where the scattering coefficients can be computed analytically, is considered in Appendix~\ref{App:gradino}. 

As we shall see, key properties of the scattering coefficients rely on the existence of turning points. Their location, and other properties of the geometrical optic approximation, are governed by the dispersion relation associated with \eq{eq:om}:
\be \label{eq:disprel}
\Omega_\om^2 \equiv (\omega -v k_\om)^2 = c^2 k_\om^2 \lp 1- \frac{1}{3} h^2 k_\om^2 \rp,
\ee 
where $c^2 = g h(x)$ gives the local 
group velocity of waves with low wave vector $k_\om(x)$ in the reference frame of the fluid. Similarly, $\Omega_\om(x)$ gives the $x$-dependent frequency in that comoving frame. As in~\cite{MacherB/W,Scott-review,ACRPFS}, we will consider only positive values of $\om$, since the potential $\phi$ of \eq{eq:waveeq} is invariant under complex conjugation.

\subsection{Subcritical and transcritical flow profiles}

\begin{figure}
\begin{center}
\includegraphics[scale=0.6]{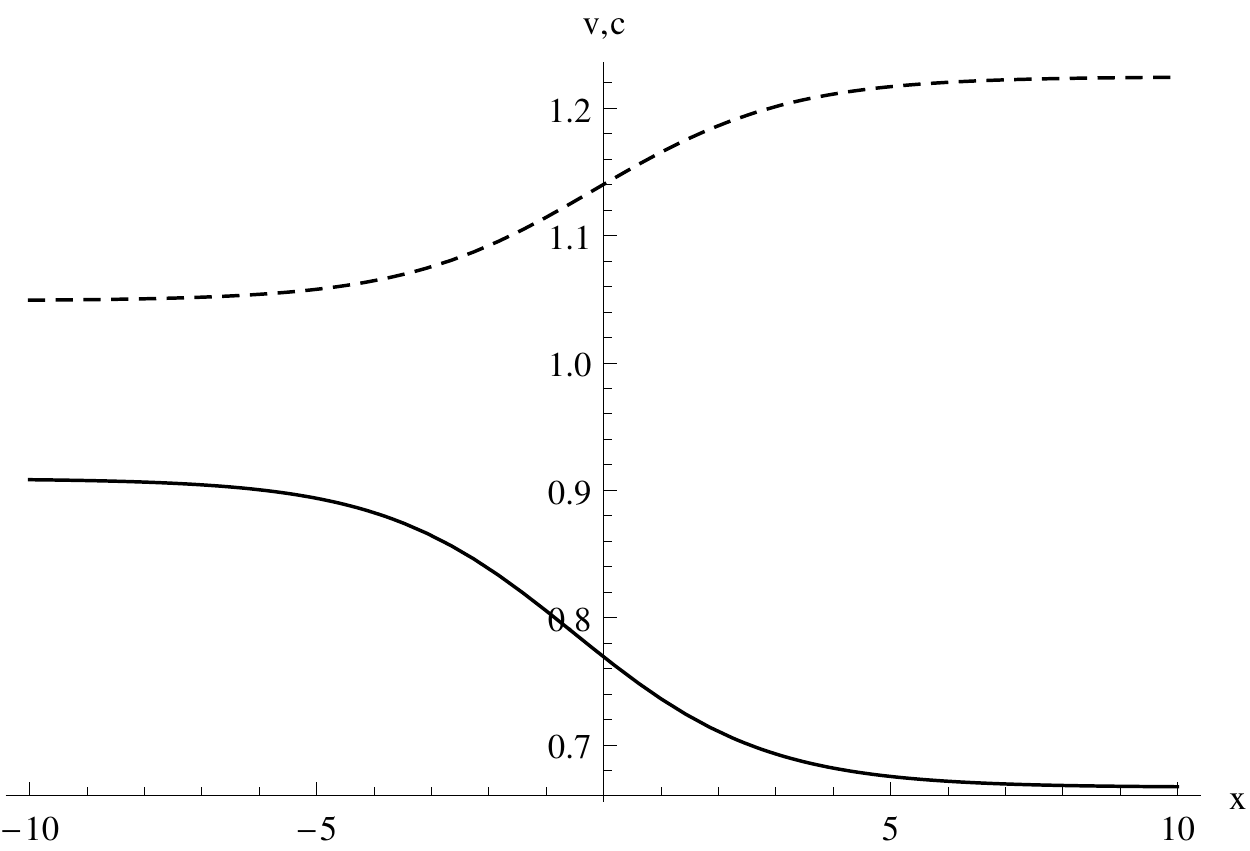}
\includegraphics[scale=0.6]{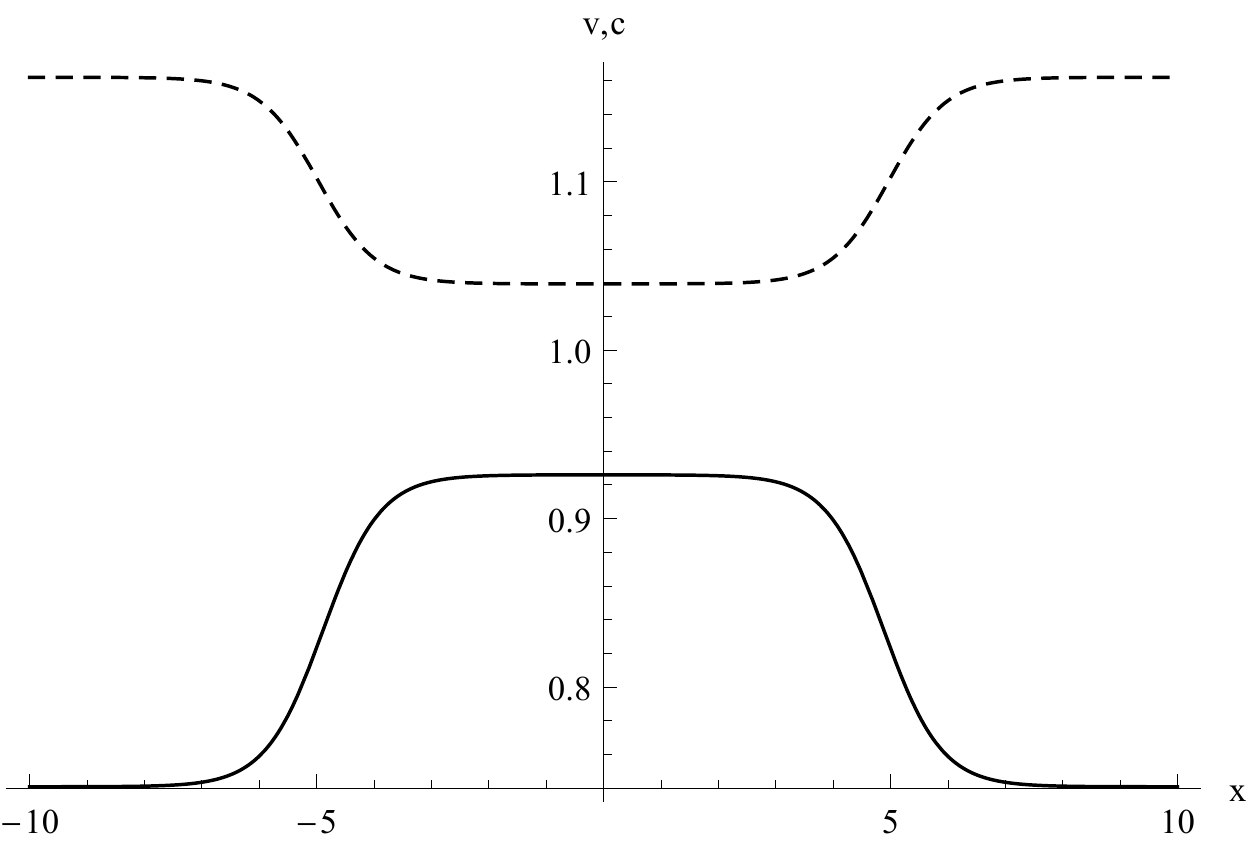}
\includegraphics[scale=0.9]{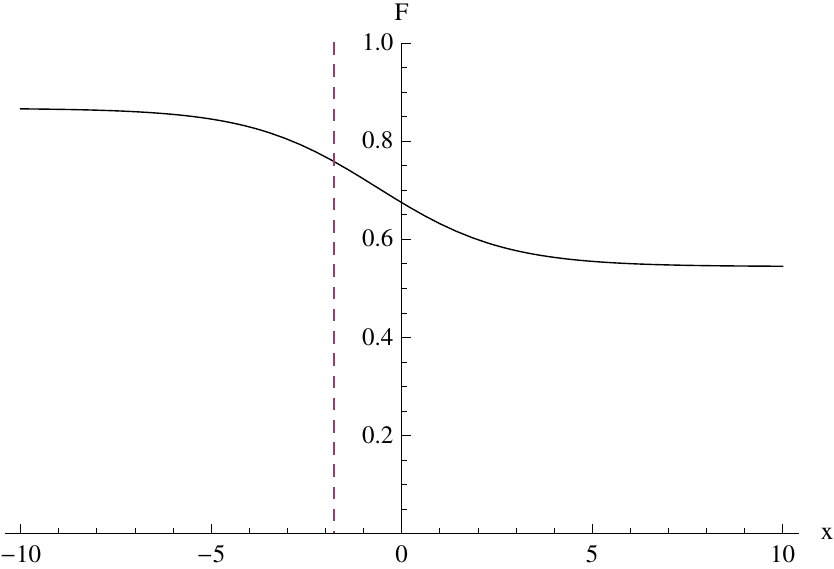}
\includegraphics[scale=0.6]{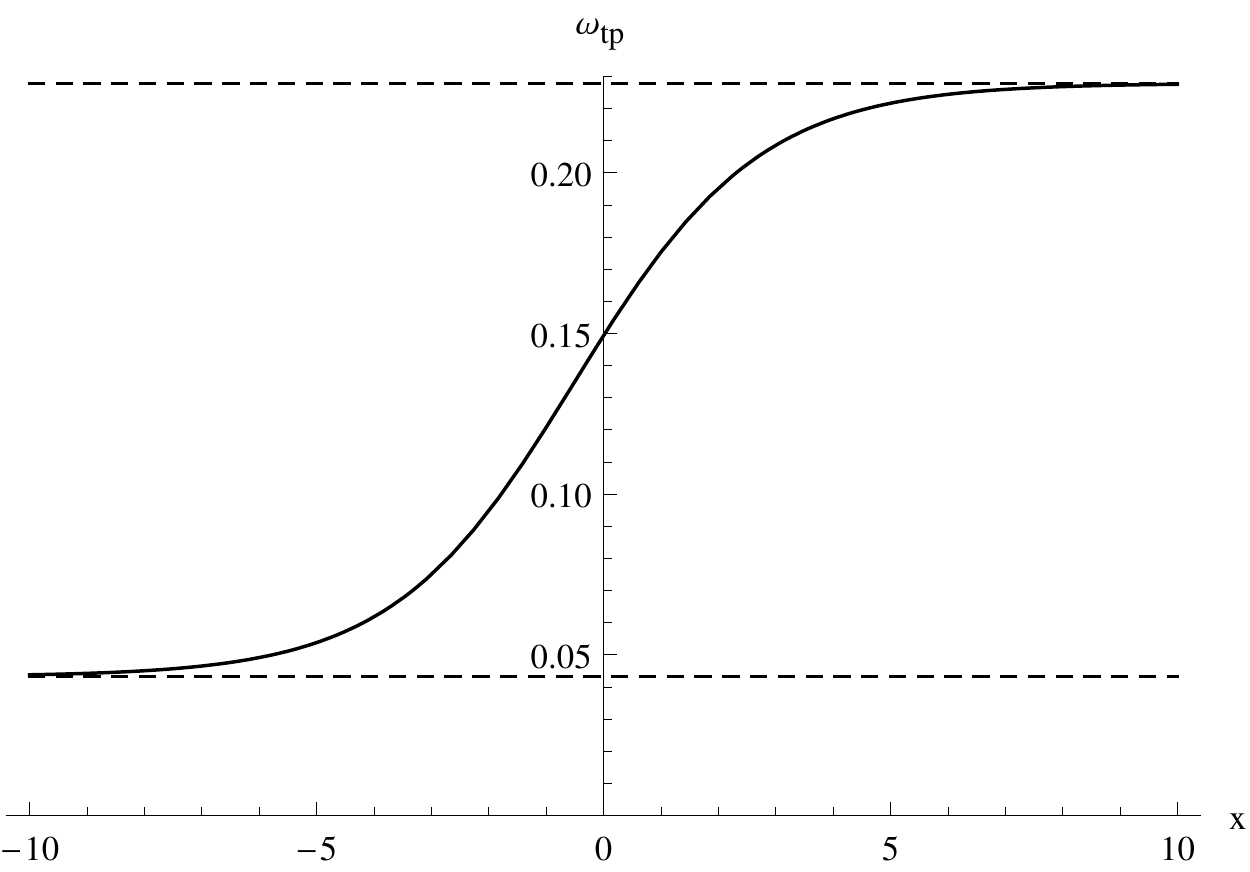}
\end{center}
\caption{
Top: Flow velocity $v$ (plain) and speed of long-wavelength waves $c$ (dashed) as functions of $x$ for a monotonic flow (left), and a nonmonotonic one (right), see \eq{eq:wdepth} and \eq{eq:wdepth_2}. The parameters are  $\sigma = \sigma_1 = \sigma_2 = 0.06$, $L=5$, $h_0 = 1.3$, and $D=0.2$, in units $g=J=1$. Both flows are subcritical since $v < c$. The transcritical cases are similar, except that $v$ and $c$ cross each other, once or twice. Bottom: As functions of $x$, Froude number $F = v/c$ (left), and angular frequency $\om_{tp}(x)$ when the turning point is located at $x$, see \eq{eq:omtp}, (right) for the flow of the top-left panel.  The vertical dashed line in the left plot shows the location of turning point when $\om = 10^{-1}$, (for the characteristics see \fig{fig:modes}). The maximal and minimal values of $\om_{tp}$ indicated by dashed horizontal lines on the right plot give the two critical frequencies of \eq{eq:om-max-min}.}\label{fig:typical_flow_1}
\end{figure}

In this paper, the sign of the flow velocity $v$ is taken positive, so that counterpropagating shallow water waves are coming from the right side. In addition, $v$ decreases in the direction of the flow, as can be seen on both upper plots of \Fig{fig:typical_flow_1}. Hence, when $F = v/c$ crosses 1, counterpropagating waves are all blocked, in analogy to what happens near a white hole horizon. The locus where $F = 1$ is sometimes referred to as a “phase velocity horizon”, as in \cite{Unruh2010}. 

Again for the sake of simplicity, in the body of the text, we use background profiles for the water depth $h(x)$ with a simple analytical description. In Appendix~\ref{sub:NL} we verify that our results remain valid for more complicate profiles which obey hydrodynamical equations. To unravel the various aspects of the scattering, we shall consider two classes of flows. The first class contains flows with monotonic $v(x)$, which are asymptotically constant on both sides. They shall be parametrized by water depths of the form
\be \label{eq:wdepth}
h(x)= h_0 + D \, \tanh \lp \frac{\sigma x}{D} \rp .
\ee
The maximum slope of $h$ is located at $x=0$, and given by ${\rm Max} \, \partial_x h= \sigma$, irrespectively of the parameter $D$ which fixes the asymptotic height change $\Delta h = 2 D$, and the asymptotic values for $x \rightarrow \pm \infty$: $h_{\rm  as.}^{\pm} = h_0 \pm D$. At fixed flux $J$, the profiles of $v$ and $c$ are respectively given by $v(x) = J/h(x)$ and $c(x)= \sqrt{g h(x)}$. Most of our results will be expressed in the system of units where $g = J = 1$. Then the Froude number is simply given by
\be 
F = h^{-3/2}. 
\ee
In these units, a “phase velocity horizon” corresponds to a point where $v = c = h = F = 1$. Notice also that the surface gravity $\kappa_G = \left\lvert \partial_x( c -v)\right\rvert_{v=c}$~\cite{Unruhprl81, Unruhprd95} is here given by 
\be  \label{eq:SG} 
\kappa_G = \left\lvert \pd_x F \right\rvert_{F=1}.
\ee
The monotonic flows of \eq{eq:wdepth} split into two subclasses. For $h_{\rm  as.}^- = h_0 - D < 1$, $F$ crosses 
$1$ and the flow is transcritical, whereas it remains globally subcritical when $h_{\rm  as.}^- > 1$. To study the transition between these two cases, we shall work with highly “asymmetric” profiles, where the minimum value $F_{\rm min} = F (x \rightarrow \infty)$ is always significantly smaller than $1$, whereas the maximum value $F_{\rm max} = F (x \rightarrow -\infty)$ is either slightly above or below 1, see~\fig{fig:typical_flow_1}.

The second class contains nonmonotonic flows where the maximal value of $F$ is reached at $x =0$, and where $F$ is asymptotically constant on both sides. These shall be parametrized by
\be \label{eq:wdepth_2}
h(x) =h_0 + 
D  \tanh \lp \frac{\sigma_1}{D} (x + L) \rp 
\tanh \lp \frac{\sigma_2}{D} (x - L) \rp, 
\ee
where $2 L$ characterizes the spatial extension of the domain where the height $h$ is minimal. When $F$ remains smaller than 1, i.e., $h_0 - D  >1$, these subcritical flows are close to those experimentally realized in Nice~\cite{Nice} and Vancouver~\cite{Unruh2010}.

\subsection{Turning points and characteristics}
\label{Tpc}

We remind the reader why the presence of a turning point directly affects the scattering of shallow water waves. On the one hand, when there is a turning point in a monotonic flow as that of \eq{eq:wdepth}, one of the solutions of \eq{eq:om} becomes exponentially divergent behind the turning point, in the “forbidden region”. On the other hand, scattering coefficients (only) relate the solutions of \eq{eq:om} which are asymptotically bound, i.e., whose modulus remains finite at $x \rightarrow \pm \infty$~\cite{MacherB/W}. As a result, for the flows of \eq{eq:wdepth}, the number of linearly independent asymptotically bound modes is three when there is one turning point, and either four, or only two, when it is absent. For definiteness, in this subsection we assume the flow is monotonic. The discussion also applies to nonmonotonic flows described by \eq{eq:wdepth_2} with minor differences. For instance, quantities evaluated at $x = -\infty$ must then be evaluated where $h$ reaches its minimum value.

Since two waves merge at a turning point, the dispersion relation has a double root, or equivalently the group velocity of the corresponding classical trajectories changes sign. 
Using the quartic law of \eq{eq:disprel}, double roots exist for the values of $\om$ 
\be \label{eq:omtp}
\om_{\rm tp} = \frac{c}{h} 
\sqrt{
\frac{6 \left(1-F^2\right)^3 \left(\left\lvert F\right\rvert + \sqrt{F^2+8}\right)}
{\left(3 \left\lvert F\right\rvert + \sqrt{F^2+8}\right)^3}
},
\ee
where $\om_{\rm tp}$, $c^2 = g h$ and $F = J/(g h^3)^{1/2}$ are functions of $x$ through the profile $h(x)$. We have adopted this writing for $\om_{\rm tp}$ to make clear that $c/h$ plays the role of the high dispersive frequency $\Lambda$ of~\cite{MacherB/W}. Notice also that $\om_{\rm tp}$ no longer exists as a real root when $F > 1$. In this paper, only real positive frequencies will be considered. 

\begin{figure}
\begin{center}
\includegraphics[scale=0.6]{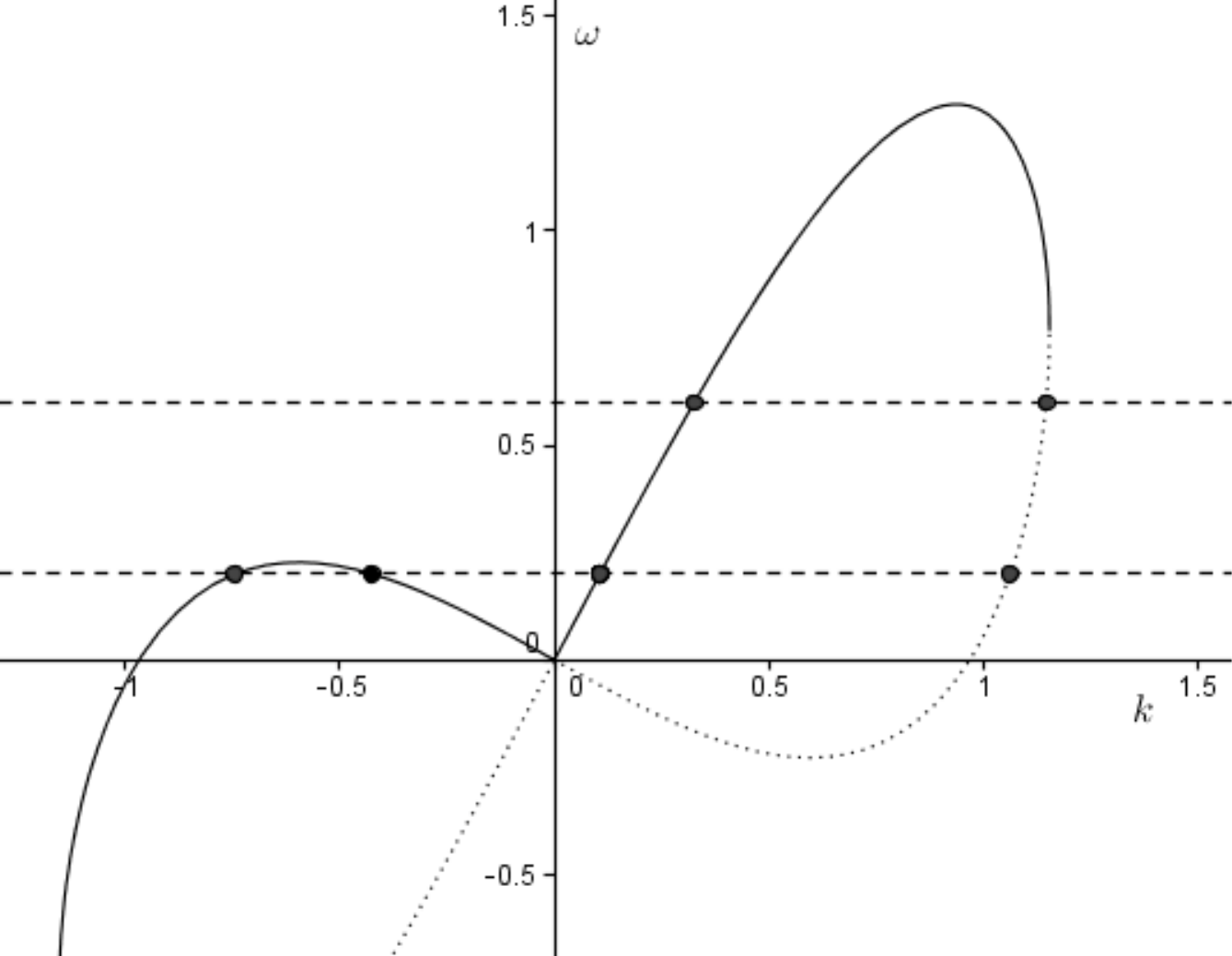}
\end{center}
\includegraphics[scale=0.8]{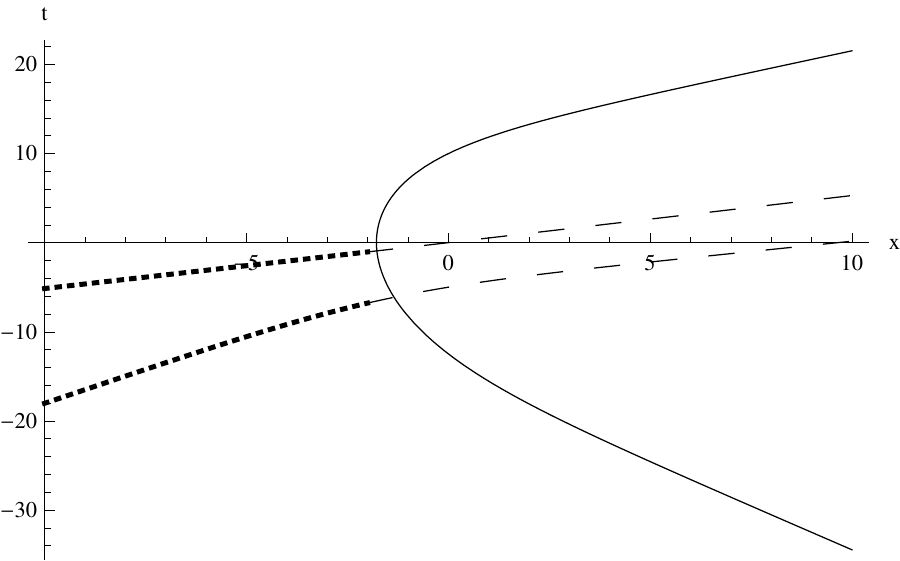}
\includegraphics[scale=0.8]{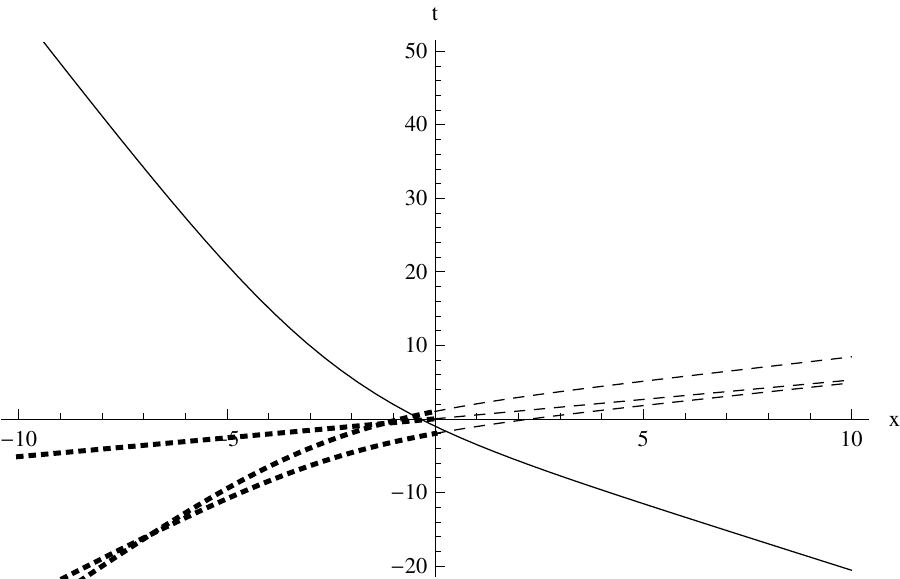}
\caption{
Top: Graphical resolution of \eq{eq:disprel} in the left asymptotic region, for a globally subcritical flow with $F_{\rm max} \approx 0.54$. The solid and dotted lines show the frequency $\om$ as a function of the wave vector $k$, for positive and negative value of the comoving frequency $\Omega$, respectively. The parameters are $g=J=1$ and $h_{\rm as}^- = 1.5$. Dashed lines show two values of $\om$ below and above $\om_{\rm min}$. Large dots show the (real) roots $k_\om$ of the dispersion relation at fixed $\om$. Bottom: Characteristics for $\om_{\rm min} < \om = 10^{-1} < \om_{\rm max}$ (left) and $\om = 10^{-2}< \om_{\rm min}$ (right). The water depth is given by \eq{eq:wdepth} with $h_0=1.3$, $D=0.2$, and $\sigma=0.06$. The solid line describes the trajectory of the low momentum incoming root propagating initially against the flow from the right side. On the left panel, there is a turning point, whereas on the right one there is none. The dashed lines indicate that these asymptotic waves are produced by the incoming mode of \eq{eq:Btrans} (left) and \eq{eq:Bsub} (right), whereas the dotted lines indicate that the waves are absent.} 
\label{fig:modes}
\end{figure}

Given $\omega$, \eq{eq:omtp} implicitly gives the location of the turning point $x_{\rm tp}$ through
\be \label{eq:xtp}
\om_{\rm tp}(x_{\rm tp}) = \om.
\ee
For monotonic flows, the minimum and maximum values of $F$ are $F_{\rm min/max} = F (x \rightarrow \pm \infty)$. Hence, for subcritical flows, the maximum and minimum values of $\om_{\rm tp}$ respectively are
\be \label{eq:om-max-min}
\om_{\rm max} &= \om_{\rm tp}(x = \infty), 
\quad
\om_{\rm min} &= \om_{\rm tp}(x = - \infty),
\ee
as clearly seen in the lower right panel of \Fig{fig:typical_flow_1}. (We can treat the subcritical and transcritical cases together by setting $\om_{\rm min} = 0$ for transcritical flows.) When $\om_{\rm min} < \om < \om_{\rm max}$, the trajectory associated with the left-moving root $k_\om^\leftarrow$ is blocked at the locus given by \eq{eq:xtp}, see \Fig{fig:modes}, lower left panel. (For more details about the calculation of these trajectories which satisfy Hamilton's equations, see~\cite{Broutetal95,Rivista05,ACRPFS}.) For frequencies  higher than $\om_{\rm max}$, there are only two real roots of the dispersion relation, see \Fig{fig:modes}, and thus only two modes. This high frequency regime will no longer be considered as it presents no relationship with the Hawking effect. 
Instead the low-frequency regime $0 < \om < \om_{\rm min}$ is much more interesting. It should be pointed out that this regime only exists when the flow is subcritical, as the root of \eq{eq:omtp} is real only for $F < 1$. In this regime, the four real roots $k_\om$ define four trajectories which are followed by the corresponding waves packets (in the WKB approximation), see \Fig{fig:modes}, bottom right panel.  

Figure~\ref{fig:modes} upper panel shows a graphical resolution of \eq{eq:disprel} in the left asymptotic region, in a subcritical flow, for two frequencies $\om$ above and below $\om_{\rm min}$. The latter is given by the lowest horizontal tangent to the curve of $\om(k)$. (For a transcritical flow, the plain and dotted curves would be more tilted so that this horizontal tangent disappears.) The solid lines correspond to positive values of comoving frequency $\Omega$ of \eq{eq:disprel}, and the dotted lines to negative values. Below $\om_{\rm min}$, the largest root $k_\om$ is the only one which lives on the negative branch. As we shall see below, the norm of the corresponding mode $\phi_\om$ has the opposite sign as that of the three other waves. Above $\om_{\rm min}$, only two real roots exist. A similar plot can be drawn for the right asymptotic region. In that case, the horizontal tangent defines $\om_{\rm max}$. 

In the lower panels of \Fig{fig:modes}, we have represented the characteristics in the monotonic subcritical flow of \Fig{fig:typical_flow_1} for $\om_{\rm min} < \om < \om_{\rm max}$ (left) and $0< \om < \om_{\rm min}$ (right). On the left, there are three characteristics, and hence three linearly independent modes. Two of them are co-propagating (in the lab frame) and the third one has a turning point. On the right, there are four characteristics (and hence four linearly independent modes) because there is no turning point. Three of them are co-propagating and the fourth one is counterpropagating. The use of dotted and dashed lines schematically represents the scattering process of a counterpropagating wave packet (represented by a continuous line) with a nearly well-defined frequency $\om$ sent from $x = \infty$: the right moving modes are initially absent (dotted) but are populated by the scattering (dashed). When the flow is monotonic the reflection of the wave packet (left panel) is total, whereas it is only partial when it is nonmonotonic as some wave can tunnel through the effective potential. Notice that both the left and the right cases of~\fig{fig:modes} are relevant for interpreting the observations of~\cite{Nice,Unruh2010}.

\subsection{Scattered modes} 
\label{sub:sm}

The above geometrical optic properties are reflected in the properties of the various modes $\phi_\om$, solutions of \eq{eq:om}. In this work, we shall be interested in the scattering coefficients relating asymptotic modes, since these coefficients should be used to test the Hawking prediction. (For a recent analysis of the local properties of the modes $\phi_\om$, we refer to~\cite{2014arXiv1402.2514C}.) 

The analytical properties of the scattering coefficients stem from the conserved scalar product~\cite{Unruhprd95} associated with \eq{eq:waveeq}. It is given by
\be \label{eq:Kprod}
\lp \phi_1, \phi_2 \rp \equiv i \int \lp \phi_1^* (\pd_t + v \pd_x) \phi_2 - \phi_2 (\pd_t + v \pd_x) \phi_1^* \rp \, dx ,
\ee 
where $\phi_1$ and $\phi_2$ are two solutions. Note that the norm $\lp \phi_1 , \phi_1 \rp$ is not positive definite. This important property allows for mode amplification (sometimes also called over-reflection~\cite{OnOverReflection} or super-radiance~\cite{GeneralSRR}) as positive-norm modes can be amplified alongside with the appearance of negative-norm ones while preserving the total norm. In fact, this over-reflection is at the root of the Hawking effect.

Since the flows we consider are asymptotically constant on both sides, the asymptotic solutions of \eq{eq:om} are plane waves (when the root $k_\om$ is real). When the asymptotic flow is subcritical, for low-frequency, the 4 wave vectors $k_\om$ are real, and the corresponding waves are, for decreasing $k_\om$, 
\begin{itemize}
\item $\phi_\om^{\rightarrow,d}$ is dispersive\footnote{The corresponding $k_\om^{\rightarrow,d}$ exists only because of the subluminal character of the dispersion relation of \eq{eq:disprel}, as can be seen in \fig{fig:modes}, top panel. We shall call a mode {\it dispersive} when this is the case, and {\it hydrodynamic} when its corresponding root still exists in the limit where the dispersion length scale is sent to 0. 
Similarly, we shall call hydrodynamic sector the set of scattering coefficients between two hydrodynamic modes, and dispersive sector the set of coefficients involving at least a dispersive mode.} 
and right moving in the laboratory frame; 
\item $\phi_\om^\leftarrow$ is hydrodynamic, and left moving;
\item $\phi_\om^\rightarrow$ is hydrodynamic, and right moving; 
\item $\lp \phi_{-\om}^{\rightarrow,d} \rp^*$ is dispersive, and right moving\footnote{When using the quartic dispersion relation of \eq{eq:disprel} in 
place of the full one, $\Omega^2 = gk \tanh (hk)$, this root becomes left moving for $\om > \sqrt{3} v / h$. 
To avoid considering this spurious effect, we restrict our analysis to frequencies smaller than $\sqrt{3} v / h$.}. 
\end{itemize}
Unlike the first three waves, the last one has a negative norm. This is in accord with what was observed in \fig{fig:modes} where the corresponding root was found on the negative $\Omega$ branch. In fact it can be easily verified that the sign of $\Omega$ and that of the scalar product match each other exactly.
{As a result, when working with a positive frequency $\om = i \partial_t$, positive-norm modes describe waves carrying positive energy in the lab frame, whereas negative norm modes describe {\it negative energy waves}, see Ref.~\cite{ACRPFS}. The latter exist only in the presence of a counterflow, and their presence signals that the system under study is energetically unstable.} Because the scalar product changes sign under complex conjugation, $\lp \phi_{-\om}^{\rightarrow,d} \rp^*$ is conventionally written as the complex conjugate of a positive-norm one with negative frequency, in virtue of the invariance of \eq{eq:om} under complex conjugation and $\om \to - \om$. 

In addition, each of the above 4 modes possesses a well-defined group velocity given by $v_{\rm gr} = 1/(\partial_\om k_\om)$. As a result, on each side, each can be identified as an incoming $in$ mode (or as a reflected $out$ mode) when $v_{\rm gr}$ is pointing towards (away from) the central region. When $F < 1$ on both asymptotic sides, for $0 < \om < \om_{\rm min}$, this identification applies at $x \rightarrow -\infty$, and at $x \rightarrow \infty$. However, because the flow is inhomogeneous, the modes mix with each other. The four globally defined incoming in modes--defined by the requirement that the asymptotic weights of the 3 other incoming modes vanish--determine 4 different superpositions of the four reflected asymptotic out modes. For instance, the incoming left-moving hydrodynamical mode is defined by the requirement that the 3 right moving waves at $x \rightarrow -\infty$ vanish, see \Fig{fig:modes}. This mode describes the counterflow shallow water waves that have been studied in~\cite{Nice,Unruh2010}. For this reason, we shall only consider this mode in what follows. For a more complete description of the scattering matrix, we refer to~\cite{MacherB/W}.

Using the scalar product to normalize all modes, the scattering of this mode is fully described by 
\be 
\label{eq:Bsub}
\phi_\omega^{\leftarrow, in} \to \alpha_\omega \phi_\om^{\rightarrow,d,out} + \beta_\omega \lp \phi_{-\om}^{\rightarrow,d,out} \rp^*
+ \tilde{A}_\omega \phi_\omega^{\leftarrow, out} + A_\omega \phi_\om^{\rightarrow, out}, 
\ee
where the coefficients obey 
\be \label{eq:4coef}
\left\lvert \alpha_\omega \right\rvert^2 - \left\lvert \beta_\omega \right\rvert^2 + \left\lvert A_\omega \right\rvert^2 + \left\lvert \tilde{A}_\omega \right\rvert^2 = 1. 
\ee
For higher frequencies, i.e., for $\om_{\rm min} < \om < \om_{\rm max}$, or if $F>1$ in the left asymptotic region, the transmitted mode $\phi_\omega^{\leftarrow, out}$ no longer exists in the left asymptotic (forbidden) region. As a result, there are only three independent asymptotically bounded modes~\cite{MacherB/W}.
In this case, \eq{eq:Bsub} becomes
\be 
\label{eq:Btrans} 
\phi_\omega^{\leftarrow, in} \to  \alpha_\omega \phi_\om^{\rightarrow,d,out} + \beta_\omega \lp \phi_{-\om}^{\rightarrow,d,out}  \rp^* + A_\omega \phi_\om^{\rightarrow, out},
\ee
and conservation of the norm now implies
\be \label{eq:3coef}
\left\lvert \alpha_\omega \right\rvert^2 - \left\lvert \beta_\omega \right\rvert^2 + \left\lvert A_\omega \right\rvert^2 = 1.
\ee

In what follows, we shall numerically compute these coefficients, with a particular attention to $\left\lvert \beta_\omega \right\rvert^2$ as this quantity allows us to test the Hawking prediction. Indeed, in quantum settings, the mean occupation number of particles spontaneously emitted (i.e., emitted when the initial state is vacuum), is given by $n_{\om}^{out} = \left\lvert \beta_\omega \right\rvert^2$. 
In the relativistic settings used by Hawking, ignoring the gray body factor~\cite{Page76,Sandro_2014}, 
one finds a Planckian spectrum: $|\beta_\om|^2 =  (e^{\om/T_H} - 1)^{-1}$, governed by the Hawking temperature, 
or better frequency\footnote{The temperature $T$ associated to a frequency 
$\om$ being $T = \hbar\om/k_B $, in units $\hbar = k_B = 1$, one has $T = \om$. In the following, 
the temperatures should be conceived as frequencies.}, $T_H = \kappa/2\pi$, where $\kappa$ is the surface gravity of \eq{eq:SG}.

\section{Numerical analysis and spectral properties}
\label{sec:SP}

Following what was done in~\cite{MacherB/W,2regimesFinazzi}, we wrote a Mathematica code which solves the wave equation \eq{eq:om} and identifies the full set of Bogoliubov coefficients, namely 16 when $0 < \om< \om_{\rm min}$, and 9 when $\om_{\rm min} < \om < \om_{\rm max}$.\footnote{We are grateful to J. Macher and S. Finazzi for providing C++ and Mathematica codes which were an appreciated source of inspiration. We also thank X. Busch for explaining us how the code written by S. Finazzi works.}   
As in these earlier works, the code computes (from right to left) a set of 4 linearly independent solutions of \eq{eq:om}, which are plane waves at the right boundary of the integration domain. For each of these solutions, it then uses the asymptotic values of $\phi_\om$ and its three first derivatives to extract the decomposition of $\phi_\om$ into plane waves at the left of the integration domain. Finally, a direct identification of the incoming and outgoing modes gives the Bogoliubov coefficients of \eq{eq:Bsub} and \eq{eq:Btrans}. When considering nonmonotonic flows of \eq{eq:wdepth_2}, since the asymptotic values of $F$ are equal to each other and smaller than 1, the four modes are plane waves on both asymptotic sides and their identification is straightforward for all frequencies $0< \om < \om_{\rm max}$. For monotonic flows, when there is a turning point, i.e., for $\om_{\rm min} < \om< \om_{\rm max}$, one should work with superpositions of solutions which do not contain the growing mode on the left side, in order to compute the three coefficients of \eq{eq:Btrans}. 

In all cases we have estimated the numerical errors by computing the “unitarity” relations of \eq{eq:4coef} and \eq{eq:3coef}. We present results where the deviations from the relevant equation is smaller than $10^{-5}$. When the coefficient $\beta$, defining the temperature, is smaller than $10^{-3}$, we imposed a better accuracy, so that the estimated relative error on $\beta$ is always smaller than $10^{-2}$. In practice, for all but a few numerical points \eq{eq:4coef} and \eq{eq:3coef} were satisfied to a much better precision, with deviations smaller than $10^{-5} \left\lvert \beta^2 \right\rvert$. As a result, the main sources of imprecision of our results 
are the approximations discussed at the end of the Introduction, and not the numerics. 

\subsection{Transcritical flows}

\subsubsection{Monotonic flows}

\begin{figure}
\begin{center}
\includegraphics[scale=0.7]{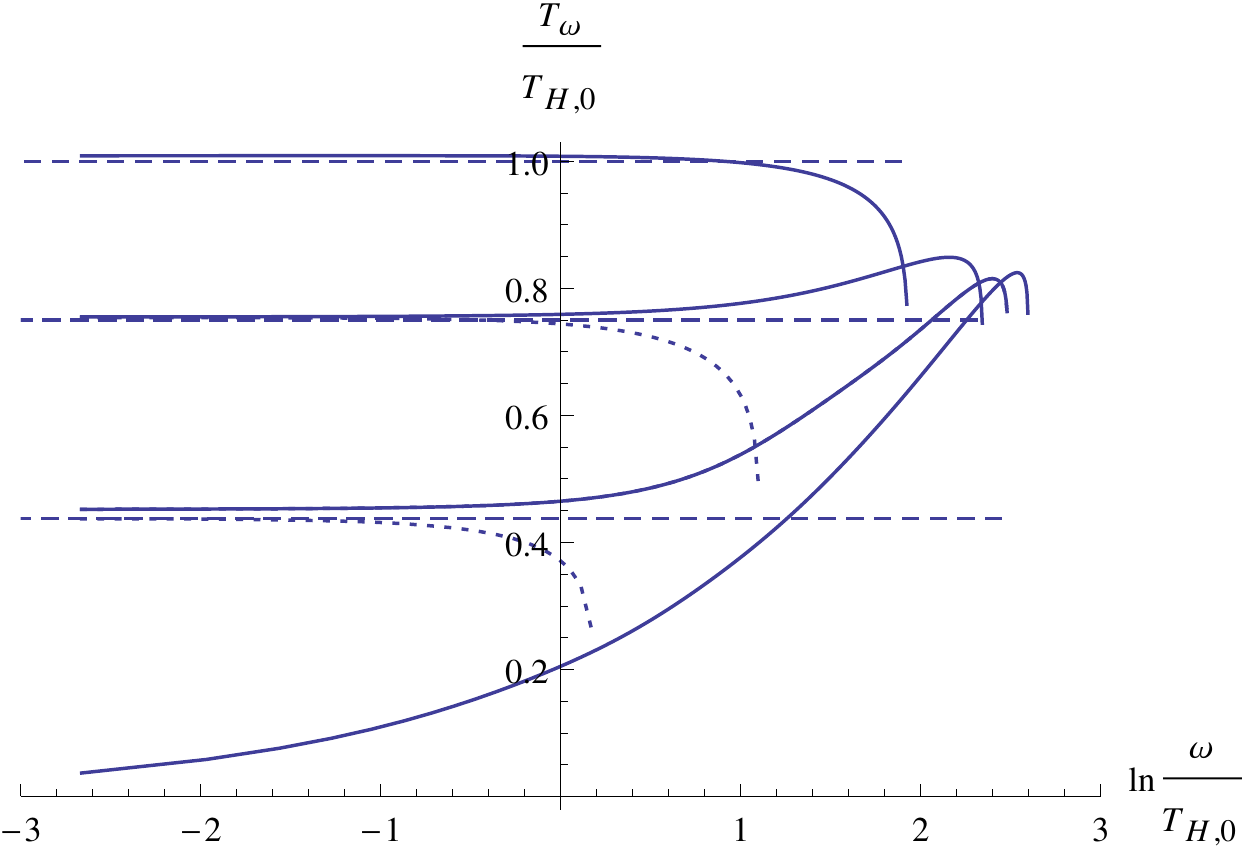}
\includegraphics[scale=0.7]{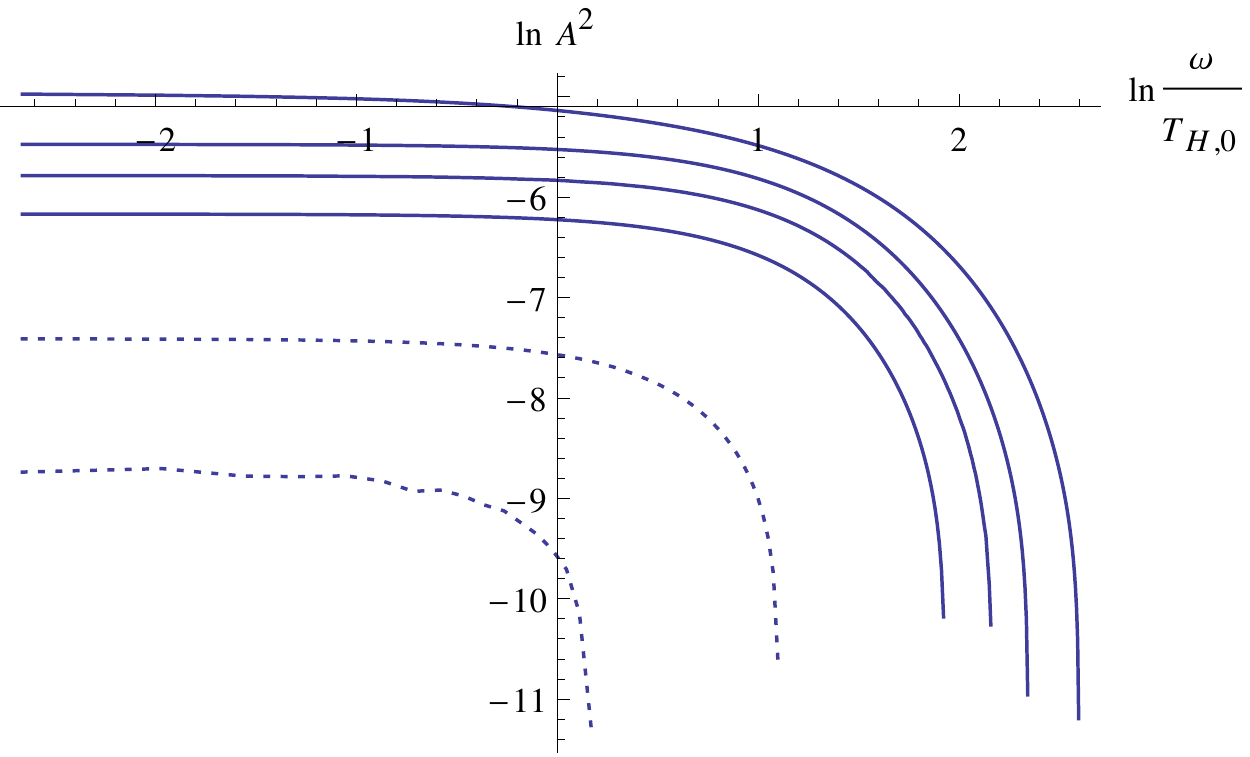}
\end{center}
\caption{Left: Effective temperature $T_\om$ of \eq{eq:effT} as a function of $\ln \om$ for a transcritical flow of the form \eq{eq:wdepth} with fixed values of $\sigma = 0.06$, and $D=0.2$, and 6 values of $h_0$. For the four plain lines, from top to bottom, the values of $F_{\rm max}$ are decreasing, and fixed by $h_0=1, 1.1, 1.15$, and $1.2$. The three horizontal dashed lines give the Hawking temperatures of \eq{eq:TH} for the corresponding flow, and $T_{H,0}$ gives the reference value defined for $h_0=1$. In units where $g=J=1$, one has $T_{H,0} \approx 0.014$. For the last flow, $T_H$ vanishes, as $F_{\rm max}= 1$. The two dotted curves correspond to flows where the maximum Froude number is larger than in the symmetric case $h_0=1$ since the values of $h_0$ are $0.9$, and $0.85$. These two curves have been included to show that the spectra still follow the thermal prediction when increasing $F_{\rm max}$. Right: Logarithm of the transmission coefficient $|A_\om|^2$ of \eq{eq:Btrans} for the same flows.} \label{fig:1}
\end{figure}

When the flow is monotonic and when $F = 1$ is crossed, \eq{eq:Btrans} applies to all frequencies $0 < \om < \om_{\rm max}$, since counterpropagating shallow water waves are blocked irrespectively of their frequency $\om$. In this case, to a good accuracy, one expects to recover the standard results for the emitted flux. To ease the comparison with the Hawking Planckian prediction, we represent, on the left panel of \fig{fig:1} the effective temperature defined by
\be \label{eq:effT}
\left\lvert \beta_\om \right\rvert^2 = \frac{1}{e^{\om/T_\om}-1}.
\ee 
In accordance with the results of \cite{MacherB/W,2regimesFinazzi,Scott-review}, when the maximum value of $F$ is significantly larger than 1, we first observe that $T_\om$ is constant in a large range of adimensional frequencies $\om/T_\om$, i.e., the spectrum is Planckian to a good accuracy. Secondly, we observe that the height of the flat plateau closely follows the Hawking prediction~\cite{Unruhprl81}
\be \label{eq:TH}
T_H = \frac{1}{2 \pi} \left\lvert \partial_x \lp v-c \rp_{v=c} \right\rvert.
\ee
The values of $T_H$ are represented by dashed lines. The agreement confirms that, for low-frequencies with respect to $\om_{\rm max}$, the effective temperature $T_\om$ only depends on the local properties of the flow where $F$ reaches unity. This is the Hawking regime~\cite{2regimesFinazzi}. A closer study reveals that the relative deviations between $\mathop{\rm lim}_{\om \rightarrow 0} T_\om$ and $T_H$ scale as $\sigma^2$, the square of the slope of $h(x)$. This observation is completed by \fig{fig:Scott} of Appendix~\ref{App:gradino} where the validity range of the Hawking regime is established when increasing the slope $\sigma$. 

For all curves in \fig{fig:1}, the range of $\om$ is $\om \in \left\lbrace 0, \om_{\rm max} \right\rbrace$. As can be seen from \eq{eq:omtp} and \eq{eq:om-max-min}, this range shrinks when $h_0$ decreases, i.e., when the minimum Froude number increases. The opposite case where the maximal value of $F$ diminishes and approaches 1 from above is much more interesting. In this limit, we clearly observe that the range of adimensional frequencies $\om/T_\om$ of the flat plateau shrinks as $F_{\rm max} \to 1$. This means that the Planckianity of the spectrum is progressively lost in this limit. We also observe that for high frequencies close to $\om_{\rm max}$, the effective temperature remains approximatively independent of the value of $F_{\rm max}$. (It is worth mentioning the similarity between the present curves and those of those obtained with a superluminal dispersion relation~\cite{finazziRP-proceedings}. The origin of this correspondence is explained in~\cite{ACRPFS}.) In conclusion, the Hawking spectrum is found only if $F_{\rm max}- 1$ is not too small. It would be interesting to determine with more precision the role of $F_{\rm max} - 1 > 0$ in limiting the validity domain of the Hawking prediction. 
Since the primary aim of this work is to study the case $F_{\rm max} < 1$, we leave a precise characterization of this domain to a future study. Interesting results can already be found in \cite{MacherB/W,2regimesFinazzi,finazziRP-proceedings}.

On the right panel of \fig{fig:1}, we plot the squared norm of the coefficient $A_\om$ of \eq{eq:Btrans} which governs the elastic scattering between the incoming mode and the spectator mode $\phi_\om^{\rightarrow, out}$. We observe that $|A_\om|^2 \lesssim e^{-5}$. 
This means that (transmission) gray body factor~\cite{Page76} is close to 1 since $\Gamma_\om^2 \sim 1 - |A_\om|^2 \sim 1$. This is unlike what is found in the case of Schwarzschild black holes, where $\Gamma^2 \propto \om^2$. 
In brief, for transcritical flows, it is legitimate to neglect $|A_\om|^2 $ as it hardly affects the unitarity relation \eq{eq:3coef}. As a result, in this regime, the Planckianity of the spectrum can be studied either from $|\beta_\om|^2$, as shown in \eq{eq:effT}, or from the ratio
\be \label{eq:R}
R_\om \equiv \left| \frac{\beta_\om}{\alpha_\om} \right|^2 , 
\ee
which gives $R_\om \approx e^{- \om/T_\om}$, since $|\alpha_\om|^2\approx 1 +|\beta_\om|^2$.

Instead, when $|A_\om|^2\ll 1$ is no longer satisfied, the meaning of $R_\om$ is no longer clear because $|\alpha_\om|^2 - |\beta_\om|^2 = 1 - |A_\om|^2 \neq 1$. On the contrary, irrespectively of the value of $|A_\om|^2$, $|\beta_\om|^2$ of \eq{eq:Bsub} (or that of \eq{eq:Btrans}) always gives $n^{BH}_\om$,  the mean number of asymptotic particles spontaneously emitted by the corresponding {\it black hole} flow. In fact, as explained in~\cite{MacherB/W}, the emission spectrum of black holes and white holes differ when the S-matrix mixes more than two 
modes.~\footnote{For this reason, it would be interesting to explicitly study the scattering of shallow water waves in black hole flows, as pointed out by S. Robertson.}  

\begin{figure}
\begin{center}
\includegraphics[scale=0.7]{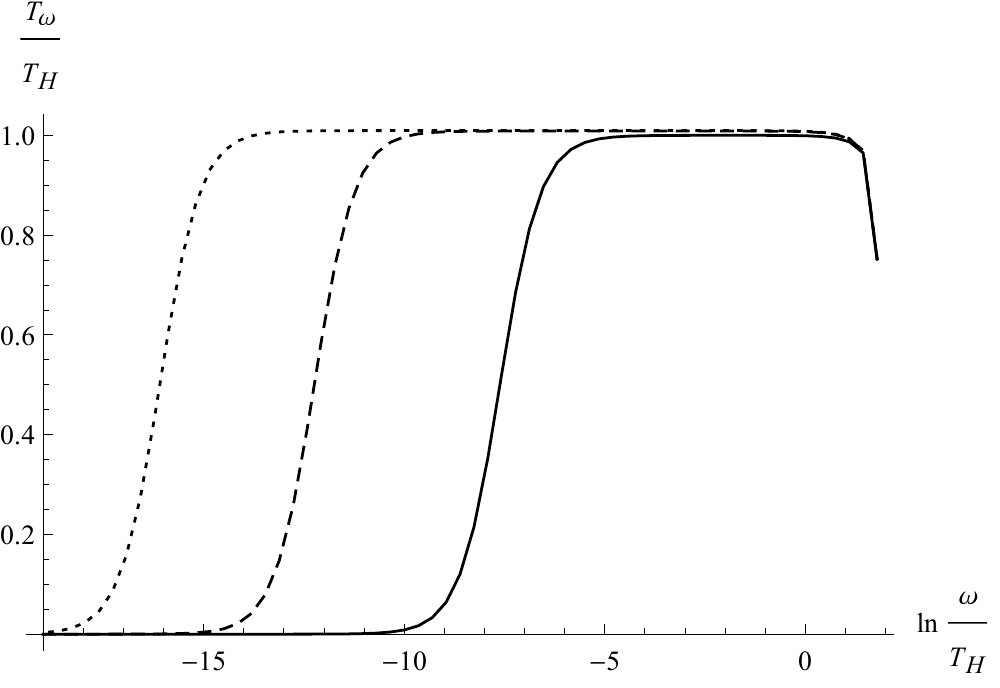}
\includegraphics[scale=0.7]{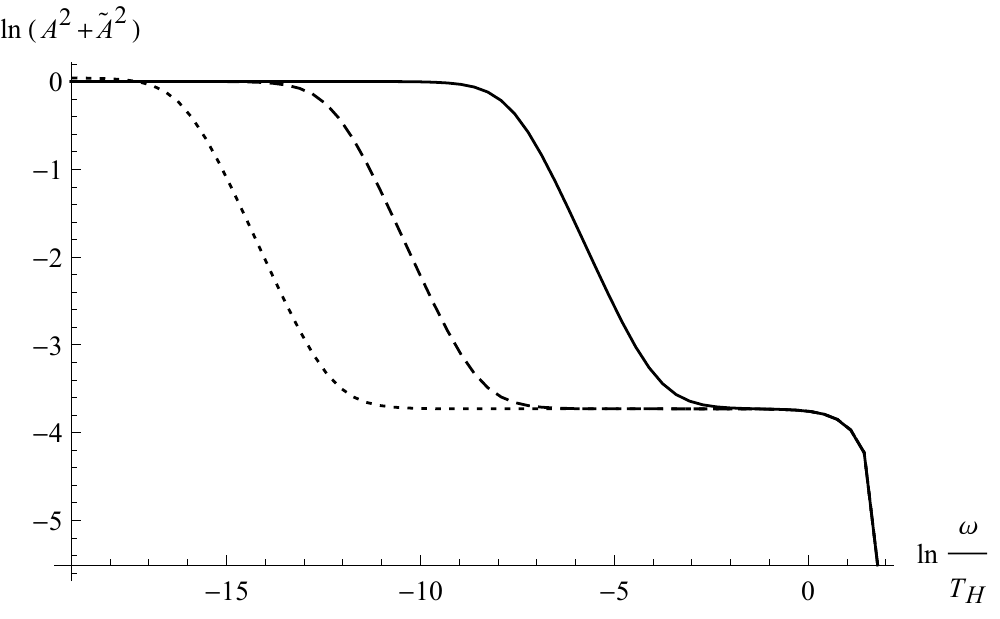}
\end{center}
\caption{Effective temperature (left) and logarithm of the squared hydrodynamic coefficients (right) as functions of the angular frequency $\om$ for a localized obstacle giving a transcritical flow, with $F_{\rm max} \sim 1.2$. (This value has been chosen so that $\om_c$ of \eq{eq:omc} is not too small.) The background water depth is given by \eq{eq:wdepth_2}, with $\sigma_1 = \sigma_2 = 0.03$, $h_0=1.0$, $D=0.1$, and for extensions $L=5$ (solid), $7$ (dashed), and $10$ (dotted), in units where  $g=J=1$. The corresponding values of $\om_c$ are $7.0 \, 10^{-4} T_H,  1.2 \, 10^{-5} T_H$, and $2.6 \, 10^{-8} T_H$ respectively, where $T_H \approx 1.4 \, 10^{-2}$. On both panels, one clearly sees that the finite extension $L$ of the transcritical flow only affects the Planckian behavior of the spectrum for ultra low-frequencies $\om \lesssim \om_c$. In the new regime, the effective temperature vanishes (left panel),  and the hydrodynamical sector completely dominates the scattering (right).} \label{fig:3}

\end{figure}

\subsubsection{nonmonotonic transcritical flows, adding a black hole horizon}
\label{App:aBHL}

We now consider profiles of the second class, see~\eq{eq:wdepth_2}, when $F_{\rm max}$ is significantly larger than 1. In these profiles, there is a second analogue horizon, which is that of a black hole since $v/c$ increases along the direction of $v$ when crossing 1. We shall be rather brief and only focus on the new aspects with respect to the former case which are brought by the presence of this second horizon.

The first important difference stems from the fact that the flows are now asymptotically subcritical on both sides. As a result, the four mode mixing of \eq{eq:Bsub} here applies for all frequencies $0 < \om < \om_{\rm max}$. However, because the flows are transcritical, in the WKB approximation, incoming modes from the right side will still be reflected for all these frequencies. As a result, we do not expect that the presence of the extra transmitted mode $\phi_\omega^\leftarrow$ will significantly modify the scattering coefficients for not too low-frequencies. To verify this, \Fig{fig:3} shows the effective temperature of \eq{eq:effT} and the sum of squared norm of the hydrodynamic coefficients for a background water depth of the form \eq{eq:wdepth_2}, with $F_{\rm max} \approx 1.2$. On the left panel we recover an approximately Planckian spectrum in a wide range of $\om$, which is still bounded by $\om_{\rm max}$ on the right, as was found in the former section. The novel feature concerns the low-frequency regime. The Hawking regime is now bounded from below by 
\be \label{eq:omc}
\om_c \sim \frac{c(0)}{h(0)} (F_{\rm max} - 1) 
(\Lambda_{\text{dec}} h(0))
e^{-2 \kappa_{\text{dec}}L},
\ee 
where the critical inverse length is 
\be
\Lambda_{\text{dec}} \equiv \frac{1}{h (0)} \sqrt{3 \left(F_{\rm max}^2-1\right)}.
\ee
The critical frequency $\om_c$ gives the value of $\om$ below which the existence of the second horizon significantly affects the scattering. The exponential factor in \eq{eq:omc} may be understood as follows. The left moving mode is exponentially decaying in the interhorizon region. Therefore the amplitude of the waves scattered on the black hole horizon back towards the white hole horizon are exponentially suppressed, unless the frequency is exponentially small, as then the matching conditions of modes on the two sides of the horizon gives an additional exponentially large factor. (Notice also that the above formula for $\om_c$ is valid provided $F_{\rm max}$ is significantly larger than unity, so that the exponential factor of \eq{eq:omc} is much smaller than unity, and $\om_c$ exponentially small. In the other limit, when $F_{\rm max}$ decreases and approaches 1, the range of frequencies in which the radiation is thermal shrinks, as $\om_c$ increases.) For $0 < \om \lesssim \om_c$, the coefficients $\alpha_\om$ and $\beta_\om$ both become proportional to $\om^{1/2}$. In addition, the reflection coefficient $A_\om$ vanishes like $\om$. So, in the limit $\om \rightarrow 0$ we have a {\it total transmission} in the hydrodynamic sector, i.e., $|\tilde{A}_\om| \to 1$.

From this brief study, we learned two important things. In the low-frequency limit, the scattering displays a new regime which, first, {\it only involves the hydrodynamic modes}, and, second, where {\it the effective temperature vanishes} because $|\beta_\om|^2$ vanishes too. These two observations will be also found below when studying subcritical flows. However, the suppression of $|\beta_\om|^2$ will be much more significant, as it will apply to a much larger range of frequencies. Indeed, when $F_{\rm max}$ is significantly smaller than $1$, $\om_{\rm min}$ of \eq{eq:om-max-min}, which will play the role of $\om_c$, is  of the same order as $\om_{\rm \max}$, and is thus in general much larger than the typical values of the gradient of $v-c$ which fixes the effective temperature. 

\subsection{Subcritical flows} 
\label{sec:sub}

We now address the main issue of this work, namely what are the properties of the scattering coefficients when $F_{\rm max}$ is lower than 1. As in the former subsection, we first consider monotonic flows described by \eq{eq:wdepth}.

\subsubsection{Monotonic subcritical flows}

To start with, we emphasize that the main modification introduced by considering subcritical flows concerns $\om_{\rm min}$ of \eq{eq:om-max-min}. As already mentioned in Sec.~\ref{Tpc}, for $0 < \om < \om_{\rm min}$, the scattering of incoming shallow water waves now involves four waves as described in \eq{eq:Bsub}.
Instead, for $\om_{\rm min} < \om < \om_{\rm max}$, one recovers the former situation involving only 3 outgoing modes, see \eq{eq:Btrans}.  As a result we expect that the scattering coefficients behave very differently below and above $\om_{\rm min}$. 

\begin{figure}
\begin{center}
\includegraphics[scale=0.9]{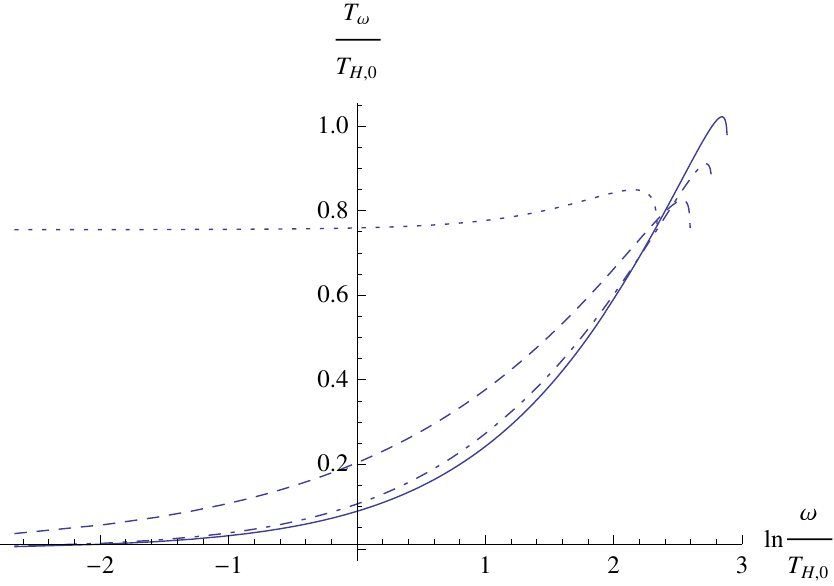}
\includegraphics[scale=0.7]{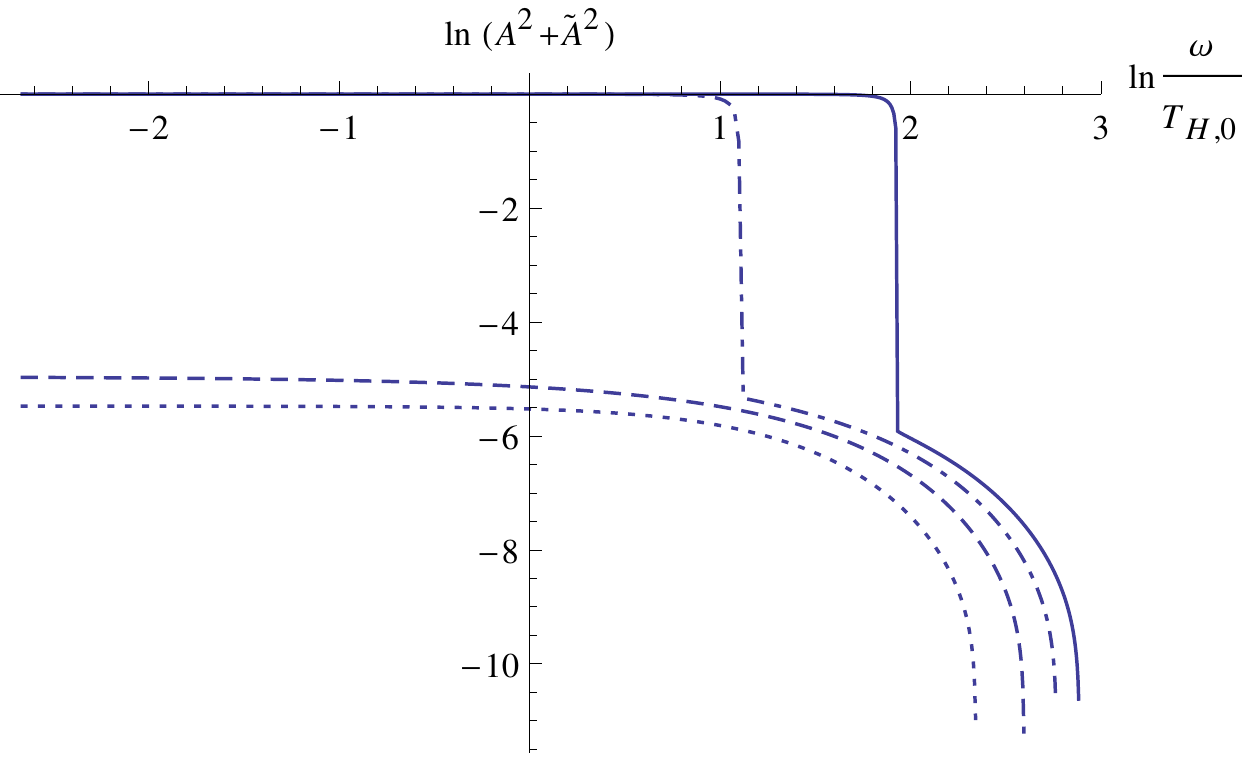}
\includegraphics[scale=0.85]{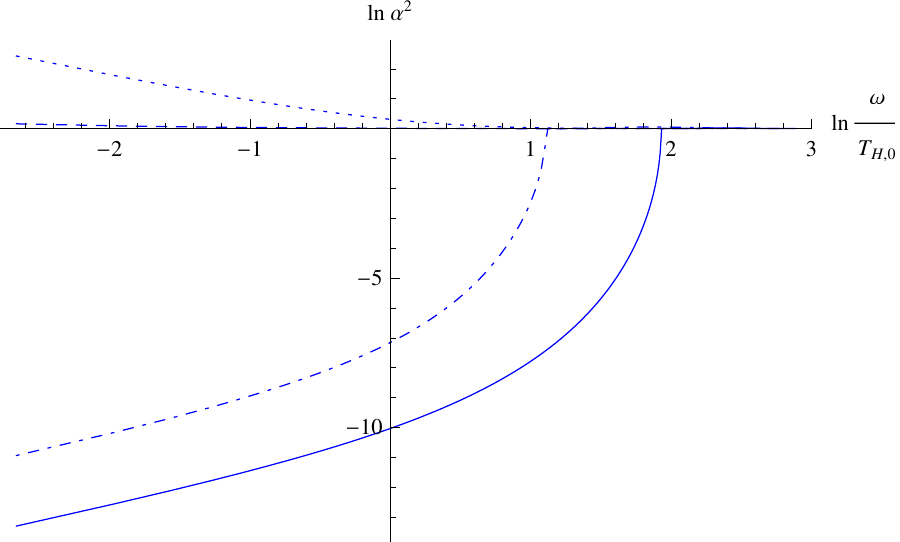}
\includegraphics[scale=0.85]{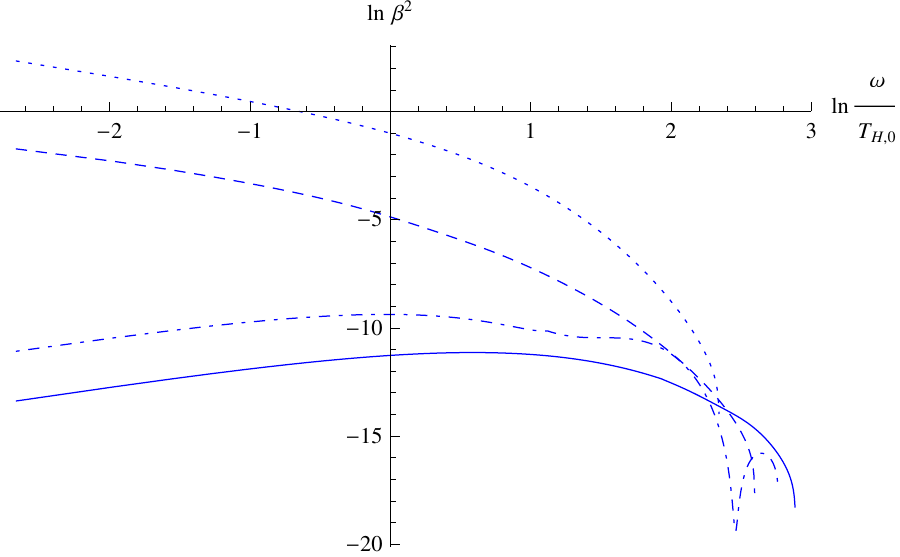}
\end{center}
\caption{Top, left: Effective temperature \eq{eq:effT} for monotonic flows of \eq{eq:wdepth} with $\sigma=0.06$, $D=0.2$, and four different values for $\Fmax$, namely, $0.75$ (solid) and $0.87$ (dot-dashed), both subcritical flows; $1.0$ critical (dashed); and $1.17$ transcritical (dotted). The temperature (frequency) $T_{H, 0}\simeq 0.014$ is here used to ease the comparison with \fig{fig:1}. In the limit $\om \rightarrow 0$, the radical change between sub- and transcritical flows is easily seen, as $T_\om$ goes to zero in the former cases, whereas it remains finite for the latter. Top, right: Logarithm of the sum of the squared transmission and reflection coefficients.  One clearly notices that these coefficients are very small above $\ommin$, i.e., when there is a turning point, but completely dominate below $\ommin$. The two values of $\ln \lp \ommin/T_{H, 0} \rp$ are $2.0$ and $1.1$. 
Bottom: Logarithm of the squared norm of the coefficients $\alpha_\om$ (left) and $\beta_\om$ (right). The sharp transition of $|\alpha_\om|^2$ occurring at $\om = \ommin$ is clearly visible for the two subcritical flows. 
For these flows, one also notices that $| \beta_\om |^2$ remains much smaller than $1$. Hence the scattering is essentially elastic, without significant mode amplification.} 
\label{fig:2}
\end{figure}

This is exactly what can be seen in the upper right panel of \fig{fig:2} where we represented the sum $|\tilde A_\om|^2 + |A_\om|^2$. Since this quantity is equal to $|\alpha_\om|^2- |\beta_\om|^2 - 1$ in virtue of \eq{eq:4coef}, it determines the relative importance of the dispersive and the hydrodynamic sectors. 
(As no counterpropagating wave exists in the left asymptotic region when $\om > \om_{\rm min}$, we extended the definition of $\tilde{A}$ by setting $\tilde{A}(\om > \om_{\rm min})=0$. $\tilde{A}$ is then {\it continuous} across $\om = \om_{\rm min}$.) 
For the two subcritical profiles, we notice a sharp transition which precisely occurs at the corresponding value of $\om_{\rm min}$. 
Above this frequency, the reflexion coefficient $|A_\om|^2$ behaves essentially like for the critical and the transcritical flows, as can also be verified by comparison with the right panel of \fig{fig:1}. This was expected from the fact that, above $\om_{\rm min}$, the characteristics of the modes possess the same structure (given in the left panel of \fig{fig:modes}) whether $\Fmax$ is greater or below 1.  Below $\om_{\rm min}$,  $|\tilde A_\om|^2 + |A_\om|^2$ tends to 1. Hence, the low-frequency regime is completely dominated by the hydrodynamical sector, which moreover is purely elastic, i.e., it involves no mode amplification. This is the first important result of this work. 

This conclusion is corroborated by the upper left panel where we observe that the effective temperature of \eq{eq:effT} vanishes as $\om \to 0$. In addition, we observe that it hardly changes when passing from $\Fmax=1$ down to $0.87$ and $0.75$. In all cases, it displays no flat plateau, which would indicate a Planckian behaviour. We therefore conclude that in these monotonic subcritical flows, the Planckianity that was present for transcritical flows, see \fig{fig:1} left panel, is completely lost. 
In fact the vanishing of the effective temperature reflects something more fundamental: whereas $|\beta_\om|^2$ 
was growing as  $\sim T_H/\om$ in transcritical flows, for subcritical ones, $|\beta_\om|^2$ remains much smaller than $1$ for all frequencies, as can be seen in the lower right panel of \fig{fig:2}. This key observation can be understood as follows. For $\om > \om_{\rm min}$, 
one still finds the exponential suppression (which is typical of nonadiabatic mode mixing~\cite{Massar:1997en}) because $\om_{\rm min}$ is typically much larger that the surface gravity scale, as it is proportional to the dispersive frequency $c/h$, see \eq{eq:omtp}. For $\om < \om_{\rm min}$, there is another mechanism in play: namely the incoming mode is essentially transmitted as there is no turning point. As a result, the deviations from the WKB approximation, which again predicts $\beta_\om = 0$, are therefore also small, hence the smallness of $|\beta_\om|$. 
 
Interestingly, the disappearance of the turning point has even a stronger consequence, namely  {\it both} $|\alpha_\om|^2$ and $|\beta_\om|^2$ vanish like 
\be
\label{eq:bato}
|\beta_\om|^2 \sim |\alpha_\om|^2 \sim \om /\om_b ,
\ee 
as can be seen from the two lower panels of \fig{fig:2}. This is our second important result.
The critical frequency $\om_b$ is found to be roughly proportional to
\be 
\label{omb} 
\om_b \propto \exp\left( \lp \sigma \frac{h_0}{D} \lp F_{\rm max} \rp^{1/3} \rp^{-2} \right), 
\ee
for large and moderate values of $\sigma_1 \approx \sigma_2 = \sigma$, and if $D \ll h_0$. As a consequence, for $\om < \om_b$, the effective temperature behaves as
\be \label{eq:Tomsub}
T_\om \approx -\frac{\om}{\ln \lp \frac{\om}{\om_b} \rp}  \lp 1+ \mathcal{O} \lp \frac{\om}{\om_b} \rp \rp .
\ee 

\begin{figure}
\begin{center}
\includegraphics[scale=0.7]{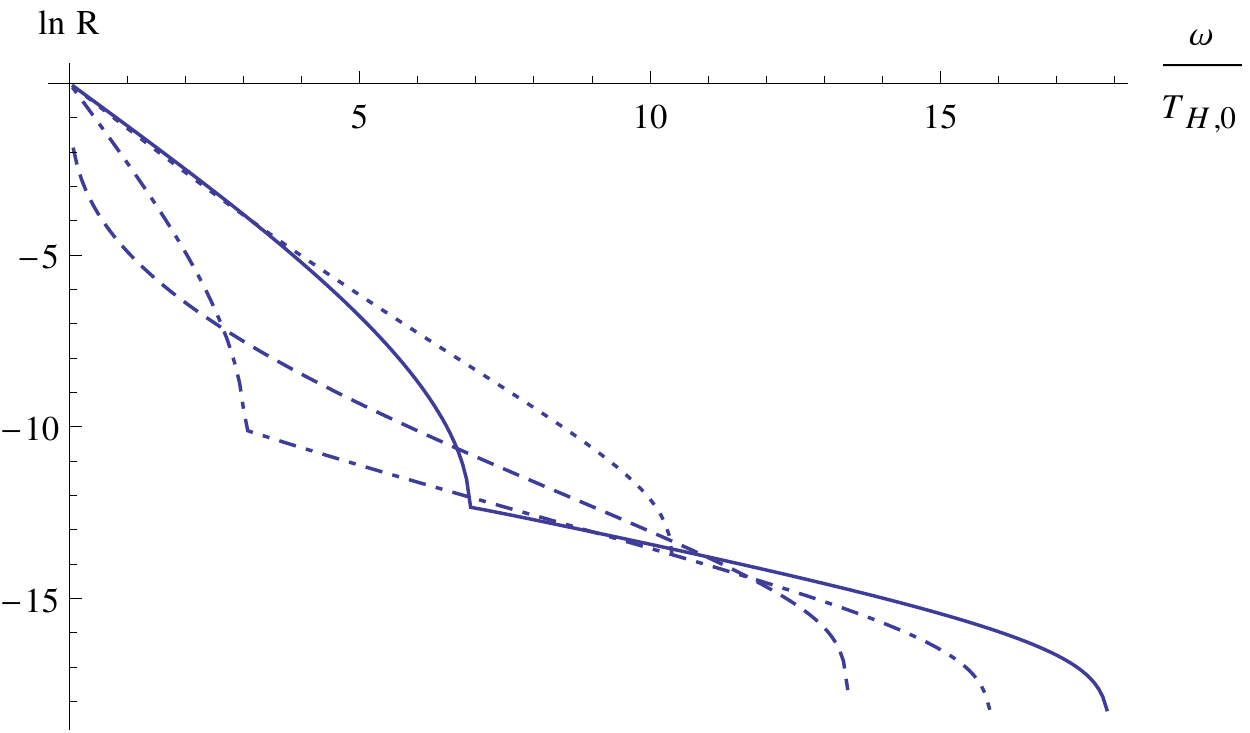}
\includegraphics[scale=0.7]{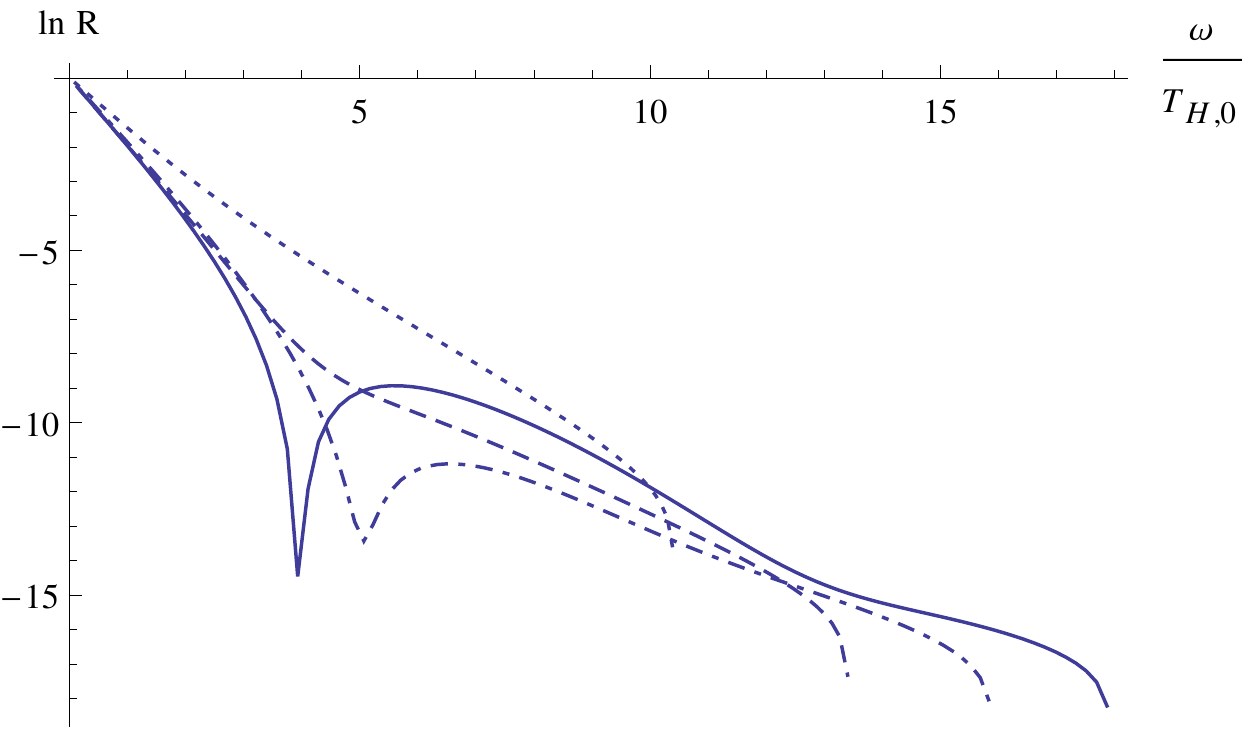}
\end{center}
\caption{Left: Logarithm of the parameter $R$ of \eq{eq:R} for the monotonic flows of \fig{fig:2}, using the conventions of \fig{fig:2} to designate the four cases. Notice that it is roughly linear in both the subcritical and transcritical cases. Notice also that the mean slopes are quite similar, above and below the values of $\om = \om_{\rm min}$ where the two curves corresponding to subcritical flows show a kink. Right: Logarithm of the parameter $R$ for the nonmonotonic flows of \eq{eq:wdepth_2} with the same parameters as those of the monotonic ones, and with $L=4$. We see that the slope of $\ln R$ remains mostly unchanged. The kinks associated with $\om = \om_{\rm min}$ have now disappeared. Notice also that the sharp peaks here are related to a different phenomenon, namely resonance effects in the cavity formed by the two would-be horizons. The slope of $\ln R$ is so robust that only a very limited amount of information can be extracted from it.}
\label{fig:lnR_subsub}
\end{figure}

We thus see that $|\alpha_\om|^2$, $|\beta_\om|^2$, and $ \left\lvert A_\om \right\rvert^2 + | \tilde{A_\om} |^2$ are all highly sensitive to the disappearance of the turning point. Surprisingly, the ratio of \eq{eq:R}, which was used in \cite{Unruh2010}, is not significantly affected by this disappearance, as can be seen in \fig{fig:lnR_subsub}, left panel. In fact the behavior of $R_\om$ is rather similar for the four flows considered in \fig{fig:2}. In particular, the limiting value of $R_\om$ when $\om \rightarrow 0$ is 1 in all cases. This requires some explanation. When $\om < 0$, the roles of $\alpha_\om$ and $\beta_\om$ are exchanged with respect to the case $\om > 0$, because of the symmetry of the wave equation \eq{eq:om} under $\om \rightarrow - \om$, $\pd_x \rightarrow -\pd_x$, which is known as “crossing symmetry”~\cite{PhysRev.130.436}. (This is a general property, see for instance the scattering of light waves on a mirror following a nonuniform trajectory, explained in Sec. 2.5.1 of \cite{Primer}). So, when $\om = 0$ the absolute values of $\alpha_\om$ and $\beta_\om$ must be equal. When there is a horizon, 
they both diverge since $|\beta_\om|^2 \sim |\alpha_\om|^2 \sim T/\om$. When there is none, we numerically observed that for $\om \to 0$,
they both vanish as $|\beta_\om|^2 \sim |\alpha_\om|^2 \sim \om /\om_b $. Hence in {\it both} cases $\ln R_\om$ is indeed linear for small values of $\om$ (if one assumes $|\beta_\om|^2 /|\alpha_\om|^2$ possess a regular Taylor expansion). It should be noticed that \eq{eq:bato} is compatible with the unitarity relations of \eq{eq:4coef}, or \eq{eq:3coef}, {\it precisely} because the hydrodynamic sector dominates the scattering ($|A_\om|^2 + |\tilde A_\om|^2 \to 1$) in the small-frequency limit. In brief, there is no contradiction between a ``Boltzmanian'' $\ln R_\om \propto - \om$ and \eq{eq:bato}. This offers a solution to the apparent contradiction (mentioned in the Introduction) between the observations of \cite{Unruh2010}, where $\ln R_\om \propto - \om$ was observed at small $\om$, and the results \cite{finazziRP-proceedings} which established that the asymptotic spectrum is nonPlanckian, 
as they follow \eq{eq:bato}.
~\footnote{{\it A priori}, \eq{eq:bato} could have been explained by some gray body factor $\Gamma_\om$. 
Indeed, for Schwarzschild black holes in four dimensions, for $\om \to 0$, one gets $\Gamma_\om \propto \om^2$~\cite{Page76}, which also gives that the asymptotic coefficient scales as $|\beta_\om|^2\sim \om$, without affecting the thermality of the Hawking radiation. For analogue white hole flows, we do not think this explanation could work because $ \phi_{-\om}^{\rightarrow,d,out}$ of \eq{eq:Bsub} cannot be elastically reflected as it is the only negative energy mode. We are grateful to W.~Unruh for interesting discussions about this issue.}


\subsubsection{nonmonotonic subcritical flows, generalities}

\begin{figure}
\begin{center}
\includegraphics[scale=0.6]{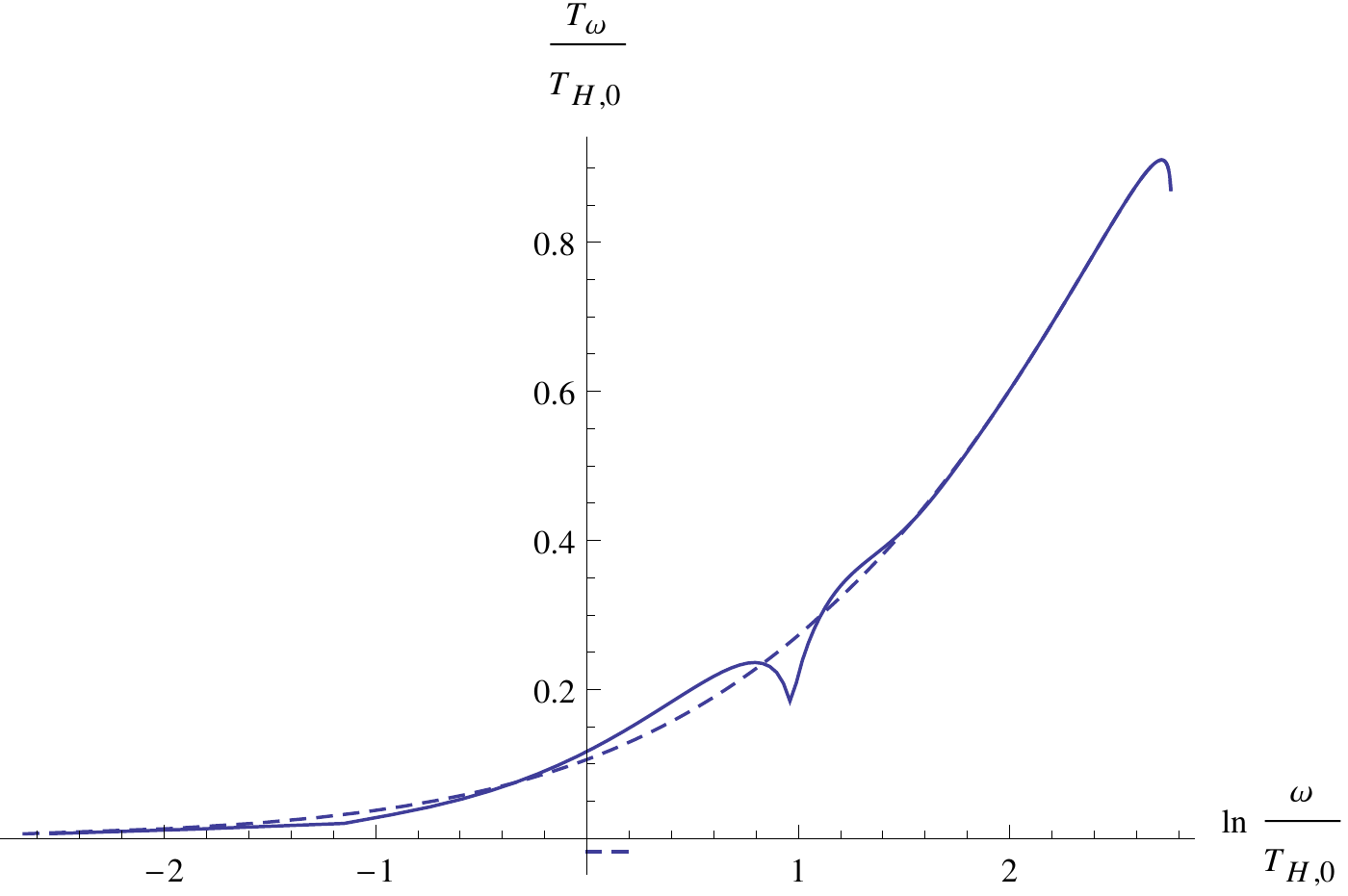}
\includegraphics[scale=0.7]{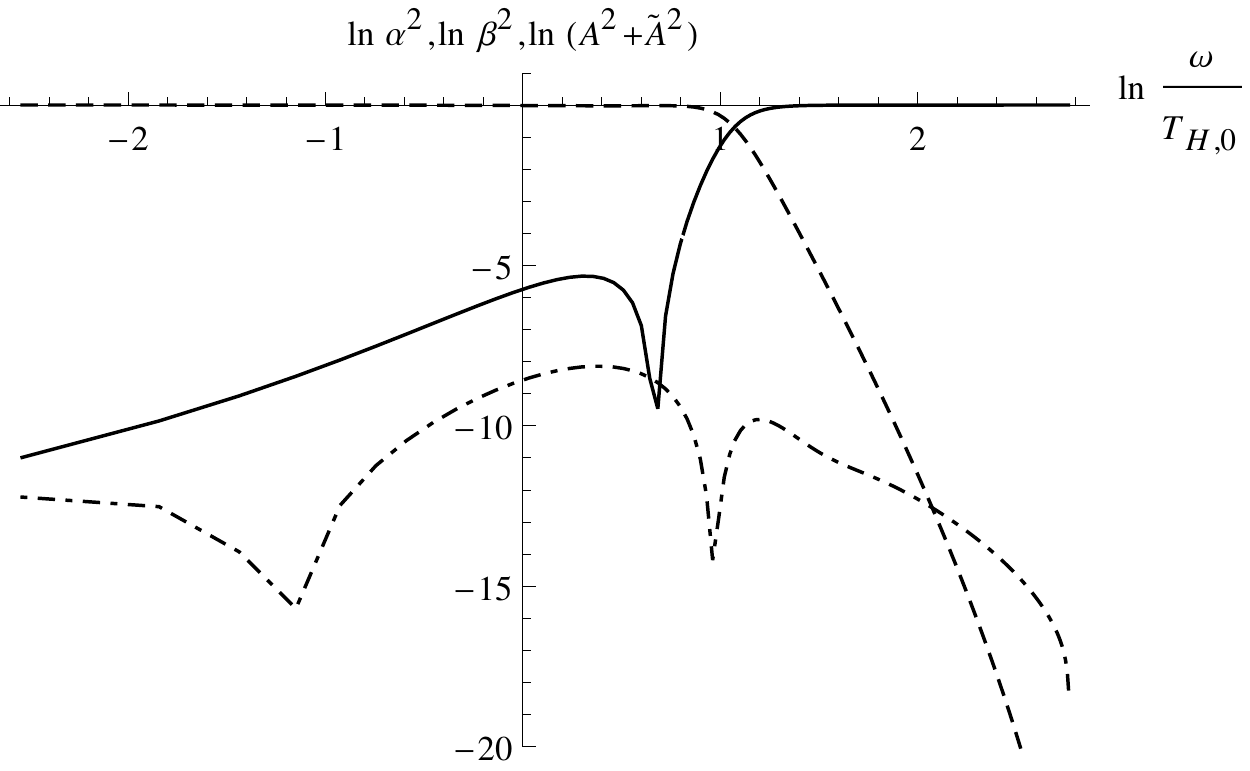}
\end{center}
\caption{Left: Effective temperature for a nonmonotonic subcritical flow \eq{eq:wdepth_2} (solid) with $F_{\rm max}= 0.87$, $\sigma_1 = \sigma_2 =0.06$ and $L = 10$, and for the corresponding monotonic one of \eq{eq:wdepth} (dashed) which coincides with the second subcritical flow of \fig{fig:2}. 
Apart from the peaks due to resonances, the effective temperature behaves in the same manner. Right: Logarithms of the squared scattering coefficients $|A_\om|^2 + |\tilde{A}_\om |^2$ (dashed), $|\alpha_\om|^2$ (solid), and $|\beta_\om|^2$ (dot-dashed) as functions of $\ln \om$, for the same nonmonotonic flow. 
At $\ln \lp \om_{\rm min} /T_{H,0} \rp \approx 1.1$, both $|\alpha_\om|^2$ and $|A_\om|^2 + |\tilde{A}_\om|^2$ display a transition.  While it was sharp in \fig{fig:2}, the transition is now smoothed out. Apart from this, the coefficients behave very much like those of the corresponding monotonic flow.} \label{fig:3bis}
\end{figure}

\begin{figure}
\begin{center}
\includegraphics[scale=0.9]{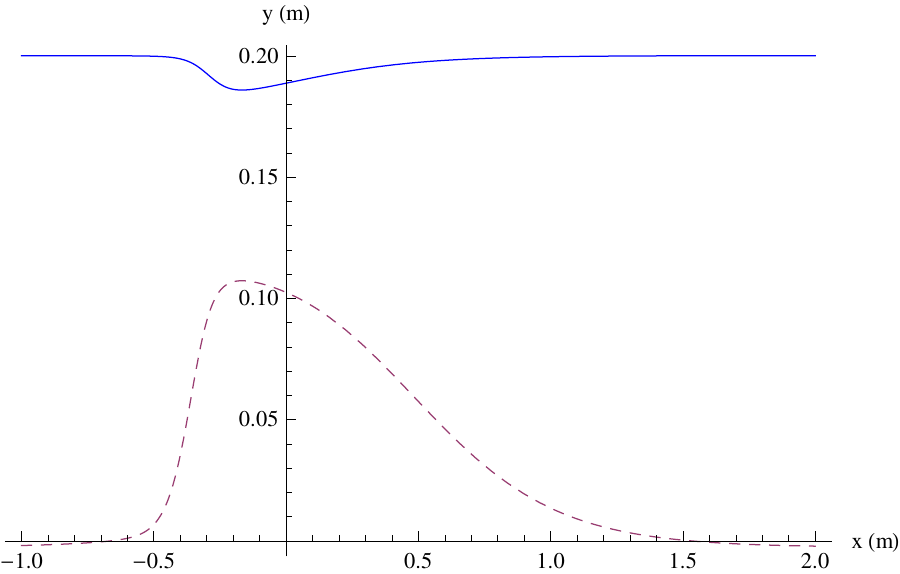}
\includegraphics[scale=0.9]{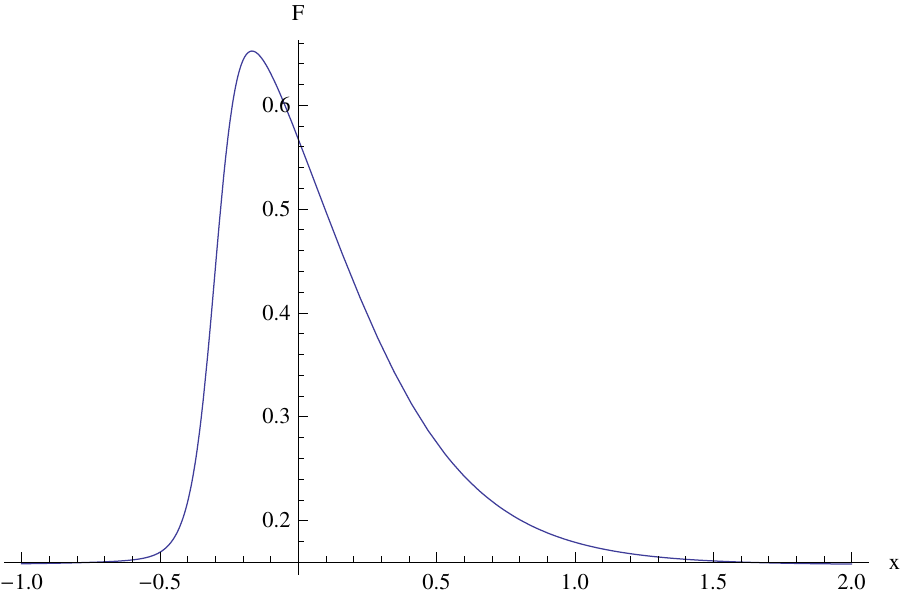}
\includegraphics[scale=0.7]{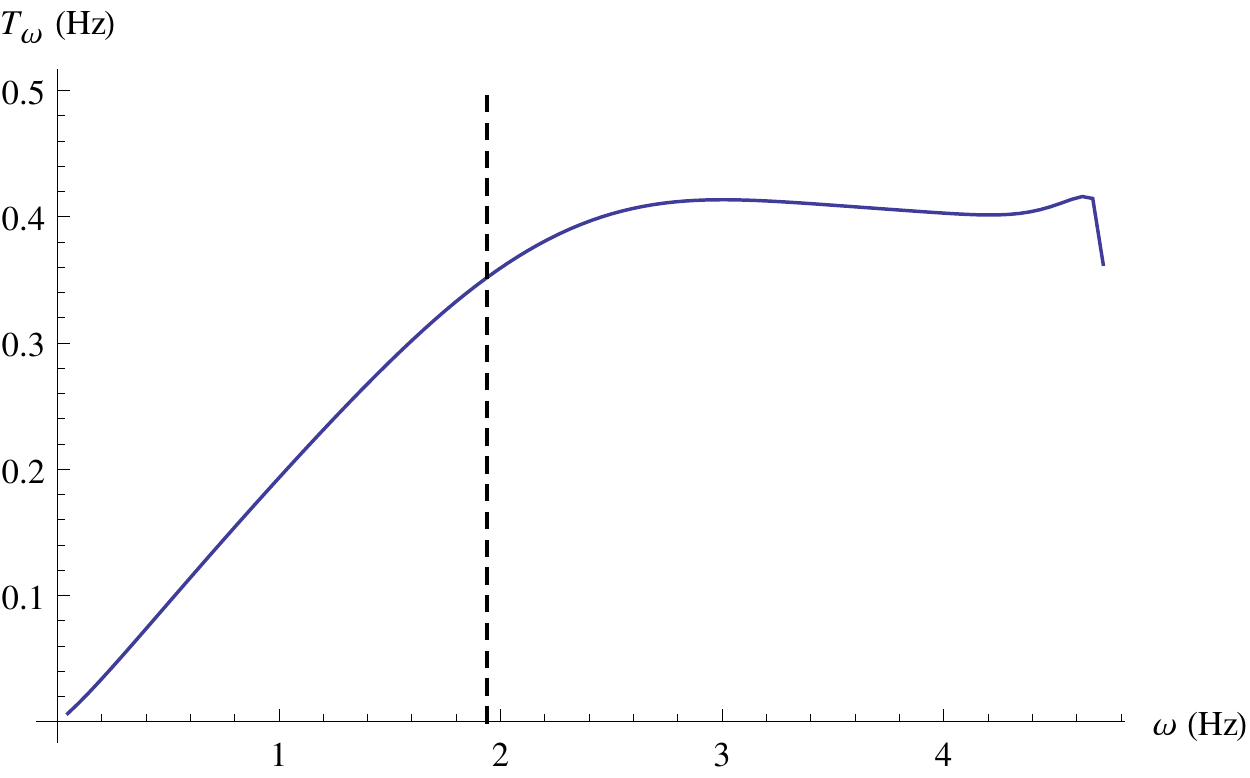}
\includegraphics[scale=0.7]{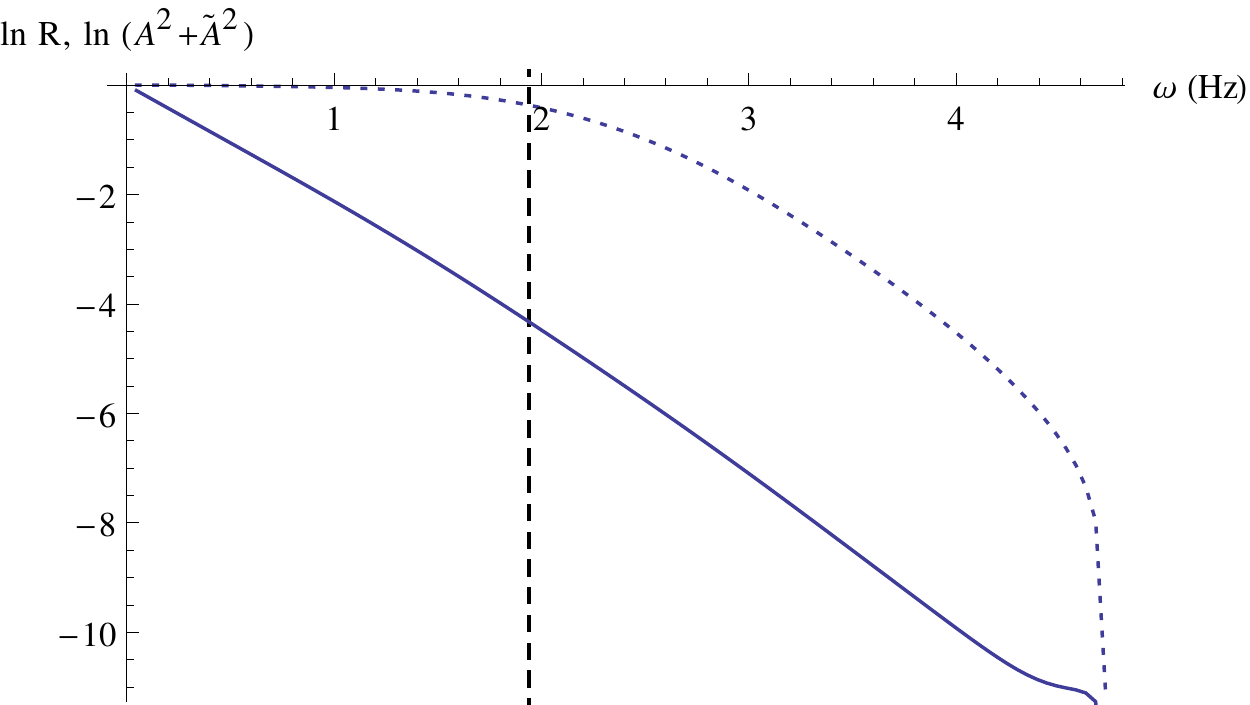}
\caption{Top, left: Free surface (plain) and obstacle (dashed) for a flow of the form \eq{eq:wdepth_2} resembling the one used in \cite{Unruh2010}. We took $g=9.8 m \cdot s^{-2}$ and $J=0.045 m^2 \cdot s^{-1}$. 
Top, right: Froude number as a function of $x$. One sees that the flow is 
subcritical. 
Bottom, left: Effective temperature of \eq{eq:effT} as a function of $\om$. Bottom, right: $\ln R_\om $ (plain) and 
logarithm of $| A_\om|^2 + |\tilde{A}_\om|^2 $ (dotted) as functions of $\om$. Vertical dashed lines indicate the value of $\om_{\rm min}$. 
For $\om < \om_{\rm min}$, we see that the effective temperature linearly vanishes, and that the hydrodynamical coefficients dominate the scattering. We also see that $\ln R_\om $ is linear to a very good approximation, as was observed in~\cite{Unruh2010}. 
}\label{fig:smoothedVancouver}
\end{center}
\end{figure}

An extra ingredient must be added to allow a clear comparison with the observations of \cite{Unruh2010}. One should indeed consider nonmonotonic flows since the flow that was used had essentially the same velocities in the upstream and downstream regions. 
Unlike what we found when studying transcritical flows, for subcritical ones, we find that the replacement of monotonic flows by the corresponding nonmonotonic one does not significantly affect the results, see \fig{fig:3bis}. Indeed, for $\om \rightarrow 0$, the behavior of the Bogoliubov coefficients is not affected, as one still finds \eq{eq:bato}, and $| \tilde{A}_\om |^2 + \left\lvert A_\om \right\rvert^2 \to 1$.~\footnote{There is a small difference between monotonic and nonmonotonic flows, in that, for the latter, the reflection coefficient $A_\om$ goes to $0$ in the limit $\om \to 0$, whereas $|{A}_\om /\tilde A_\om |$ is finite in the former flows.} 
The absence of major difference with respect to the monotonic case reflects the fact that it is the value of $\ommin$ that matters and not the shape of the profile $h(x)$ in the upstream region far from the would-be white hole horizon. This can also be understood as follows. For transcritical flows, a qualitative change of behavior occurs as the left asymptotic region is supercritical for a flow of the form \eq{eq:wdepth}, but subcritical for a flow of the form \eq{eq:wdepth_2}. For subcritical flows instead, no such qualitative change can possibly occur when going from \eq{eq:wdepth} to \eq{eq:wdepth_2}. In particular, a closer study reveals that \eq{eq:Tomsub} and \eq{eq:bato} are still valid for nonmonotonic subcritical flows.

Two relatively minor differences between monotonic and nonmonotonic flows are nevertheless worth mentioning. First, in~\Fig{fig:3bis} we observe hollows in $T_\om$, $|\alpha_\om|$ and $|\beta_\om|$, which correspond to resonances. The presence of these hollows is to be expected, as the high velocity central region acts as a resonant cavity, see~\cite{ZP11}. In fact, their frequency strongly depends on $L$, which defines the length of the effective cavity. In particular, they disappear when $2 L \lesssim {\rm Min} \lp D/\sigma_1, D / \sigma_2 \rp$, as can be verified in \Fig{fig:smoothedVancouver}. The second difference can be seen on the right panel of \fig{fig:lnR_subsub}. It concerns the disappearance of the sharp kinks observed for monotonic flows (see left panel), and associated with the presence of $\om_{\rm min}$. 
This disappearance can be understood from the fact that the transmission coefficients progressively vanish above $\om_{\rm min}$ when the flow is nonmonotonic, whereas they completely vanish for monotonic flows.
 
\subsubsection{Comparison with the Vancouver experiment} 
 
Having clarified all these points, we now consider a profile similar to the one used in \cite{Unruh2010}, save for the fact that we do not include the zero-frequency mode which modulated their background flow. At the end of this section, we shall briefly consider its impact on the scattering coefficients, and show that the modifications are not significant. 

Ignoring the undulation, the water depth has the form \eq{eq:wdepth_2} where the parameters are chosen to fit the profile of~\cite{Unruh2010} using a least-square method. In the international system of units, the optimum parameters are
\be 
\label{opt}
h_0 \approx 0.13 \, m , \quad 
D \approx 0.07 \, m , \quad 
\sigma_1 \approx 0.13,\quad
\sigma_2 \approx 0.76, \quad
2 L \approx 0.79 \, m.
\ee 
We did not use the exact description of the profile because its slope is discontinuous, making the numerical integration difficult. We believe this replacement has no significant consequences on our main results. 

Our description of the profile is plotted in the left upper panel of \fig{fig:smoothedVancouver}. On the right upper one, we represent the associated profile of the Froude number $F(x)$. 
In the lower plots, we represent the effective temperature, the squared norm of the hydrodynamic coefficients, and $\ln R$ of \eq{eq:R} as functions of $\om$. Vertical dashed lines indicate the value of $\om_{\rm min}$. 
By making series of simulations we observed that the values of the scattering coefficients can significantly depend on the exact shape of the profile, so we do not expect a good quantitative agreement. However, more importantly, we did observe that three important features are not sensitive to the profile shape. 

\begin{figure}
\begin{center}
\includegraphics[scale=0.8]{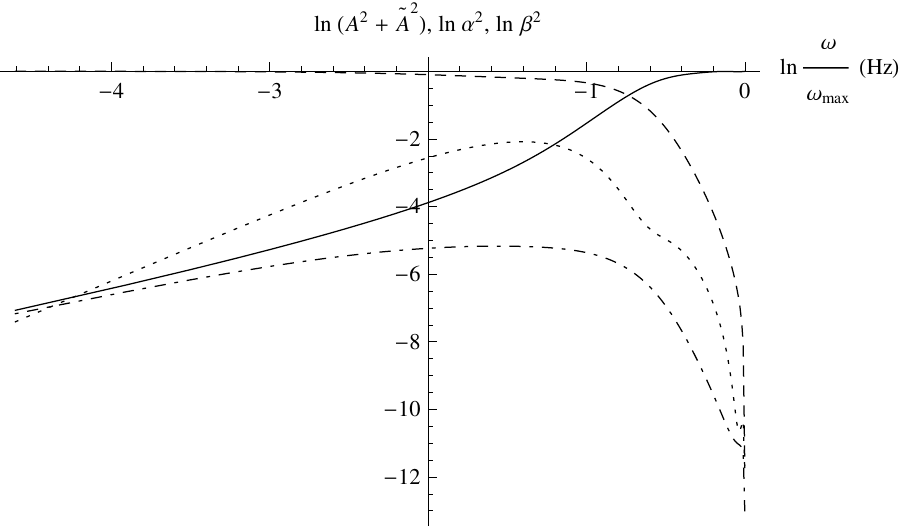}
\includegraphics[scale=0.6]{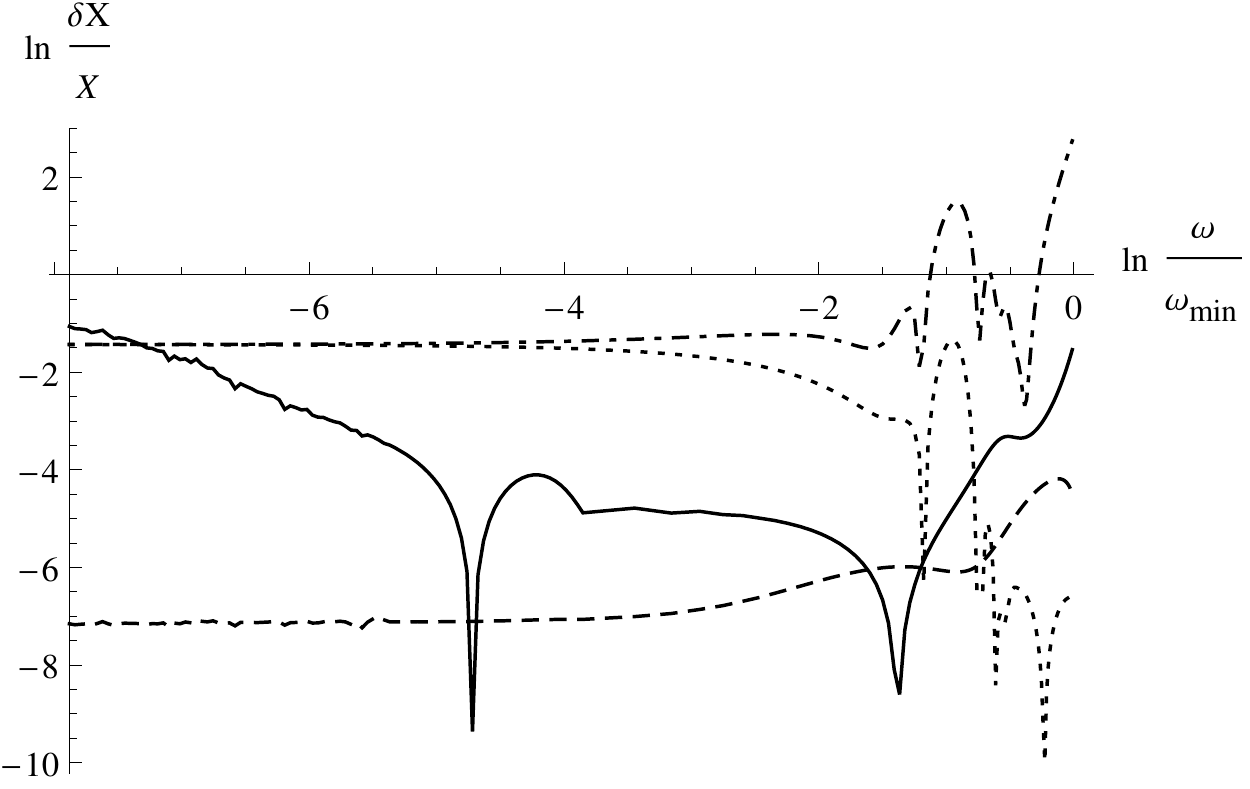}
\caption{Left: Logarithms of  $|\alpha_\om|^2$ (solid), $|\tilde{A}_\om|^2$ (dashed), $|A_\om|^2$ (dotted),
and $|\beta_\om|^2$ (dot-dashed), as functions of $\om$ for the flow of \fig{fig:smoothedVancouver}. 
We see that $|\alpha_\om|^2$ becomes smaller than $|\tilde{A}_\om|^2$ 
for frequencies smaller than $\om_{\rm min} \sim 2 \, Hz$, when there is no turning point, 
exactly as was seen in the lower right panel of \fig{fig:3bis}. 
Right: Logarithm of the relative differences of the norm of the 4 Bogoliubov coefficients  introduced by including an undulation of the form of \eq{eq:undul}, with the parameters of \eq{eq:paraun}. The rapid oscillations seen for $\ln \frac{\om}{\om_{\rm min}} < -6$ are due to numerical errors. 
We observe that the relative modification of the transmission coefficient $\tilde{A}_\om$ is extremely small, whereas those of the other three coefficients remain of the order of $2 \delta h_u/(h_0-D)\sim 0.06$, the relative change of the water height due to the undulation. 
}\label{fig:coefVanc}
\end{center}
\end{figure}

First, the hydrodynamic channels always dominate the scattering for $\om < \om_{\rm min}$, as can be seen in the bottom right panel. When observing the left plot of \Fig{fig:coefVanc}, left panel, we find that it is the transmission coefficient which dominates, 
\be
\left\lvert \tilde{A}_\om \right\rvert^2 \mathop{=}_{\om \rightarrow 0} 1 + O \lp \frac{\om}{\om_{\rm min}} \rp .
\label{eq:blo}\ee
Using the experimental data available, we estimate that $\om_{\rm min} \approx 2.7 \, Hz$ for the setup of~\cite{Unruh2010}\footnote{This value is computed with the full dispersion relation $\Omega_\om^2 = g k \tanh \lp h k \rp$, whereas in \fig{fig:smoothedVancouver} $\om_{\rm min}$ follows from the quartic law of \eq{eq:disprel}. We are currently improving Eq.~(\ref{eq:om}) so as to reduce this discrepancy.}, corresponding to a linear frequency $f_{\rm min} \approx 0.42 \, Hz$. 
The second feature concerns the vanishing of $T_\om$ as $\om \to 0$, as can be seen in the left lower panel. In fact, we found that $\beta_\om$ and $\alpha_\om$ always obey \eq{eq:bato}. The third feature is a consequence of \eq{eq:bato}, and concerns the linearity of $\ln R_\om$ for low-frequency. In fact, $\ln R_\om$ is remarkably linear throughout the domain $\om \in \left\lbrace 0, \om_{\rm max} \right\rbrace$.

We now discuss a potentially important aspect that we so far neglected. It concerns the zero-frequency mode with a large amplitude that was observed in the downstream region. To investigate its effects on the scattering, we added various undulations to our profiles, along the lines of Sec. IV.B. of~\cite{FPBroad}. To be able to distinguish the asymptotic modes on the left side, the amplitude of the undulation is exponentially suppressed at large values of $x$. To make the numerical integration simple, and to incorporate the information on the undulation we possess, its profile was taken of the form
\be \label{eq:undul}
\delta h_u = \delta h_{u,0} \, \cos \lp k_u x \rp \lp 1 - \tanh \lp \kappa_l ( x-x_l) \rp \tanh \lp \kappa_r ( x-x_r) \rp \rp,
\ee
where $k_u$ is the asymptotic wave number of the zero-frequency mode
\be 
k_u = \frac{\sqrt{ 3 (1-F_{\rm min}^2) }}{h_0+D},
\ee
and where 
\be \label{eq:paraun}
\delta h_{u,0} = 0.002 \, m, \, \kappa_l = 1.0 \, m^{-1}, \, \kappa_r = 0.1 \, m^{-1}, \, x_l = 1.0 \, m, \, x_r = 50.0 \, m.
\ee
To illustrate the various effects introduced by undulations, we present, in the left panel of \fig{fig:coefVanc}, the norm of the four coefficients of \eq{eq:Bsub} for the flow of \eq{opt} without undulation, and, in the right panel, the relative variations of these coefficients when including the undulation parametrized by \eq{eq:undul} and \eq{eq:paraun}.
As can be seen in the figure, the relatives corrections are smaller than $e^{-2}$ for $\om < e^{-2} \om_{\rm min}$. So, the undulation does not change the qualitative behavior of the 4 scattering coefficients, in particular the key behaviors of \eq{eq:blo} and \eq{eq:bato} are still found. 
Having done series of simulations with different amplitudes for the undulation, we found the relative deviations are linear in the amplitude. These relative differences, evaluated for $\alpha$ and $\beta$ in the limit $\om \rightarrow 0$, are of the order of $\frac{2 \delta h_{u,0}}{h_0 - D}$. We believe that a more systematic study of the effects of zero-modes is beyond the scope of this paper. 
In collaboration with X.~Busch, we are currently completing this analysis in a separate work.

When reporting the above estimation of $\om_{\rm min}$, we notice that more than half the data points shown in Fig.~5 of \cite{Unruh2010}, left panel, correspond to frequencies \textit{below} $\om_{\rm min}$, for which the squared norm of the transmission coefficient, $| \tilde{A}_\om |^2 $ is close to 1, since there is no turning point. Hence, we conclude that the linearity of $\ln R_\om$ observed in the Vancouver experiment is probably not due to the fact that the incoming waves were blocked.~\footnote{Since the transmission coefficient $\tilde A_\om$ cannot be neglected for $\om < \om_{\rm min}$, we do not think it is legitimate to use, even as an approximation, $|\alpha_\om|^2 - |\beta_\om|^2 = 1$, as done between Eqs. (15) and (16) in~\cite{Unruh2014}. We are grateful to the referee to suggest us to discuss this recent work.}
Together with the absence of blocking, it would be interesting to see whether the low-frequency behavior of \eq{eq:bato} can be validated (or invalidated) by the experimental data of \cite{Unruh2010}, or some new data. (The behavior of the norms of $\beta_\om$ and $\alpha_\om$ in the left panel of Fig.~5 of this reference indicates that \eq{eq:bato} could apply.) It should be stressed that \eq{eq:bato} and \eq{eq:blo} will not be easily accessible when measuring the changes of the free surface associated with the 4 outgoing waves resulting by sending shallow water waves, see \eq{eq:Bsub}. Indeed, if we denote as $\delta h_\om^{\rm hydro}$, the variation associated with the transmitted wave, and $\delta h_\om^{\rm disp}$ that associated with the dispersive reflected wave of with negative $k_\om$ (or positive $k_\om$), their ratio scales as
\be
\frac{\delta h_\om^{\rm hydro}}{\delta h_\om^{\rm disp}} \sim  \om^{0},
\ee
in spite of the fact that the ratio of the corresponding coefficients diverges as $|\tilde A_\om /\alpha_\om | \sim \om^{- 1/2}$, as it is implied by \eq{eq:bato} and \eq{eq:blo}.
The origin of the additional factor of $\om^{1/2}$ comes from, first, the action of the derivative operators in \eq{eq:dh}
which brings a factor of $\om - vk \propto \om$, and, second from the normalization factors, see for instance~\eq{eq:B8}.  
There is another property which probably further complicates the measurement of the transmitted wave, namely that its wave length diverges like $1/\om$: for $\om = \om_{\rm min} \sim 2 \, Hz$, it is equal to $\sim 5 m$, and can become larger than the length of the flume used in~\cite{Unruh2010} if $\om$ is decreased. 

To conclude this section, we would like to discuss the status of the relationship between the effective temperature and the surface gravity, as this is a key feature of the Hawking effect. 
Since the limit $\om \to 0$ of the effective temperature $T_\om$ of \eq{eq:effT} vanishes, there is no unique way to associate a well-defined temperature to the system. A first possibility is to use the value of $T_\om$ at the plateau seen in the bottom left panel of \Fig{fig:smoothedVancouver}. This gives an effective temperature of approximately $0.4 \, Hz$. A second possibility is to use the inverse slope of $\ln R$ as a function of $\om$, giving a temperature of $0.5 Hz$, which is rather close to the previous one. 
(For comparison, the temperature obtained from the inverse slope of $\ln R$ in Fig. 5 of~\cite{Unruh2010} is $T_H \approx 0.70 Hz$. The relative good agreement between these numbers confirms that our numerical simulations correctly captures the key properties of the observations made in Vancouver.) 
The third possibility is to use the gradient of $h$ to define a pseudo-Hawking temperature as 
\be \label{eq:Tps}
T_{\rm pseudo-H} = \frac{1}{2 \pi} {\rm Max} \left\lvert \pd_x (v-c) \right\rvert. 
\ee
 If the maximum is taken over the descending slope, where the scattering is supposed to occur, and where one would find the white hole horizon if $F$ crossed 1, we find $T_{\rm pseudo-H} \approx 0.15  \, Hz$, smaller  than the previous ones by a factor 3. If the maximum is taken instead on the steeper ascending slope, we find $T_{\rm pseudo-H} \approx 0.77  \, Hz$, significantly larger than the previous ones.  
We believe this discussion gives a fair idea of the difficulties to relate the gradient of $v - c$ to an effective temperature. It seems to us that it is pointless to try to identify a precise relationship in subcritical flows. On the contrary, when the flow is sufficiently transcritical, the standard relationship of \eq{eq:TH} works very well, as can be seen in \Fig{fig:iVhF:r}.

\section{Conclusion}
\label{sec:concl}

In this work, we first recalled the basics elements governing the scattering of shallow water waves; and showed that, when the Froude number is significantly larger than 1, in which case the analogue of a relativistic (Killing) horizon is clearly present, the scattering coefficients quantitatively follow Hawking's thermal prediction, and this despite the fact that dispersion is included in the wave equation and strongly affects the characteristics of the waves. 

Turning to subcritical flows, we explained the important roles played by the critical frequency $\om_{\rm min}$ in governing the behavior of the scattering coefficients. For frequencies above $\om_{\rm min}$, incoming counter propagating modes are blocked, and one essentially recovers the behaviour found for {\it slightly transcritical flows}, in particular, the Planckianity of the spectrum is already lost. For frequencies below $\om_{\rm min}$, we observed a decrease of the effective temperature, which vanishes when $\om \to 0$. This reflects the fact that the square norms of both $|\beta_\om|^2$ and $|\alpha_\om|^2$, which were the dominant coefficients for  significantly transcritical flows, now both linearly decrease to 0 in the low-frequency limit. At the same time, we saw that the sum of the hydrodynamic (elastic) coefficients  $|A_\om|^2 + |\tilde A_\om|^2$ tends to $1$, which means that they dominate the scattering in this low-frequency regime. We then showed, and explained, how these facts imply that the logarithm of $R_\om = |\beta_\om/\alpha_\om|^2$ is linear in $\om$ for small $\om$, as if the spectrum were still Planckian.

Besides comparing the scattering in sub- and transcritical flows, we also identified the consequences of considering nonmonotonic flows which are subcritical on both sides of the obstacle. For transcritical flows, this amounts to adding an analogue black hole horizon. The effects are very clear: whereas the high frequency regime is hardly affected, there is a new critical frequency $\om_c$ which governs the “tunneling” across the region where $F > 1$. When the latter is long enough, $\om_c$ is very small. Below $\om_c$ a new regime is found where $|\beta_\om|^2$ and $|\alpha_\om|^2$ again linearly decrease to 0 as $\om \to 0$. 

We combined these aspects by considering nonmonotonic subcritical flows. We found that the nonmonotonic character of the flow does not significantly modify the scattering coefficients. Hence the spectral properties are similar to those found for monotonic flows. In particular for $\om < \om_{\rm min}$, the saturation of $|A_\om|^2 + |\tilde A_\om|^2 \to 1$, the vanishing $|\beta_\om|^2 \sim |\alpha_\om|^2\sim \om$, and the linearity of $\ln R_\om$  appear to be very robust features of the scattering. Moreover, the three features have also been found when including an undulation with a macroscopic amplitude, and when considering a subcritical nonmonotonic flow, solution of the nonlinear hydrodynamical equations, see Appendix A. We therefore conclude that these properties should apply to the experiment of Ref.~\cite{Unruh2010}. In fact, when comparing the observed behavior of 
$R_\om = |\beta_\om/\alpha_\om|^2$ to that predicted by our analysis, we found a good qualitative agreement in that both the linearity of its logarithm, and the value of the slope 
are well approximated. 
It would therefore be interesting to conceive new experiments to validate the other two predictions, which, to our knowledge,
have not been reported by any experimental group,
namely the saturation of the transmission coefficient $\tilde A_\om$, 
and the vanishing of $|\beta_\om|^2$ and $|\alpha_\om|^2$
for $\om < \om_{\rm min}$.

We should also remind the reader that our predictions have been derived using a slightly simplified version of the wave equation derived in~\cite{Coutant_on_Undulations,Unruh2012}. Therefore a comparison with detailed experimental data might allow one to determine the validity range of this simplified wave equation. 

Finally, in Appendix A, we considered a transcritical nonmonotonic flow over an obstacle which is a solution of the nonlinear hydrodynamical equations. Our aim was to show that in this more "realistic" case the scattering coefficients closely follow, in quantitative terms, Hawking's prediction, i.e. $|\beta_\om|^2 \sim |\alpha_\om|^2\sim T/\om$ for low-frequencies. This indicates that, by a careful choice of the obstacle, one could engender a  {\it transcritical} background flow hardly contaminated by an undulation, which could then be used to experimentally test the thermal prediction.%
~\footnote{Notice that a trans-critical flow was clearly realized in the settings of Ref.~\cite{PhysRevE.83.056312} involving a circular jump. What is unclear to us is how to generate stationary waves in a controlled way so as to probe the mode mixing at the sonic horizon.}
We hope that this analysis may persuade an experimental team to pick up the gauntlet. 

\section*{ACKNOWLEDGEMENTS}

We are grateful to G. Rousseaux for many discussions which motivated the present work. 
We thank X. Busch, A. Coutant, and S. Robertson for interesting discussions, and a careful reading of an early version of the manuscript.
We also thank W. G. Unruh and T. Jacobson for helpful comments on this manuscript. 
We gratefully acknowledge partial financial support from the French CNRS interdisciplinary project PTI-2014 {\it DEMRATNOS} (D\'etection et Mesures de Radiation des Trous Noirs en gravit\'e analogue par utilisation d’Ondes de Surface dans un canal \`a houle).

\appendix

\section{Link with the nonlinear hydrodynamic equations}
\label{sub:NL}

So far our analysis was restricted to the linear wave equation \eq{eq:waveeq} in a background flow specified from the outset by the profile of the 
water depth $h(x)$. Since \eq{eq:waveeq} comes from the linearization of nonlinear hydrodynamical equations~\cite{LLhydro,Batchelor,Unruh2012}, it is worth verifying that our results still apply to background flows which solve these equations. To this end, we use the hodograph transform method described in~\cite{Unruh2012}. Given a flow with a prescribed free surface, asymptotic water depth, and velocity, this method allows to find an explicit parametrization of the obstacle shape. We shall consider two typical examples, one transcritical and one subcritical, so as to be able to compare the resulting scattering coefficients with those obtained in the body of the text. We stress that these two examples may not be suitable for an experimental realization. They were chosen to show that the results of the main text apply when using solutions of the hydrodynamic equations with a simple shape of the bottom. In particular, their descending slopes may well be too large to maintain a laminar flow. However, the general method that we present here can be applied to find smoother obstacles, with smaller slopes.

We remind the reader that an ideal, incompressible, inviscid, irrotational, 2D flow may be described using the velocity potential $\varphi$,~\footnote{Here $\varphi$ is the “full” velocity potential, whose gradient gives the velocity of the background flow, while in the main text we denoted $\phi$ the linear perturbation on this potential describing waves, see \eq{eq:waveeq}.} 
defined as
\be 
\nabla \varphi = \vec{v},
\ee
and the stream function $\psi$
\be 
\nabla \psi = \vec{e}_z \wedge \vec{v}.
\ee
Here, $\vec{e}_z$ is the unit vector in the horizontal direction orthogonal to the mean flow velocity. In order to find a localized obstacle shape centered close to the origin for a flow with $F(\infty) < 1$, the free surface must have a hollow. A particularly simple choice is
\be \label{eq:Fs}
y(\varphi) = \frac{h_0}{(1+A e^{-\sigma^2(\varphi-\varphi_0)^2}) (1+A e^{-\sigma^2 (\varphi+\varphi_0)^2})},
\ee 
where $A, \sigma, \varphi_0, h_0$ are real numbers. The parametric representations of the free surface and the obstacle in real space are then obtained once the asymptotic velocity is chosen, assuming the height of the obstacle goes to zero at infinity, as we now briefly explain. More details can be found in \cite{Unruh2012}. 

The two potentials $\varphi, \psi$ can be used as coordinates. Then the former Cartesian $x$ and $y$ coordinates are seen as functions of $\varphi$ and $\psi$. It is convenient to unite them in a single complex-valued function $X \equiv x + i \, y$. It can be shown that for inviscid, irrotational flows, $X$ satisfies the Laplace equation 
\be 
\lp \pd_\varphi + i \pd_\psi \rp \lp \pd_\varphi - i \pd_\psi \rp X=0 .
\ee
Hence, $X$ may be expressed as the sum of a holomorphic function of $\Phi \equiv \varphi + i \, \psi$, and an antiholomorphic function. Performing the change of coordinates from $(x,y)$ to $(\varphi, \psi)$, one finds 
\be 
\pd_\varphi x = \pd_\psi y, \nn
\pd_\psi x = - \pd_\varphi y.
\ee
These are just the Cauchy-Riemann conditions, showing that $X$ is actually a holomorphic function of $\Phi$. The stream function $\psi$ being constant along the free surface (since the latter is a streamline), an ansatz of the form \eq{eq:Fs} entirely defines the imaginary part of $X$ at $\psi = \psi_s$, where $\psi_s$ is the value of $\psi$ at the free surface. We choose the convention that $\psi = 0$ at the bottom. Then $\psi_s$ is equal to the 2D conserved current $J$ \cite{Unruh2012,Coutant_on_Undulations}. The real part of $X$ at $\psi = \psi_s$ is found using the Bernouilli boundary condition, which reads 
\be 
\pd_\varphi \lp g y\left(\varphi ,\psi _s\right)+\frac{1}{2 \left(\left(\frac{\partial x}{\partial \varphi }\right)^2+\left(\frac{\partial y}{\partial \varphi }\right)^2\right)}\rp = 0.
\ee
This gives a first-order ordinary differential equation on $\varphi \mapsto x(\varphi, \psi_s)$
\be 
\partial _{\varphi }x\left(\varphi ,\psi _s\right)=\sqrt{\frac{1}{v_0^2+2g \left(H_0-y\left(\varphi ,\psi _s\right)\right)}-\left(\partial _{\varphi }y\left(\varphi ,\psi _s\right)\right)^2},
\ee
where $H_0$ is the asymptotic water depth and $v_0$ the asymptotic velocity, so that $\Re X(\varphi, \psi_s)$ is uniquely determined up to a constant. Changing this constant has the effect of translating the free surface and the obstacle in $x$ by the same amount. The obstacle can then be parametrized by making use of the holomorphic properties of $X$:
\be 
X(\varphi,0) = X(\varphi-i \psi_s,\psi_s), \nn
x_{bottom}(\varphi) = \Re X(\varphi,0), \nn
y_{bottom}(\varphi) = \Im X(\varphi,0).
\ee

\begin{figure}
\begin{center}
\includegraphics[scale=0.6]{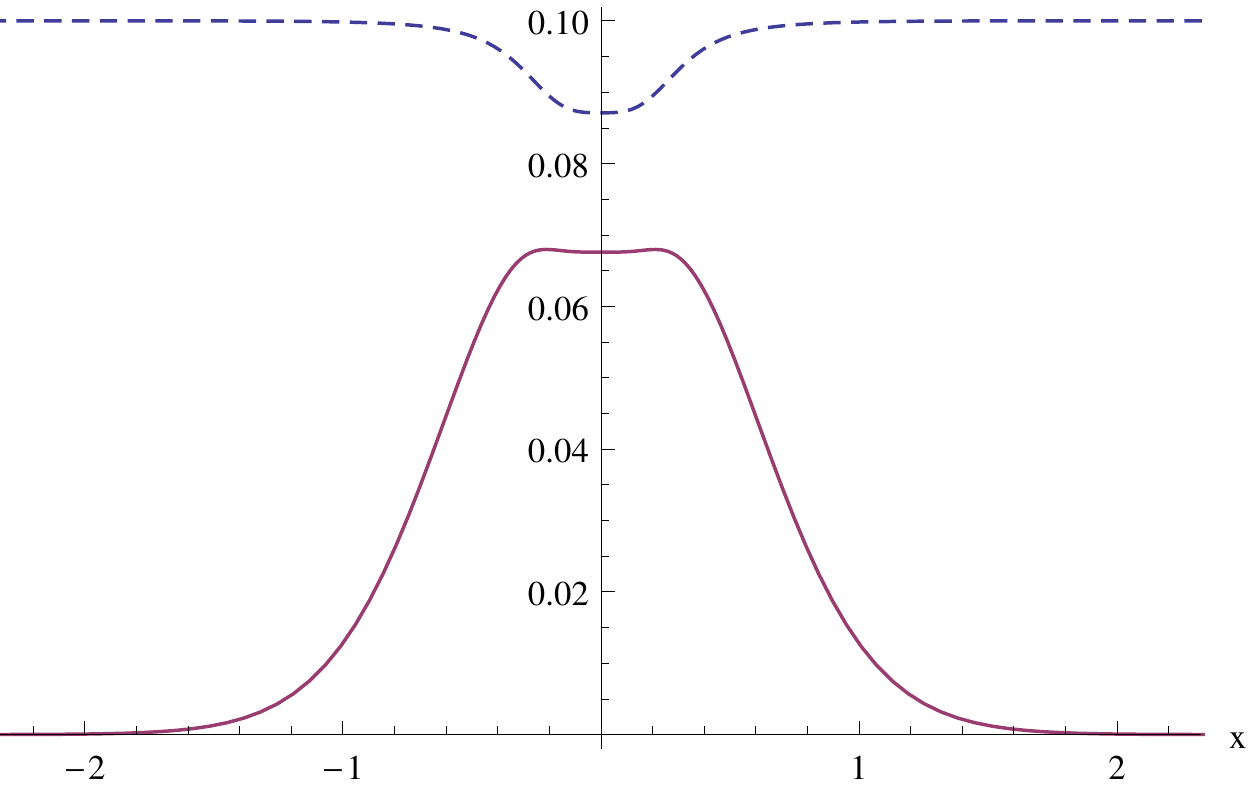}
\includegraphics[scale=0.6]{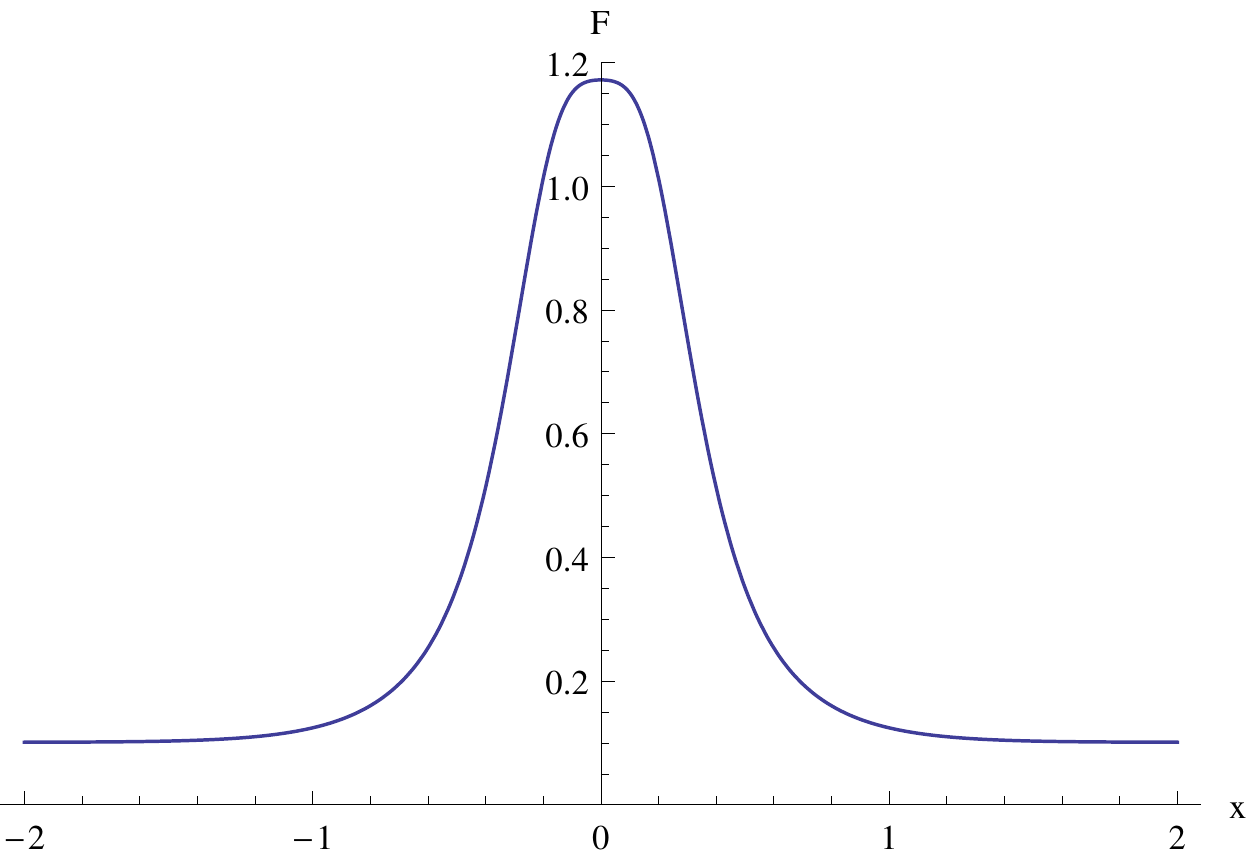}
\caption{Left: Heights of the free surface (dashed) and of the obstacle (solid) as functions of $x$ for the supercritical flow obtained by solving the hydrodynamical equations with the free surface specified by \eq{eq:Fs}. The units of both axes are meters. Right: Froude number for the same flow. The maximum value of $F$ is $1.17$ and the length of the supercritical region is $0.41$ meters.}
\label{fig:iVhF}
\end{center}
\end{figure}

\begin{figure}
\begin{center}
\includegraphics[scale=0.6]{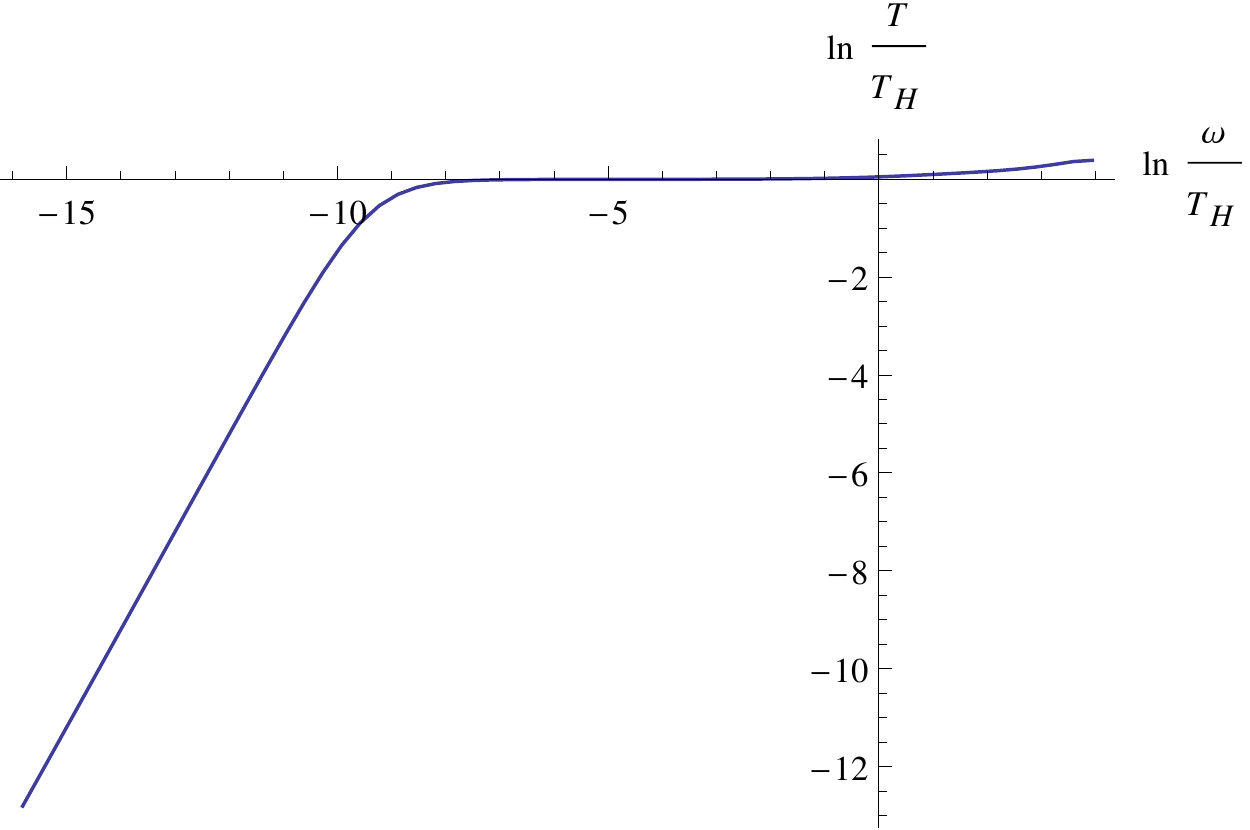}
\includegraphics[scale=0.6]{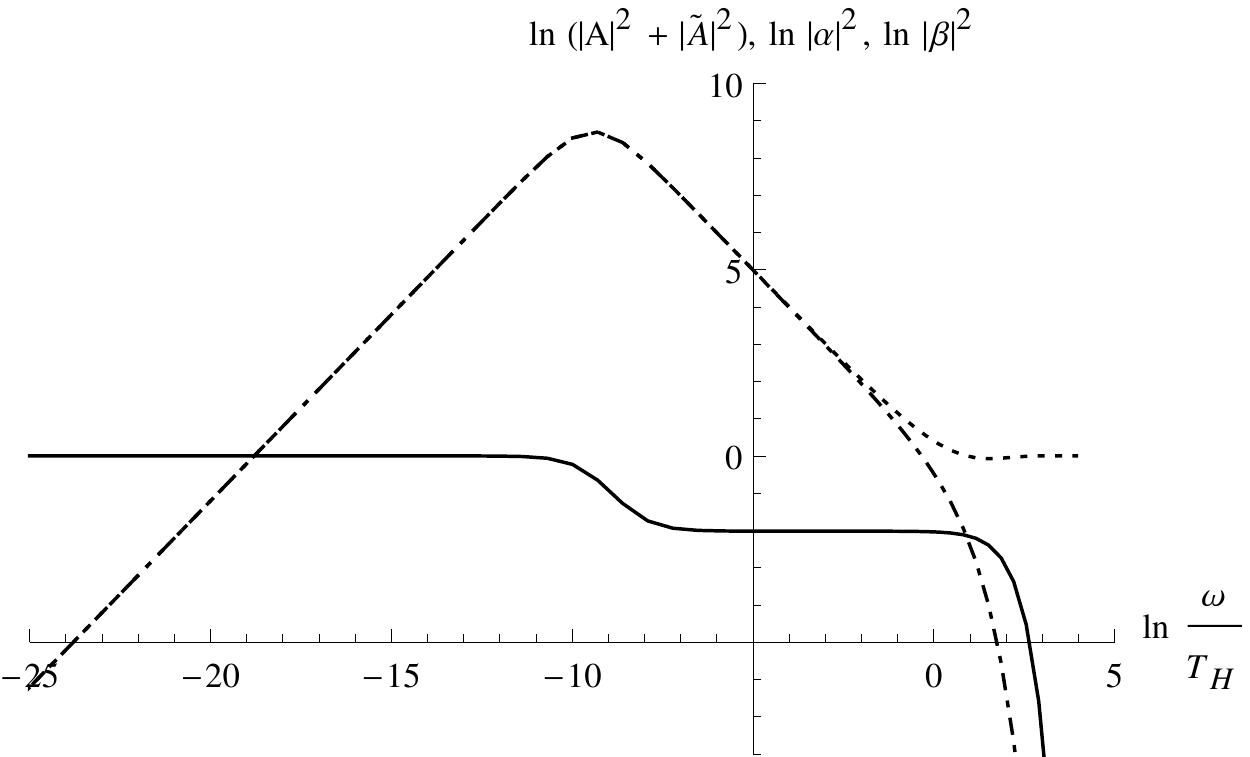}
\caption{On the left panel, we represent the logarithm of the effective temperature as a function of $\ln (\om/T_H)$ for the flow of \fig{fig:iVhF}. The Hawking frequency $T_H$ is approximately $0.164\, Hz$. The good agreement with Hawking's prediction 
is clearly visible by the long extension of the plateau of relative height equal to 1. 
The plateau is bordered by the lower critical frequency $\om_c \approx e^{-9} T_H$ of \eq{eq:omc} and the higher one $\om_{\rm max}$. On the right panel, we represent the logarithm of the squared norms of the Bogoliubov coefficients $|\alpha_\om|^2$ (dashed),  $|\beta_\om|^2$ (dot-dashed), and $|A_\om|^2 + |\tilde{A}_\om|^2$ for the same flow.
It is clear that the hydrodynamic coefficients can be completely neglected for all frequencies larger than  $\om_c \approx e^{-9} T_H$, thereby confirming the Hawkingness of this regime. 
}\label{fig:iVhF:r}
\end{center}
\end{figure}

The first example we considered describes a nonmonotonic transcritical flow. To be explicit, we now express quantities in the international system of units. The flow is characterized by $A=0.12$, $\sigma=10 \, s \cdot m^{-2}$, $\varphi_0=0.072 \, m^2 \cdot s^{-1}$, and an asymptotic velocity $v_0=0.1 \, m \cdot s^{-1}$. The resulting water depth 
 and the Froude number are shown in \fig{fig:iVhF}. 
The two small bumps at the top of the obstacle are fine-tuned to prevent the appearance of the undulation. One verifies that the flow is transcritical, since $F_{\rm max} \simeq 1.17$. The main properties of the scattering coefficients are shown in \fig{fig:iVhF:r}. The comparison with \Fig{fig:3} shows a good correspondence of the two cases. In particular, for the left plot, we recover the extended flat plateau indicating a Planckian spectrum, with a value of the effective temperature $T_\om$ close to the Hawking frequency of \eq{eq:TH}, 
here given by $T_H = 0.143 Hz$. We also observe the signature of the high frequency cutoff $\om_{\rm max}$ of \eq{eq:om-max-min}, and that of the low-frequency one, $\om_c$ of \eq{eq:omc}. The approximate values of these critical frequencies are respectively $7.4 Hz$, and $5. \, 10^{-6} Hz$, which is very low.  From the right panel, we also verify that below $\om_c$ the scattering is dominated by the hydrodynamic coefficients $A_\om$ and $\tilde A_\om$ of \eq{eq:Bsub}. 

\begin{figure}
\begin{center}
\includegraphics[scale=0.6]{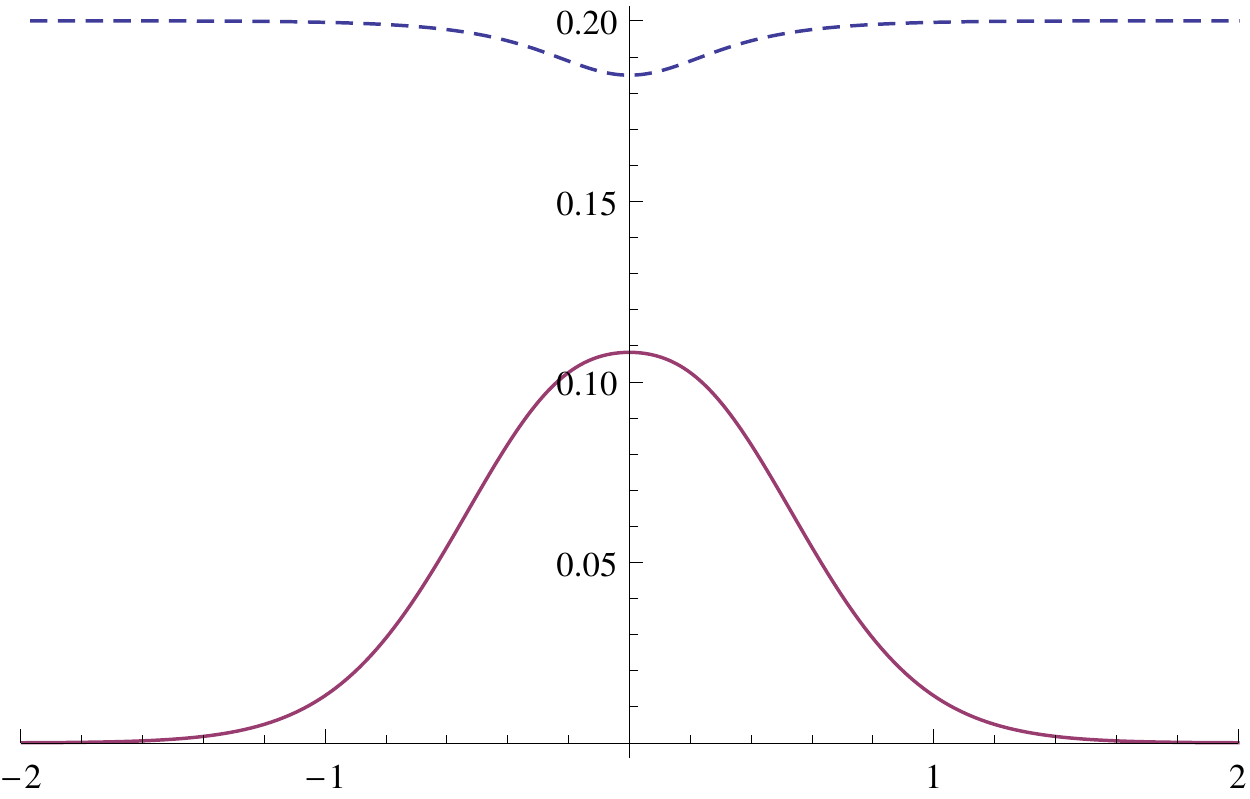}
\includegraphics[scale=0.6]{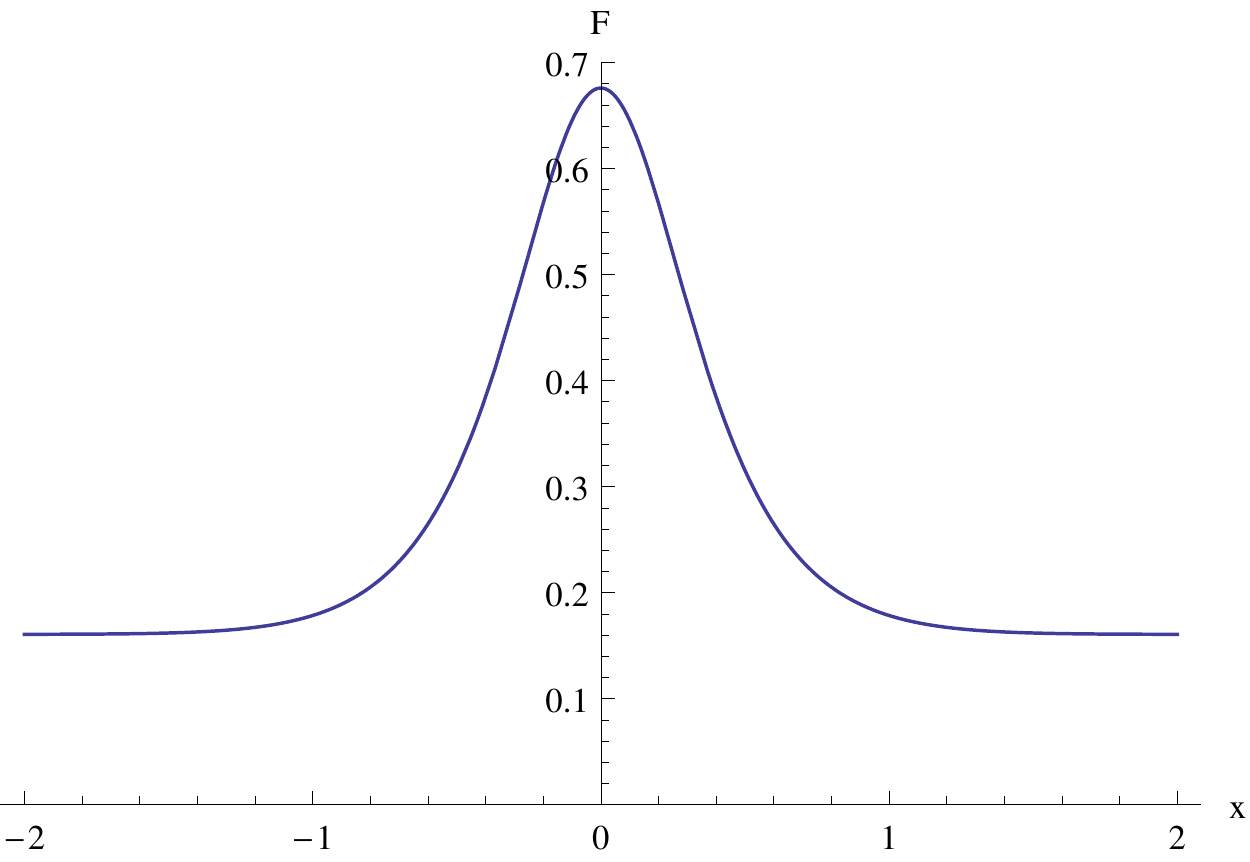}
\caption{Left: Free surface and obstacle for the second flow we obtained by solving the hydrodynamical equations with a known free surface \eq{eq:Fs}. Right: Froude number for the same flow.}\label{fig:iVhFbis}
\end{center}
\end{figure}

Our second example describes a subcritical flow. The parameters are $A=0.04$, $H_0 = 0.2 m$, $v_0 = 0.0225 m \cdot s^{-1}$, $\sigma = 5 s \cdot m^{-2}$, and $\varphi_0 = 0.01 m^2 \cdot s^{-1}$. They have been chosen to give a profile relatively close to the one used in~\cite{Unruh2010}, at least for the downstream part $x > 0$ where the scattering and wave blocking occur. 
The water depth and the Froude number are shown in \fig{fig:iVhFbis}.
The maximum Froude number for this profile is equal to $0.68$, the critical frequency $\om_{\rm min} =  1.9 \, Hz$, 
and the effective temperature of \eq{eq:Tps} is $0.21\,  Hz$. The profile and the Froude number are similar those of \Fig{fig:iVhF} as far as the downstream right side of the flow is concerned. At this point, we consider that trying to reproduce more precisely the profile of~\cite{Unruh2010} is unjustified, as we have neither a good enough control of the various approximations we used, nor enough experimental data. 

\begin{figure}
\begin{center}
\includegraphics[scale=0.5]{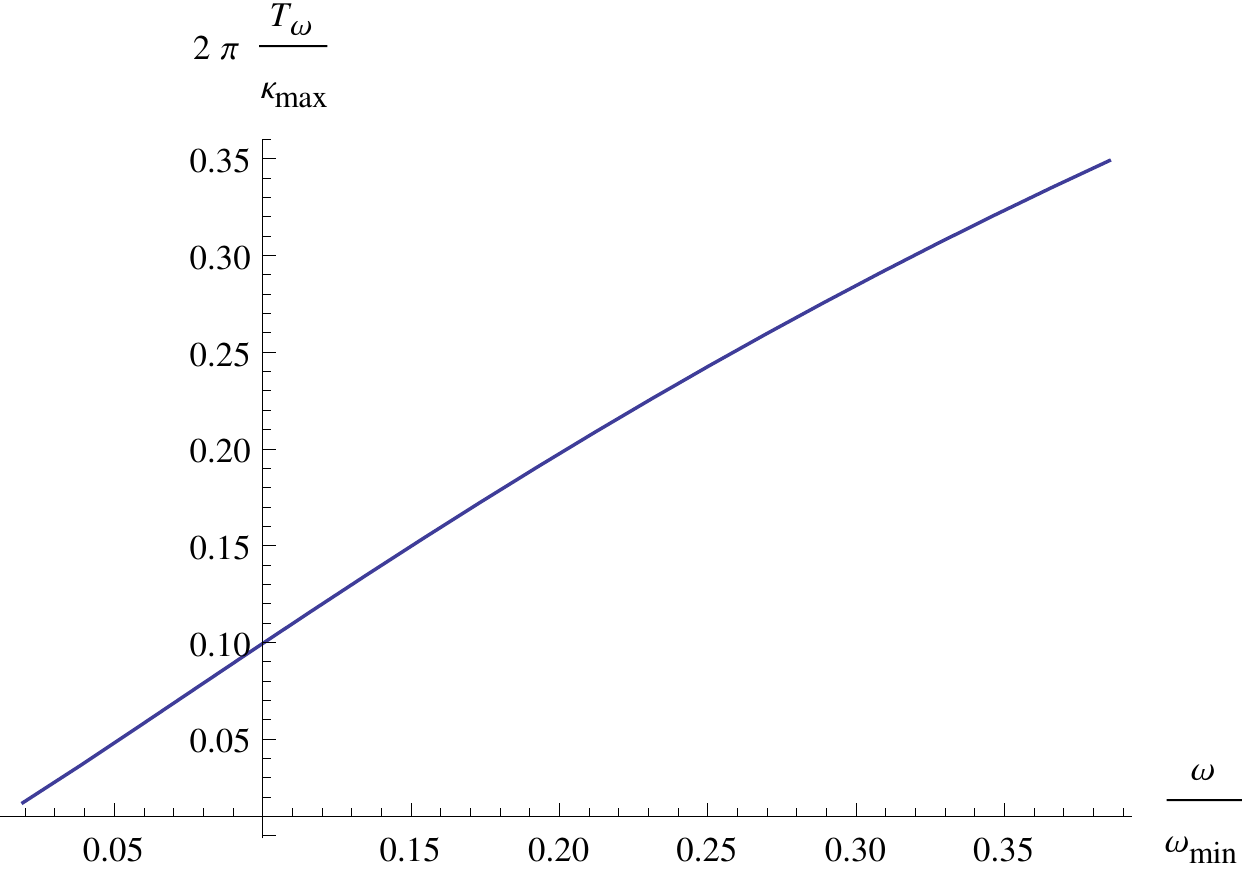}
\includegraphics[scale=0.6]{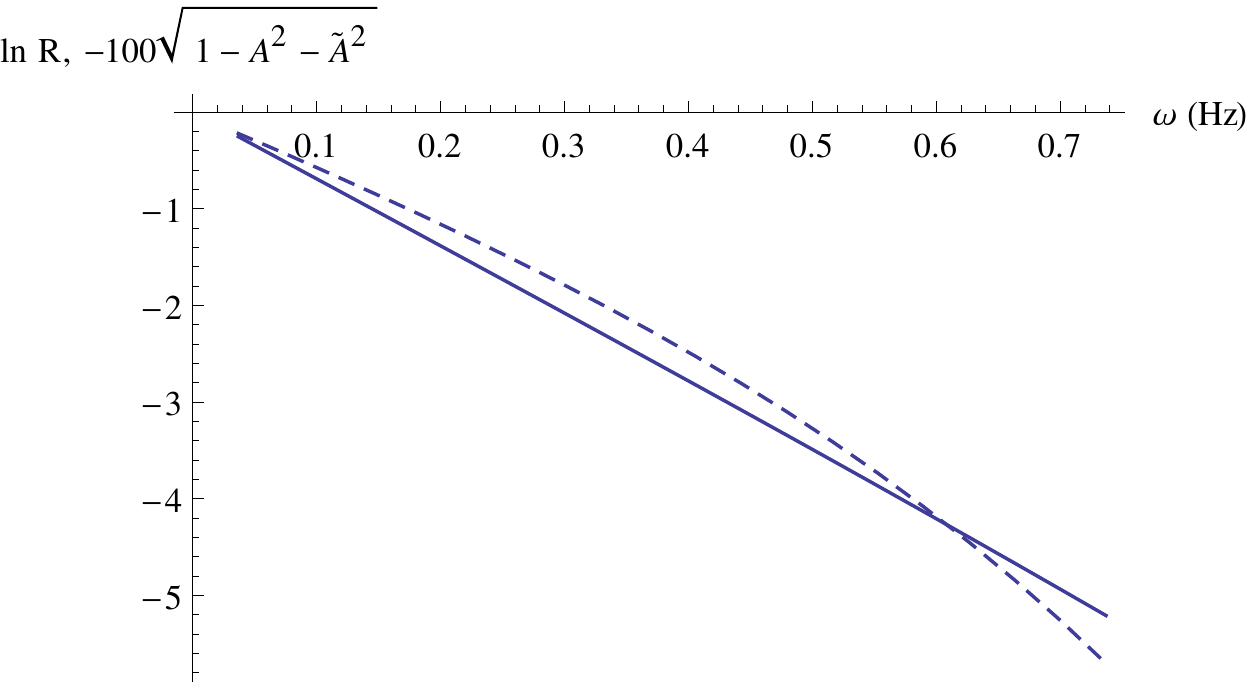}
\caption{Left: Effective temperature adimentionalized by making use of the parameter of \eq{eq:Tps} as a function of the adimensional frequency $\om/\om_{\rm min}$ for the flow of \fig{fig:iVhFbis}. Right: $\ln R_\om$ 
(plain) and $\sqrt{|A_\om|^2 - |\tilde{A}_\om|^2 - 1}$ (dashed) for the same flow. The square root and the factor $100$  have been used so as to clearly see the linear behaviors of both quantities for small $\om$.} 
\label{fig:iVhF:rbis}
\end{center}
\end{figure}

The properties of the scattering coefficients of our second flow are shown in \fig{fig:iVhF:rbis}. 
We again see a good correspondence with those of \Fig{fig:2} and those of \fig{fig:smoothedVancouver}. 
Namely, first, the effective temperature goes to 0 as $\om \to 0$, which confirms that Planckianity is lost; and second, the hydrodynamic elastic coefficients  $A_\om$ and $\tilde A_\om$ dominate the scattering for low-frequencies. A few comments are in order. First, the range of frequencies we represented is smaller than the one in \Fig{fig:smoothedVancouver}. The reason is that obtaining a good numerical accuracy is more difficult in the present case because we no longer have a closed analytical formula for $h(x)$. Although our code can provide accurate results at higher values of $\om$, this becomes time consuming. We thus only present here results for small values of $\om$. For the same reason, the deviations from \eq{eq:4coef} are larger than in the other cases, going from $10^{-3} \left\lvert \beta \right\rvert^2$ to $10^{-1} \left\lvert \beta \right\rvert^2$. 

From the right panel, we verify that the slope of ${2 \pi T_\om}/{\kappa_{\rm max}}$ versus $\om / \om_{\rm \min}$ is close to one. In addition, we also computed the effective temperature $T_\om$ for a few larger values of $\om$ and checked that the
qualitative agreement with \Fig{fig:smoothedVancouver} remains. 
Hence, we expect to get a rough plateau for $T_\om$ with a height close to the pseudo-Hawking temperature of \eq{eq:Tps}.
We checked that it is indeed the case: this plateau is at $T_\om \approx 0.17 \, Hz$ while \eq{eq:Tps} gives $\sim 0.21 \, Hz$. 

\section{Analytical calculation in the steplike limit}
\label{App:gradino}

In this appendix, following \cite{2regimesFinazzi, Scott-review}, we consider the limit where the background water depth is piecewise constant, with one single discontinuity at $x=0$. This limit is rather unrealistic as in a real fluid the effects of viscosity, vorticity and compressibility are expected to become important when the slope of the obstacle is large~\cite{LLhydro,Batchelor}. Its interest lies in its mathematical simplicity, allowing a straightforward calculation of the spectrum. In spite of this, interestingly, one recovers the following important results of the main text
\begin{itemize}
\item in a transcritical flow, the effective temperature goes to a finite constant when $\om \rightarrow 0$;
\item in a subcritical flow, it goes to zero like $\frac{\om}{\ln \, \om}$;
\item still for $F_{\rm max} < 1$, the coefficients $A$ and $\tilde{A}$ dominate the scattering for $\om < \om_{\rm min}$, but become small before $\alpha$ and $\beta$ for $\om > \om_{\rm min}$. 
\end{itemize} 

In each asymptotic region $x < 0$ or $x > 0$, the solutions are proportional to $e^{i k_\om x}$, where $k_\om$ is a root of the dispersion relation \eq{eq:disprel}. Equation~\ref{eq:disprel} in general has 4 solutions in $k$ at fixed $\om$. We denote them as $k_1, k_2, k_3 , k_4$. If they are all real, we order them as $k_1 \leq k_2 \leq k_3 \leq k_4$, see \fig{fig:modes}. 
If two of them are real and two are complex, we call $k_1 \leq k_2$ the two real roots, $k_3$ the root with a positive imaginary part, and $k_4$ the root with a negative imaginary part. 
We restrict our attention to these two cases, i.e., to $\om < \om_{\rm max}$, see \eq{eq:om-max-min}.

The modes computed in the two regions are matched at $x = 0$. To derive the matching conditions, it is most convenient to use the variable $\xi$ defined by
\be 
\xi \equiv \int_0^x \frac{dx}{h}.
\ee
instead of $x$. The wave equation \eq{eq:om} then takes the simpler form
\be 
\left(-i \om+\frac{1}{h}\partial _{\xi }\frac{J}{h}\right)\left(-i \om+\frac{J}{h^2}\partial _{\xi }\right)\phi - \frac{g}{h} \partial _{\xi }^2\phi -\frac{g}{3 h}\partial _{\xi }^4\phi = 0,
\ee
where $J \equiv v h$.
The worst singularities are now delta functions from $\partial_\xi$ acting on $h$. So, $\phi$ and its first and second derivatives with respect to $\xi$ are continuous across $\xi = 0$. The discontinuity in $\partial_\xi^3 \phi$ is given by
\be 
\left[\partial _{\xi }^3\phi \right]_{0^-}^{0^+}=3 i \frac{\omega }{g} [v]_{0^-}^{0^+}\phi (0)+3 \left[\frac{v^2}{c^2}\right]_{0^-}^{0^+}\partial _{\xi }\phi (0).
\ee
We consider modes of the form
\be 
\phi (t,x)=e^{-i \omega  t}
\left\lbrace
\begin{array}{ll}
 L_1e^{i k_{1,L}x}+L_2e^{i k_{2,L}x}+L_3e^{i k_{3,L}x}+L_4e^{i k_{4,L}x} & x<0 \\
 R_1e^{i k_{1,R}x}+R_2e^{i k_{2,R}x}+R_3e^{i k_{3,R}x}+R_4e^{i k_{4,R}x} & x>0
\end{array}
\right. ,
\ee
where a subscript $L$ (respectively $R$) indicates a quantity evaluated for $x < 0$ (respectively $x > 0$).
The matching conditions at $x=0$ give a system of 4 linear equations on the coefficients $L_1$, $L_2$, $L_3$, $L_4$, $R_1$, $R_2$, $R_3$, and $R_4$. So, in general there are 4 linearly independent solutions. 
 
We now compute these coefficients, then the Bogoliubov coefficients, for white hole and subcritical flows. We restrict our attention to the left moving incoming mode, with
$\om < \sqrt{3}\frac{v}{h}$ in each region, so that the sign of the group velocity computed with \eq{eq:disprel} agrees with that computed from the full dispersion relation $(\om _ v k)^2 = g k \tanh(hk)$. 
We first assume there is a turning point, i.e.,  $\om_{\rm min} < \om < \om_{\rm max}$. The left moving in mode satisfies 
\be 
\phi _{\text{in},v}
\text{: }
L_1=L_2=L_4=0.
\ee
We find
\be \label{eq:bigformula}
\left\lbrace
\begin{array}{ll}
 R_1=\frac{\left(h_Rk_{2,R}-h_Rk_{3,R}\right)\left(h_Rk_{4,R}-h_Rk_{2,R}\right)+\left(h_Lk_{3,L}-h_Rk_{4,R}\right)\left(h_Lk_{3,L}-h_Rk_{3,R}\right) L_3}{\left(h_Rk_{3,R}-h_Rk_{1,R}\right)\left(h_Rk_{4,R}-h_Rk_{1,R}\right)} R_2 &   \\
 R_3=\frac{\left(h_Rk_{2,R}-h_Rk_{1,R}\right)\left(h_Rk_{4,R}-h_Rk_{2,R}\right)+\left(h_Lk_{3,L}-h_Rk_{4,R}\right)\left(h_Lk_{3,L}-h_Rk_{1,R}\right) L_3}{\left(h_Rk_{1,R}-h_Rk_{3,R}\right)\left(h_Rk_{4,R}-h_Rk_{3,R}\right)} R_2 &   \\
 R_4=\frac{\left(h_Rk_{2,R}-h_Rk_{1,R}\right)\left(h_Rk_{3,R}-h_Rk_{2,R}\right)+\left(h_Lk_{3,L}-h_Rk_{1,R}\right)\left(h_Lk_{3,L}-h_Rk_{3,R}\right) L_3}{\left(h_Rk_{4,R}-h_Rk_{1,R}\right)\left(h_Rk_{4,R}-h_Rk_{3,R}\right)} R_2 &   \\
 L_3=\frac{\left(h_Rk_{2,R}-h_Rk_{1,R}\right)\left(h_Rk_{2,R}-h_Rk_{3,R}\right)\left(h_Rk_{2,R}-h_Rk_{4,R}\right)}{\left(h_Lk_{3,L}-h_Rk_{1,R}\right)\left(h_Lk_{3,L}-h_Rk_{3,R}\right)\left(h_Lk_{3,L}-h_Rk_{4,R}\right)-3 \left[\frac{v^2}{c^2}\right]_{0^-}^{0^+}h_Lk_{3,L}+3 \frac{\omega  }{g}[v]_{0^-}^{0^+}} R_2 &  
\end{array}
\right. .
\ee
When the flow is transcritical, $\om_{\rm min} = 0$ and the limit $\om \rightarrow 0$ can be taken. In this limit, all coefficients remain finite. We denote as $\varphi_{in,v}$ the normalized in mode, and $\varphi_{out,u,1}$, $\varphi_{out,u,3}$, and $\varphi_{out,u,4}$ the three out modes. The numeral denotes the wave whose coefficient is unity for the corresponding naively normalized mode. 
In agreement with \eq{eq:Btrans}, we define the Bogoliubov coefficients as 
\be 
\varphi_{in,v} = \alpha_\om \varphi_{out,u,1} + \beta_\om \varphi_{out,u,4} + A_\om \varphi_{out,u,3}.
\ee
Then,
\be \label{eq:B8}
\alpha_\om &= \sqrt{\left|\frac{\left(\omega -k_{R,1}\right)v_{g,R}\left(k_{R,1}\right)}{\left(\omega -k_{R,2}\right)v_{g,R}\left(k_{R,2}\right)}\right|} R_1, \nn
\beta_\om &= \sqrt{\left|\frac{\left(\omega -k_{R,4}\right)v_{g,R}\left(k_{R,4}\right)}{\left(\omega -k_{R,2}\right)v_{g,R}\left(k_{R,2}\right)}\right|} R_4 , \nn
A_\om &= \sqrt{\left|\frac{\left(\omega -k_{R,3}\right)v_{g,R}\left(k_{R,3}\right)}{\left(\omega -k_{R,2}\right)v_{g,R}\left(k_{R,2}\right)}\right|} R_3 . \nn
\ee
where $v_g$ denotes the corresponding group velocity. Note in particular that $\alpha$ and $\beta$ diverge like $\om^{-1/2}$ in the limit $\om \rightarrow 0$. 
In the transcritical case where $\om_{\rm min} = 0$
the $\om \rightarrow 0$ limit of the effective temperature of \eq{eq:effT} is
\be \label{eq:Tgradino}
T^{\rm step}_{\om = 0}= \frac{\sqrt{3}\left(c_R^2-v_R^2\right)^{3/2}\left(v_L^2-c_L^2\right) h_R}{\left(\frac{v_L^2c_R^2}{c_L^2}-v_Rc_R+v_L\left(c_R-v_R\right)\right)^2\left(c_R+v_R\right)^2 h_L}\left|\sqrt{\frac{c_L^2-v_L^2}{h_L}}+\sqrt{\frac{c_R^2-v_R^2}{ h_R}}\right|^2.
\ee

We now turn to subcritical flows. In this case, there is no turning point for $\om < \om_{\rm min}$. As a result, the left moving in mode is defined by
\be 
\phi _{\text{in},v}
\text{: }
L_1=L_3=L_4=0.
\ee
We find that \eq{eq:bigformula} is modified only through the replacements of $L_3$ by $L_2$ and $k_{3,L}$ by $k_{2,L}$. We define the Bogoliubov coefficients as 
\be 
\varphi_{in,v} = A_\om \varphi_{out,u,3} + \tilde{A}_\om \varphi_{out,v} + \alpha_\om \varphi_{out,u,1} + \beta_\om \varphi_{out,u,4}.
\ee
Taking the normalization into account, $\om \rightarrow 0$, the effective temperature behaves as
\be 
T^{\rm step}_\om = -\frac{\om}{\ln \lp \frac{\om}{\om_b} \rp} \lp 1+ \mathcal{O} \lp \frac{\om}{\om_b} \rp \rp,
\ee
where 
\be \label{eq:omb}
\om_b = \left(\left(\frac{h_L}{v_L-c_L}-\frac{h_R}{v_R+c_R}\right)\frac{ c_R}{c_R+c_L}+\frac{h_Rc_R}{c_R^2-v_R^2}\right)^{-2}\frac{\sqrt{3} h_R}{\left(c_R^2-v_R^2\right)^{1/2}}.
\ee
The limiting values of $A$ and $\tilde{A}$ for $\om \rightarrow 0$ are 
\be 
A \mathop{\rightarrow}_{\om \rightarrow 0} \frac{c_R - c_L}{c_R + c_L}, \quad \tilde{A} \mathop{\rightarrow}_{\om \rightarrow 0} \frac{2 \sqrt{c_R c_L}}{c_R + c_L}.
\ee
As found for smooth flows in the body of the text, the hydrodynamical sectors dominate the scattering as $\left\lvert A_\om \right\rvert^2 + \left\lvert A_\om \right\rvert^2  \to 1$ for $\om \to 0$, while $\alpha_\om$ and $\beta_\om$ both go to zero like $\om^{1/2}$. The only important difference is that for a steplike discontinuity $\tilde{A}_\om$ does not vanish at $\om = \om_{\rm min}$, even though it still vanishes above $\om_{\rm min}$.

\begin{figure}
\begin{center}
\includegraphics[scale=1.0]{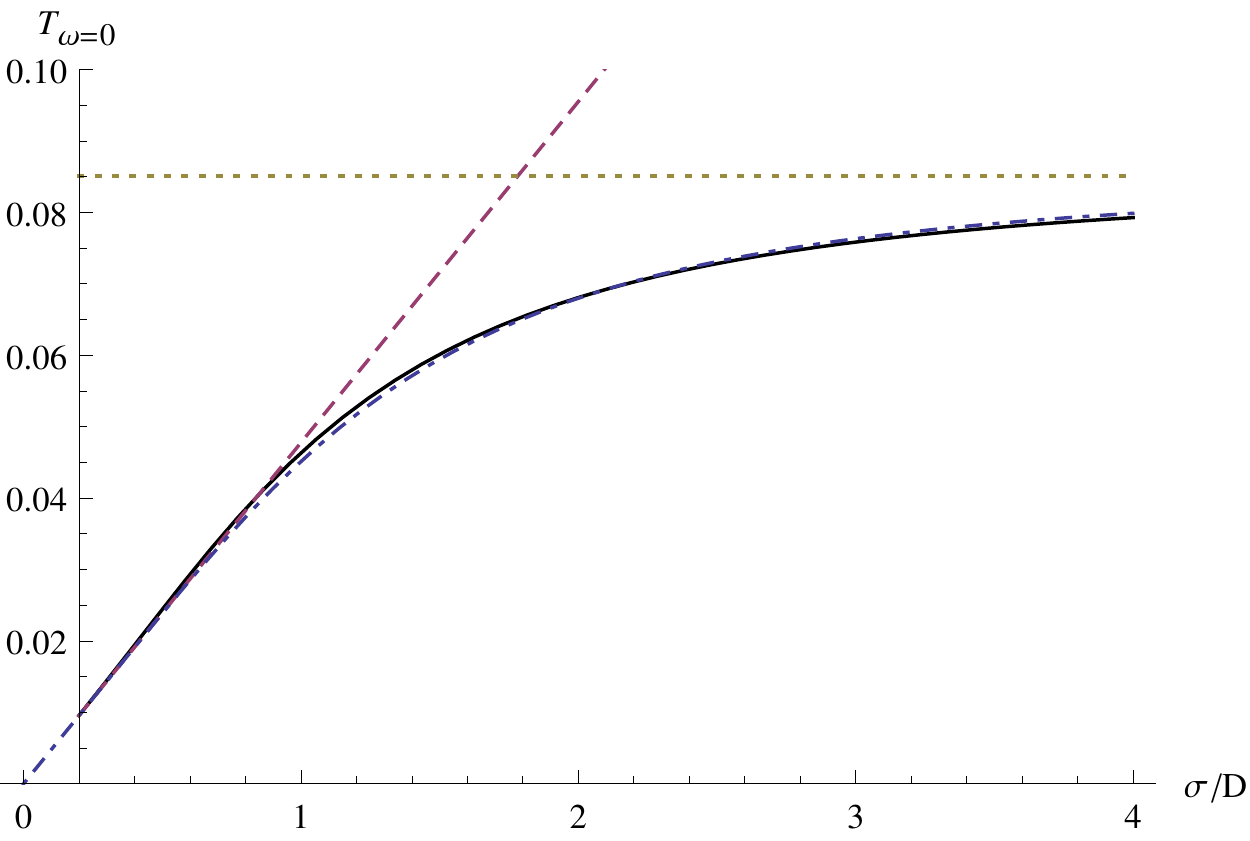}
\end{center}
\caption{Limit $\om \rightarrow 0$ of the effective temperature (solid), Hawking temperature (dashed), and steplike result of \eq{eq:Tgradino} (dotted) as functions of $\sigma$, for $g=J=1$, $h_0=1$, and $D=0.2$. The dot-dashed line shows the prediction \eq{eq:Scott_interpolation}. Note the very good agreement between the exact numerical calculation and \eq{eq:Scott_interpolation}.} \label{fig:Scott}
\end{figure}

To complete the analysis, we studied the transition between the Hawking regime when the surface gravity is small enough in units of the dispersive scale, and the steplike regime studied above, for $\om \rightarrow 0$. 
Although we work in a slightly different case since the ordering of $h(x)$ and $\partial_x$ in the dispersive term 
of the wave equation is different from that of our \eq{eq:om}, we found a very good agreement with the formula given in \cite{Scott-thesis}:
\be \label{eq:Scott_interpolation}
T_{\om = 0} \approx T_H \, \tanh \lp \frac{T^{\rm step}_{\om = 0} }{T_H} \rp,
\ee
where $T_{\om= 0}$ is the zero-frequency limit of the temperature of \eq{eq:effT} numerically evaluated in the
smooth flow, $T_H$ is the corresponding Hawking temperature, 
and $T^{\rm step}_{\om = 0}$ the 
zero-frequency limit of the
temperature for the corresponding steplike profile, given by \eq{eq:Tgradino}. The agreement is illustrated in \fig{fig:Scott}.

\bibliographystyle{apsrev4-1}
\bibliography{biblio}

\begin{thebibliography}{35}%
\makeatletter
\providecommand \@ifxundefined [1]{%
 \@ifx{#1\undefined}
}%
\providecommand \@ifnum [1]{%
 \ifnum #1\expandafter \@firstoftwo
 \else \expandafter \@secondoftwo
 \fi
}%
\providecommand \@ifx [1]{%
 \ifx #1\expandafter \@firstoftwo
 \else \expandafter \@secondoftwo
 \fi
}%
\providecommand \natexlab [1]{#1}%
\providecommand \enquote  [1]{``#1''}%
\providecommand \bibnamefont  [1]{#1}%
\providecommand \bibfnamefont [1]{#1}%
\providecommand \citenamefont [1]{#1}%
\providecommand \href@noop [0]{\@secondoftwo}%
\providecommand \href [0]{\begingroup \@sanitize@url \@href}%
\providecommand \@href[1]{\@@startlink{#1}\@@href}%
\providecommand \@@href[1]{\endgroup#1\@@endlink}%
\providecommand \@sanitize@url [0]{\catcode `\\12\catcode `\$12\catcode
  `\&12\catcode `\#12\catcode `\^12\catcode `\_12\catcode `\%12\relax}%
\providecommand \@@startlink[1]{}%
\providecommand \@@endlink[0]{}%
\providecommand \url  [0]{\begingroup\@sanitize@url \@url }%
\providecommand \@url [1]{\endgroup\@href {#1}{\urlprefix }}%
\providecommand \urlprefix  [0]{URL }%
\providecommand \Eprint [0]{\href }%
\providecommand \doibase [0]{http://dx.doi.org/}%
\providecommand \selectlanguage [0]{\@gobble}%
\providecommand \bibinfo  [0]{\@secondoftwo}%
\providecommand \bibfield  [0]{\@secondoftwo}%
\providecommand \translation [1]{[#1]}%
\providecommand \BibitemOpen [0]{}%
\providecommand \bibitemStop [0]{}%
\providecommand \bibitemNoStop [0]{.\EOS\space}%
\providecommand \EOS [0]{\spacefactor3000\relax}%
\providecommand \BibitemShut  [1]{\csname bibitem#1\endcsname}%
\let\auto@bib@innerbib\@empty
\bibitem [{\citenamefont {Unruh}(1981)}]{Unruhprl81}%
  \BibitemOpen
  \bibfield  {author} {\bibinfo {author} {\bibfnamefont {W.~G.}\ \bibnamefont
  {Unruh}},\ }\href {\doibase 10.1103/PhysRevLett.46.1351} {\bibfield
  {journal} {\bibinfo  {journal} {Phys. Rev. Lett.}\ }\textbf {\bibinfo
  {volume} {46}},\ \bibinfo {pages} {1351} (\bibinfo {year}
  {1981})}\BibitemShut {NoStop}%
\bibitem [{\citenamefont {Hawking}(1975)}]{Hawking75}%
  \BibitemOpen
  \bibfield  {author} {\bibinfo {author} {\bibfnamefont {S.}~\bibnamefont
  {Hawking}},\ }\href {\doibase 10.1007/BF02345020} {\bibfield  {journal}
  {\bibinfo  {journal} {Commun.Math.Phys.}\ }\textbf {\bibinfo {volume} {43}},\
  \bibinfo {pages} {199} (\bibinfo {year} {1975})}\BibitemShut {NoStop}%
\bibitem [{\citenamefont {Jacobson}(1991)}]{Jacobson-prd91}%
  \BibitemOpen
  \bibfield  {author} {\bibinfo {author} {\bibfnamefont {T.}~\bibnamefont
  {Jacobson}},\ }\href {\doibase 10.1103/PhysRevD.44.1731} {\bibfield
  {journal} {\bibinfo  {journal} {Phys. Rev. D}\ }\textbf {\bibinfo {volume}
  {44}},\ \bibinfo {pages} {1731} (\bibinfo {year} {1991})}\BibitemShut
  {NoStop}%
\bibitem [{\citenamefont {Unruh}(1995)}]{Unruhprd95}%
  \BibitemOpen
  \bibfield  {author} {\bibinfo {author} {\bibfnamefont {W.~G.}\ \bibnamefont
  {Unruh}},\ }\href {\doibase 10.1103/PhysRevD.51.2827} {\bibfield  {journal}
  {\bibinfo  {journal} {Phys. Rev. D}\ }\textbf {\bibinfo {volume} {51}},\
  \bibinfo {pages} {2827} (\bibinfo {year} {1995})}\BibitemShut {NoStop}%
\bibitem [{\citenamefont {Sch\"utzhold}\ and\ \citenamefont
  {Unruh}(2002)}]{Schutzhold-U_2002}%
  \BibitemOpen
  \bibfield  {author} {\bibinfo {author} {\bibfnamefont {R.}~\bibnamefont
  {Sch\"utzhold}}\ and\ \bibinfo {author} {\bibfnamefont {W.~G.}\ \bibnamefont
  {Unruh}},\ }\href {\doibase 10.1103/PhysRevD.66.044019} {\bibfield  {journal}
  {\bibinfo  {journal} {Phys. Rev. D}\ }\textbf {\bibinfo {volume} {66}},\
  \bibinfo {pages} {044019} (\bibinfo {year} {2002})}\BibitemShut {NoStop}%
\bibitem [{\citenamefont {Rousseaux}\ \emph {et~al.}(2008)\citenamefont
  {Rousseaux}, \citenamefont {Mathis}, \citenamefont {Maissa}, \citenamefont
  {Philbin},\ and\ \citenamefont {Leonhardt}}]{Nice}%
  \BibitemOpen
  \bibfield  {author} {\bibinfo {author} {\bibfnamefont {G.}~\bibnamefont
  {Rousseaux}}, \bibinfo {author} {\bibfnamefont {C.}~\bibnamefont {Mathis}},
  \bibinfo {author} {\bibfnamefont {P.}~\bibnamefont {Maissa}}, \bibinfo
  {author} {\bibfnamefont {T.~G.}\ \bibnamefont {Philbin}}, \ and\ \bibinfo
  {author} {\bibfnamefont {U.}~\bibnamefont {Leonhardt}},\ }\href {\doibase
  10.1088/1367-2630/10/5/053015} {\bibfield  {journal} {\bibinfo  {journal}
  {New J.Phys.}\ }\textbf {\bibinfo {volume} {10}},\ \bibinfo {pages} {053015}
  (\bibinfo {year} {2008})}\BibitemShut {NoStop}%
\bibitem [{\citenamefont {Weinfurtner}\ \emph {et~al.}(2011)\citenamefont
  {Weinfurtner}, \citenamefont {Tedford}, \citenamefont {Penrice},
  \citenamefont {Unruh},\ and\ \citenamefont {Lawrence}}]{Unruh2010}%
  \BibitemOpen
  \bibfield  {author} {\bibinfo {author} {\bibfnamefont {S.}~\bibnamefont
  {Weinfurtner}}, \bibinfo {author} {\bibfnamefont {E.~W.}\ \bibnamefont
  {Tedford}}, \bibinfo {author} {\bibfnamefont {M.~C.~J.}\ \bibnamefont
  {Penrice}}, \bibinfo {author} {\bibfnamefont {W.~G.}\ \bibnamefont {Unruh}},
  \ and\ \bibinfo {author} {\bibfnamefont {G.~A.}\ \bibnamefont {Lawrence}},\
  }\href {\doibase 10.1103/PhysRevLett.106.021302} {\bibfield  {journal}
  {\bibinfo  {journal} {Phys. Rev. Lett.}\ }\textbf {\bibinfo {volume} {106}},\
  \bibinfo {pages} {021302} (\bibinfo {year} {2011})}\BibitemShut {NoStop}%
\bibitem [{\citenamefont {Brout}\ \emph
  {et~al.}(1995{\natexlab{a}})\citenamefont {Brout}, \citenamefont {Massar},
  \citenamefont {Parentani},\ and\ \citenamefont {Spindel}}]{Broutetal95}%
  \BibitemOpen
  \bibfield  {author} {\bibinfo {author} {\bibfnamefont {R.}~\bibnamefont
  {Brout}}, \bibinfo {author} {\bibfnamefont {S.}~\bibnamefont {Massar}},
  \bibinfo {author} {\bibfnamefont {R.}~\bibnamefont {Parentani}}, \ and\
  \bibinfo {author} {\bibfnamefont {P.}~\bibnamefont {Spindel}},\ }\href
  {\doibase 10.1103/PhysRevD.52.4559} {\bibfield  {journal} {\bibinfo
  {journal} {Phys. Rev. D}\ }\textbf {\bibinfo {volume} {52}},\ \bibinfo
  {pages} {4559} (\bibinfo {year} {1995}{\natexlab{a}})}\BibitemShut {NoStop}%
\bibitem [{\citenamefont {Corley}\ and\ \citenamefont
  {Jacobson}(1996)}]{CorleyJac96}%
  \BibitemOpen
  \bibfield  {author} {\bibinfo {author} {\bibfnamefont {S.}~\bibnamefont
  {Corley}}\ and\ \bibinfo {author} {\bibfnamefont {T.}~\bibnamefont
  {Jacobson}},\ }\href {\doibase 10.1103/PhysRevD.54.1568} {\bibfield
  {journal} {\bibinfo  {journal} {Phys. Rev. D}\ }\textbf {\bibinfo {volume}
  {54}},\ \bibinfo {pages} {1568} (\bibinfo {year} {1996})}\BibitemShut
  {NoStop}%
\bibitem [{\citenamefont {Corley}(1998)}]{Corley97}%
  \BibitemOpen
  \bibfield  {author} {\bibinfo {author} {\bibfnamefont {S.}~\bibnamefont
  {Corley}},\ }\href {\doibase 10.1103/PhysRevD.57.6280} {\bibfield  {journal}
  {\bibinfo  {journal} {Phys. Rev. D}\ }\textbf {\bibinfo {volume} {57}},\
  \bibinfo {pages} {6280} (\bibinfo {year} {1998})}\BibitemShut {NoStop}%
\bibitem [{\citenamefont {Balbinot}\ \emph {et~al.}(2005)\citenamefont
  {Balbinot}, \citenamefont {Fabbri}, \citenamefont {Fagnocchi},\ and\
  \citenamefont {Parentani}}]{Rivista05}%
  \BibitemOpen
  \bibfield  {author} {\bibinfo {author} {\bibfnamefont {R.}~\bibnamefont
  {Balbinot}}, \bibinfo {author} {\bibfnamefont {A.}~\bibnamefont {Fabbri}},
  \bibinfo {author} {\bibfnamefont {S.}~\bibnamefont {Fagnocchi}}, \ and\
  \bibinfo {author} {\bibfnamefont {R.}~\bibnamefont {Parentani}},\ }\href@noop
  {} {\bibfield  {journal} {\bibinfo  {journal} {Riv.Nuovo Cim.}\ }\textbf
  {\bibinfo {volume} {28}},\ \bibinfo {pages} {1} (\bibinfo {year}
  {2005})}\BibitemShut {NoStop}%
\bibitem [{\citenamefont {Macher}\ and\ \citenamefont
  {Parentani}(2009)}]{MacherB/W}%
  \BibitemOpen
  \bibfield  {author} {\bibinfo {author} {\bibfnamefont {J.}~\bibnamefont
  {Macher}}\ and\ \bibinfo {author} {\bibfnamefont {R.}~\bibnamefont
  {Parentani}},\ }\href {\doibase 10.1103/PhysRevD.79.124008} {\bibfield
  {journal} {\bibinfo  {journal} {Phys. Rev. D}\ }\textbf {\bibinfo {volume}
  {79}},\ \bibinfo {pages} {124008} (\bibinfo {year} {2009})}\BibitemShut
  {NoStop}%
\bibitem [{\citenamefont {Robertson}(2012)}]{Scott-review}%
  \BibitemOpen
  \bibfield  {author} {\bibinfo {author} {\bibfnamefont {S.~J.}\ \bibnamefont
  {Robertson}},\ }\href {\doibase 10.1088/0953-4075/45/16/163001} {\bibfield
  {journal} {\bibinfo  {journal} {J. Phys. B}\ }\textbf {\bibinfo {volume}
  {45}},\ \bibinfo {pages} {163001} (\bibinfo {year} {2012})}\BibitemShut
  {NoStop}%
\bibitem [{\citenamefont {Finazzi}\ and\ \citenamefont
  {Parentani}(2011{\natexlab{a}})}]{finazziRP-proceedings}%
  \BibitemOpen
  \bibfield  {author} {\bibinfo {author} {\bibfnamefont {S.}~\bibnamefont
  {Finazzi}}\ and\ \bibinfo {author} {\bibfnamefont {R.}~\bibnamefont
  {Parentani}},\ }\href {\doibase 10.1088/1742-6596/314/1/012030} {\bibfield
  {journal} {\bibinfo  {journal} {J.Phys.Conf.Ser.}\ }\textbf {\bibinfo
  {volume} {314}},\ \bibinfo {pages} {012030} (\bibinfo {year}
  {2011}{\natexlab{a}})}\BibitemShut {NoStop}%
\bibitem [{\citenamefont {Coutant}\ \emph {et~al.}(2012)\citenamefont
  {Coutant}, \citenamefont {Parentani},\ and\ \citenamefont
  {Finazzi}}]{ACRPFS}%
  \BibitemOpen
  \bibfield  {author} {\bibinfo {author} {\bibfnamefont {A.}~\bibnamefont
  {Coutant}}, \bibinfo {author} {\bibfnamefont {R.}~\bibnamefont {Parentani}},
  \ and\ \bibinfo {author} {\bibfnamefont {S.}~\bibnamefont {Finazzi}},\ }\href
  {\doibase 10.1103/PhysRevD.85.024021} {\bibfield  {journal} {\bibinfo
  {journal} {Phys. Rev. D}\ }\textbf {\bibinfo {volume} {85}},\ \bibinfo
  {pages} {024021} (\bibinfo {year} {2012})}\BibitemShut {NoStop}%
\bibitem [{\citenamefont {{Robertson}}(2011)}]{Scott-thesis}%
  \BibitemOpen
  \bibfield  {author} {\bibinfo {author} {\bibfnamefont {S.~J.}\ \bibnamefont
  {{Robertson}}},\ }\emph {\bibinfo {title} {{Hawking Radiation in Dispersive
  Media}}},\ \href@noop {} {Ph.D. thesis} (\bibinfo {year} {2011}),\ \Eprint
  {http://arxiv.org/abs/1106.1805} {arXiv:1106.1805 [gr-qc]} \BibitemShut
  {NoStop}%
\bibitem [{\citenamefont {Finazzi}\ and\ \citenamefont
  {Parentani}(2012)}]{2regimesFinazzi}%
  \BibitemOpen
  \bibfield  {author} {\bibinfo {author} {\bibfnamefont {S.}~\bibnamefont
  {Finazzi}}\ and\ \bibinfo {author} {\bibfnamefont {R.}~\bibnamefont
  {Parentani}},\ }\href {\doibase 10.1103/PhysRevD.85.124027} {\bibfield
  {journal} {\bibinfo  {journal} {Phys. Rev. D}\ }\textbf {\bibinfo {volume}
  {85}},\ \bibinfo {pages} {124027} (\bibinfo {year} {2012})}\BibitemShut
  {NoStop}%
\bibitem [{\citenamefont {Coutant}\ and\ \citenamefont
  {Parentani}(2014{\natexlab{a}})}]{Coutant_on_Undulations}%
  \BibitemOpen
  \bibfield  {author} {\bibinfo {author} {\bibfnamefont {A.}~\bibnamefont
  {Coutant}}\ and\ \bibinfo {author} {\bibfnamefont {R.}~\bibnamefont
  {Parentani}},\ }\href {\doibase 10.1063/1.4872025} {\bibfield  {journal}
  {\bibinfo  {journal} {Phys.Fluids}\ }\textbf {\bibinfo {volume} {26}},\
  \bibinfo {pages} {044106} (\bibinfo {year} {2014}{\natexlab{a}})}\BibitemShut
  {NoStop}%
\bibitem [{\citenamefont {Rousseaux}(2013)}]{Germain5}%
  \BibitemOpen
  \bibfield  {author} {\bibinfo {author} {\bibfnamefont {G.}~\bibnamefont
  {Rousseaux}},\ }\enquote {\bibinfo {title} {Analogue gravity
  phenomenology},}\ \ (\bibinfo  {publisher} {Springer},\ \bibinfo {year}
  {2013})\ Chap.~\bibinfo {chapter} {5}\BibitemShut {NoStop}%
\bibitem [{\citenamefont {Chaline}\ \emph {et~al.}(2013)\citenamefont
  {Chaline}, \citenamefont {Jannes}, \citenamefont {Maïssa},\ and\
  \citenamefont {Rousseaux}}]{Germain7}%
  \BibitemOpen
  \bibfield  {author} {\bibinfo {author} {\bibfnamefont {J.}~\bibnamefont
  {Chaline}}, \bibinfo {author} {\bibfnamefont {G.}~\bibnamefont {Jannes}},
  \bibinfo {author} {\bibfnamefont {P.}~\bibnamefont {Maïssa}}, \ and\
  \bibinfo {author} {\bibfnamefont {G.}~\bibnamefont {Rousseaux}},\ }\enquote
  {\bibinfo {title} {Analogue gravity phenomenology},}\ \ (\bibinfo
  {publisher} {Springer},\ \bibinfo {year} {2013})\ Chap.~\bibinfo {chapter}
  {7}\BibitemShut {NoStop}%
\bibitem [{\citenamefont {Unruh}(2013)}]{Unruh2012}%
  \BibitemOpen
  \bibfield  {author} {\bibinfo {author} {\bibfnamefont {W.~G.}\ \bibnamefont
  {Unruh}},\ }\href {\doibase 10.1007/978-3-319-00266-8_4} {\bibfield
  {journal} {\bibinfo  {journal} {Lect.Notes Phys.}\ }\textbf {\bibinfo
  {volume} {870}},\ \bibinfo {pages} {63} (\bibinfo {year} {2013})}\BibitemShut
  {NoStop}%
\bibitem [{\citenamefont {Coutant}\ and\ \citenamefont
  {Parentani}(2014{\natexlab{b}})}]{2014arXiv1402.2514C}%
  \BibitemOpen
  \bibfield  {author} {\bibinfo {author} {\bibfnamefont {A.}~\bibnamefont
  {Coutant}}\ and\ \bibinfo {author} {\bibfnamefont {R.}~\bibnamefont
  {Parentani}},\ }\href@noop {} {\  (\bibinfo {year} {2014}{\natexlab{b}})},\
  \Eprint {http://arxiv.org/abs/1402.2514} {arXiv:1402.2514 [gr-qc]}
  \BibitemShut {NoStop}%
\bibitem [{\citenamefont {{Acheson}}(1976)}]{OnOverReflection}%
  \BibitemOpen
  \bibfield  {author} {\bibinfo {author} {\bibfnamefont {D.~J.}\ \bibnamefont
  {{Acheson}}},\ }\href {\doibase 10.1017/S0022112076002206} {\bibfield
  {journal} {\bibinfo  {journal} {Journal of Fluid Mechanics}\ }\textbf
  {\bibinfo {volume} {77}},\ \bibinfo {pages} {433} (\bibinfo {year}
  {1976})}\BibitemShut {NoStop}%
\bibitem [{\citenamefont {Richartz}\ \emph {et~al.}(2009)\citenamefont
  {Richartz}, \citenamefont {Weinfurtner}, \citenamefont {Penner},\ and\
  \citenamefont {Unruh}}]{GeneralSRR}%
  \BibitemOpen
  \bibfield  {author} {\bibinfo {author} {\bibfnamefont {M.}~\bibnamefont
  {Richartz}}, \bibinfo {author} {\bibfnamefont {S.}~\bibnamefont
  {Weinfurtner}}, \bibinfo {author} {\bibfnamefont {A.~J.}\ \bibnamefont
  {Penner}}, \ and\ \bibinfo {author} {\bibfnamefont {W.~G.}\ \bibnamefont
  {Unruh}},\ }\href {\doibase 10.1103/PhysRevD.80.124016} {\bibfield  {journal}
  {\bibinfo  {journal} {Phys.Rev.D}\ }\textbf {\bibinfo {volume} {80}},\
  \bibinfo {pages} {124016} (\bibinfo {year} {2009})}\BibitemShut {NoStop}%
\bibitem [{\citenamefont {Page}(1976)}]{Page76}%
  \BibitemOpen
  \bibfield  {author} {\bibinfo {author} {\bibfnamefont {D.~N.}\ \bibnamefont
  {Page}},\ }\href {\doibase 10.1103/PhysRevD.13.198} {\bibfield  {journal}
  {\bibinfo  {journal} {Phys. Rev. D}\ }\textbf {\bibinfo {volume} {13}},\
  \bibinfo {pages} {198} (\bibinfo {year} {1976})}\BibitemShut {NoStop}%
\bibitem [{\citenamefont {Anderson}\ \emph {et~al.}(2014)\citenamefont
  {Anderson}, \citenamefont {Balbinot}, \citenamefont {Fabbri},\ and\
  \citenamefont {Parentani}}]{Sandro_2014}%
  \BibitemOpen
  \bibfield  {author} {\bibinfo {author} {\bibfnamefont {P.}~\bibnamefont
  {Anderson}}, \bibinfo {author} {\bibfnamefont {R.}~\bibnamefont {Balbinot}},
  \bibinfo {author} {\bibfnamefont {A.}~\bibnamefont {Fabbri}}, \ and\ \bibinfo
  {author} {\bibfnamefont {R.}~\bibnamefont {Parentani}},\ }\href@noop {} {\
  (\bibinfo {year} {2014})},\ \Eprint {http://arxiv.org/abs/1404.3224}
  {arXiv:1404.3224 [gr-qc]} \BibitemShut {NoStop}%
\bibitem [{\citenamefont {Massar}\ and\ \citenamefont
  {Parentani}(1998)}]{Massar:1997en}%
  \BibitemOpen
  \bibfield  {author} {\bibinfo {author} {\bibfnamefont {S.}~\bibnamefont
  {Massar}}\ and\ \bibinfo {author} {\bibfnamefont {R.}~\bibnamefont
  {Parentani}},\ }\href {\doibase 10.1016/S0550-3213(97)00718-9} {\bibfield
  {journal} {\bibinfo  {journal} {Nucl.Phys.}\ }\textbf {\bibinfo {volume}
  {B513}},\ \bibinfo {pages} {375} (\bibinfo {year} {1998})}\BibitemShut
  {NoStop}%
\bibitem [{\citenamefont {Barut}(1963)}]{PhysRev.130.436}%
  \BibitemOpen
  \bibfield  {author} {\bibinfo {author} {\bibfnamefont {A.~O.}\ \bibnamefont
  {Barut}},\ }\href {\doibase 10.1103/PhysRev.130.436} {\bibfield  {journal}
  {\bibinfo  {journal} {Phys. Rev.}\ }\textbf {\bibinfo {volume} {130}},\
  \bibinfo {pages} {436} (\bibinfo {year} {1963})}\BibitemShut {NoStop}%
\bibitem [{\citenamefont {Brout}\ \emph
  {et~al.}(1995{\natexlab{b}})\citenamefont {Brout}, \citenamefont {Maissar},
  \citenamefont {Parentani},\ and\ \citenamefont {Spindel}}]{Primer}%
  \BibitemOpen
  \bibfield  {author} {\bibinfo {author} {\bibfnamefont {R.}~\bibnamefont
  {Brout}}, \bibinfo {author} {\bibfnamefont {S.}~\bibnamefont {Maissar}},
  \bibinfo {author} {\bibfnamefont {R.}~\bibnamefont {Parentani}}, \ and\
  \bibinfo {author} {\bibfnamefont {P.}~\bibnamefont {Spindel}},\ }\href
  {\doibase 10.1016/0370-1573(95)00008-5} {\bibfield  {journal} {\bibinfo
  {journal} {Phys.Rep.}\ }\textbf {\bibinfo {volume} {260}},\ \bibinfo {pages}
  {329} (\bibinfo {year} {1995}{\natexlab{b}})}\BibitemShut {NoStop}%
\bibitem [{\citenamefont {Zapata}\ \emph {et~al.}(2011)\citenamefont {Zapata},
  \citenamefont {Albert}, \citenamefont {Parentani},\ and\ \citenamefont
  {Sols}}]{ZP11}%
  \BibitemOpen
  \bibfield  {author} {\bibinfo {author} {\bibfnamefont {I.}~\bibnamefont
  {Zapata}}, \bibinfo {author} {\bibfnamefont {M.}~\bibnamefont {Albert}},
  \bibinfo {author} {\bibfnamefont {R.}~\bibnamefont {Parentani}}, \ and\
  \bibinfo {author} {\bibfnamefont {F.}~\bibnamefont {Sols}},\ }\href {\doibase
  10.1088/1367-2630/13/6/063048} {\bibfield  {journal} {\bibinfo  {journal}
  {New J.Phys.}\ }\textbf {\bibinfo {volume} {13}},\ \bibinfo {pages} {063048}
  (\bibinfo {year} {2011})}\BibitemShut {NoStop}%
\bibitem [{\citenamefont {Finazzi}\ and\ \citenamefont
  {Parentani}(2011{\natexlab{b}})}]{FPBroad}%
  \BibitemOpen
  \bibfield  {author} {\bibinfo {author} {\bibfnamefont {S.}~\bibnamefont
  {Finazzi}}\ and\ \bibinfo {author} {\bibfnamefont {R.}~\bibnamefont
  {Parentani}},\ }\href {\doibase 10.1103/PhysRevD.83.084010} {\bibfield
  {journal} {\bibinfo  {journal} {Phys.Rev.D}\ }\textbf {\bibinfo {volume}
  {83}},\ \bibinfo {pages} {084010} (\bibinfo {year}
  {2011}{\natexlab{b}})}\BibitemShut {NoStop}%
\bibitem [{\citenamefont {Unruh}(2014)}]{Unruh2014}%
  \BibitemOpen
  \bibfield  {author} {\bibinfo {author} {\bibfnamefont {W.}~\bibnamefont
  {Unruh}},\ }\href {\doibase 10.1007/s10701-014-9778-0} {\bibfield  {journal}
  {\bibinfo  {journal} {Found.Phys.}\ }\textbf {\bibinfo {volume} {44}},\
  \bibinfo {pages} {532} (\bibinfo {year} {2014})}\BibitemShut {NoStop}%
\bibitem [{\citenamefont {Jannes}\ \emph {et~al.}(2011)\citenamefont {Jannes},
  \citenamefont {Piquet}, \citenamefont {Maïssa}, \citenamefont {Mathis},\
  and\ \citenamefont {Rousseaux}}]{PhysRevE.83.056312}%
  \BibitemOpen
  \bibfield  {author} {\bibinfo {author} {\bibfnamefont {G.}~\bibnamefont
  {Jannes}}, \bibinfo {author} {\bibfnamefont {R.}~\bibnamefont {Piquet}},
  \bibinfo {author} {\bibfnamefont {P.}~\bibnamefont {Maïssa}}, \bibinfo
  {author} {\bibfnamefont {C.}~\bibnamefont {Mathis}}, \ and\ \bibinfo {author}
  {\bibfnamefont {G.}~\bibnamefont {Rousseaux}},\ }\href {\doibase
  10.1103/PhysRevE.83.056312} {\bibfield  {journal} {\bibinfo  {journal} {Phys.
  Rev. E}\ }\textbf {\bibinfo {volume} {83}},\ \bibinfo {pages} {056312}
  (\bibinfo {year} {2011})}\BibitemShut {NoStop}%
\bibitem [{\citenamefont {Landau}\ and\ \citenamefont
  {Lifschitz}(1959)}]{LLhydro}%
  \BibitemOpen
  \bibfield  {author} {\bibinfo {author} {\bibfnamefont {L.}~\bibnamefont
  {Landau}}\ and\ \bibinfo {author} {\bibfnamefont {E.}~\bibnamefont
  {Lifschitz}},\ }\href@noop {} {\emph {\bibinfo {title} {Fluid Mechanics}}}\
  (\bibinfo  {publisher} {Pergamon Press},\ \bibinfo {year} {1959})\BibitemShut
  {NoStop}%
\bibitem [{\citenamefont {Batchelor}(1967)}]{Batchelor}%
  \BibitemOpen
  \bibfield  {author} {\bibinfo {author} {\bibfnamefont {G.}~\bibnamefont
  {Batchelor}},\ }\href@noop {} {\emph {\bibinfo {title} {An Introduction to
  Fluid Mechanics}}}\ (\bibinfo  {publisher} {Cambridge University Press},\
  \bibinfo {year} {1967})\BibitemShut {NoStop}%
\end{thebibliography}%

\end{document}